\documentclass[12pt]{article}
\usepackage{epsf}
\usepackage{graphicx}
\usepackage{amssymb}
\def\uu{\langle \bar u u \rangle}
\def\dd{\langle \bar d d \rangle}
\def\ss{\langle \bar s s \rangle}
\def\qq{\langle \bar q q \rangle}

\title{ Radiative Decays of the  Heavy Flavored  Baryons  in  Light Cone QCD Sum Rules }
\author{
  T. M. Aliev \footnote{Permanent address: Institute of Physics, Baku,
 Azerbaijan} \thanks{ e-mail: taliev@metu.edu.tr},
  K. Azizi\thanks {e-mail:e146342@metu.edu.tr},
  A. Ozpineci \thanks {e-mail:ozpineci@metu.edu.tr} \\
  \small Physics
Department, Middle East Technical University, 06531, Ankara, Turkey }
 \date{}
\begin{document}
\setlength{\baselineskip}{26pt} \maketitle
\setlength{\baselineskip}{7mm}
\begin{abstract}
The transition magnetic dipole and electric quadrupole   moments of  the  radiative decays of the sextet heavy flavored spin $\frac{3}{2}$  to the heavy spin $\frac{1}{2}$ baryons  are calculated  within the light cone QCD
  sum rules approach. Using the obtained results, the decay rate for these  transitions are also computed and compared with the existing predictions of the other approaches.
\end{abstract}
\thispagestyle{empty}
\newpage
\setcounter{page}{1}
\section{Introduction}
In the last years, considerable experimental progress has been made
in the identification  of new bottom baryon states. The CDF
Collaboration \cite{Aaltonen1} reported the first observation of the
$\Sigma^{\pm}_{b}$ and $\Sigma^{\ast\pm}_{b}$. After this
discovery, the DO \cite{Abazov} and CDF \cite{Aaltonen2}
Collaborations announced the observation of the  $\Xi_{b^{-}}$ state
(for a review see \cite{dort}). The BaBar Collaboration discovered the $\Omega^{\ast}_{c}$ state \cite{Aubert} and observed
the $\Omega_{c}$ state in B decays  
\cite{alti}. Lastly, the BaBar and BELLE Collaborations observed $\Xi_{c}$
state \cite{yedi}.

Heavy baryons containing c and b quarks have been subject of the
intensive theoretical studies (see \cite{sekiz} and references
therein). A theoretical  study of experimental results  can give
essential information about the structure of these baryons. In this
sense, study of the electromagnetic properties of the heavy baryons
receives special attention. One of the main static electromagnetic
quantities of the baryons is their magnetic moments. The magnetic
moments of the heavy baryons have been discussed  within different
approaches in the literature (see \cite{kazem1} and references
therein).

In the present work, we investigate  the electromagnetic decays of
the ground state heavy baryons containing single heavy quark with total angular momentum
$J=\frac{3}{2}$ to the heavy baryons with $J=\frac{1}{2}$
 in the  framework of the light cone
QCD sum rules method. Note that, some of the considered decays have
been studied  within heavy hadron chiral perturbation theory
\cite{on}, heavy quark and chiral symmetries \cite{onbir,oniki}, in
the relativistic quark model \cite{onuc} and light cone QCD sum
rules at leading order in HQET in \cite{ondort}. Here, we also
emphasize that the radiative decays of the light decuplet baryons to the
octet baryons have been studied in the framework of the light cone
QCD sum rules in  \cite{onbes}.

The work is organized as follows. In section 2, the light cone QCD
sum rules for the   form factors describing electromagnetic
transition of the heavy $J=\frac{3}{2}$, to the heavy baryons with
$J=\frac{1}{2}$  are  obtained.  Section 3 encompasses the numerical
analysis of the transition magnetic dipole and electric quadrupole
moments as well  as the radiative decay rates. A    comparison of our
results for the total decay width with the existing predictions of
the other approaches  is also presented in section 3.
\section{Light cone QCD sum rules for the electromagnetic form factors  of the heavy flavored  baryons  }
We start this section with a few remarks about the classification of the
heavy baryons.   Heavy baryons with single
heavy quark belong to either SU(3) antisymmetric $\bar3_{F}$ or
symmetric $6_{F}$ flavor representations. For the baryons containing single heavy quark, in the
$m_{Q}\rightarrow \infty$ limit, the angular momentum of the light
quarks is a good quantum number. The spin of the
 light diquark is either $S=1$ for $6_{F}$ or $S=0$ for
$\bar3_{F}$. The ground state will have angular momentum $l=0$. Therefore, the  spin of the ground state  is $1/2$ for
$\bar3_{F}$ representing the $\Lambda_{Q}$ and $\Xi_{Q}$ baryons,
while it can be both $3/2$ or $1/2$ for $6_{F}$, corresponding  to
$\Sigma_{Q}$, $\Sigma^{*}_{Q}$, $\Xi'_{Q}$, $\Xi^{*}_{Q}$,
$\Omega_{Q}$ and $\Omega^{*}_{Q}$ states, where $*$ indicates spin
3/2 states.

After this  remark, let us calculate the electromagnetic transition form factors  of the heavy baryons.
  For this aim,
 consider the following correlation function, which is the main tool of the light cone QCD sum rules:
\begin{equation}\label{T}
T_{\mu}(p,q)=i\int d^{4}xe^{ipx}\langle0\mid T\{\eta(x)\bar{\eta}_{\mu}(0) \}\mid0\rangle_{\gamma},
\end{equation}
where $\eta$ and $\eta_{\mu}$ are  the generic interpolating quark currents of the heavy flavored baryons with $J=1/2$ and 3/2, respectively and $\gamma$ means the external electromagnetic field. In this work we will discuss the following electromagnetic transitions
\begin{eqnarray}\label{electransitions}
\Sigma_{Q}^{*}\rightarrow\Sigma_{Q}\gamma,\nonumber\\
\Xi_{Q}^{*}\rightarrow\Xi_{Q}\gamma,\nonumber\\
\Sigma_{Q}^{*}\rightarrow\Lambda_{Q}\gamma,\nonumber\\
\end{eqnarray}
where $Q=$ c or b quark.

In QCD sum rules approach, this
correlation function is calculated in two different ways: from one side, it is calculated in terms of the quarks and gluons interacting
in QCD vacuum. In the phenomenological side, on the other hand, it is saturated by a tower
of hadrons with the same  quantum numbers as the interpolating currents. The physical quantities are
determined by matching these two different representations of the
correlation function.

The hadronic representation  of the correlation function can be
obtained  inserting  the complete set of states with the same
quantum numbers as the interpolating currents.
\begin{eqnarray}\label{T2}
T_{\mu}(p,q)&=&\frac{\langle0\mid \eta\mid
2(p)\rangle}{p^{2}-m_{2}^{2}}\langle 2(p)\mid
1(p+q)\rangle_\gamma\frac{\langle 1(p_+q)\mid
\bar{\eta}_{\mu}\mid 0\rangle}{(p+q)^{2}-m_{1}^{2}}+...,
\end{eqnarray}
where $\langle1(p+q)|$ and   $\langle2(p)|$ denote heavy spin 3/2
and 1/2 states and $m_{1}$ and $m_{2}$ represent their masses,
respectively and q is the photon momentum. In the above equation,
the dots  correspond to the contributions of the higher states and  continuum.
For the calculation of the phenomenological part, it follows from Eq.
(\ref{T2}) that we need to know the  matrix elements of the
interpolating currents between the  vacuum and baryon states.  They
are defined as
\begin{eqnarray}\label{lambdabey}
\langle1(p+q,s)\mid \bar \eta_{\mu}(0)\mid 0\rangle&=&\lambda_{1}\bar u_{\mu}(p+q,s),\nonumber\\
\langle 0 \mid \eta (0)\mid 2(p,s')\rangle &=&\lambda_{2} u(p,s'),
\end{eqnarray}
where $\lambda_{1}$ and  $\lambda_{2}$ are the   residues of the
heavy baryons,  $u_{\mu}(p,s)$ is the Rarita-Schwinger spinor and s
and $s'$ are the polarizations of the spin 3/2 and 1/2 states,
respectively.  The electromagnetic
vertex $\langle 2(p)\mid
1(p+q)\rangle_\gamma$ of the spin 3/2 to spin 1/2 transition is parameterized in
terms of the three form factors in the following way
\cite{onalti,onyedi}
\begin{eqnarray}\label{matelpar}
\langle 2(p)\mid
1(p+q)\rangle_\gamma&=&e\bar u(p,s')\left\{\vphantom{\int_0^{x_2}}G_{1}(q_{\mu}\not\!\varepsilon-\varepsilon_{\mu}\not\!q)\right.
+G_{2}[({\cal P}\varepsilon)q_{\mu}-({\cal P}q)\varepsilon_{\mu}]\gamma_{5}
\nonumber\\&+&G_{3}[(q\varepsilon)q_{\mu}-q^{2}\varepsilon_{\mu}]\gamma_{5}\left.\vphantom{\int_0^{x_2}}\right\} u_{\mu}(p+q),\nonumber\\
\end{eqnarray}
where ${\cal P}=\frac{p+(p+q)}{2}$ and $\varepsilon_{\mu}$ is the
photon polarization vector. Since for the considered decays the
photon is real, the terms proportional to $G_{3}$ are exactly
zero, and for analysis of these decays, we need to know the values of the form factors 
$G_{1}(q^{2})$ and $G_{2}(q^{2})$ only at $q^{2}=0$. From the
experimental point of view, more convenient form factors are
magnetic dipole  $G_{M}$, electric quadrupole $G_{E}$ and Coulomb
quadrupole $G_{C}$ which are linear combinations of the form factors
$G_{1}$ and $G_{2}$ (  see \cite{onbes}). At $q^2=0$, these relations  are 
\begin{eqnarray}\label{acayip}
 G_{M}=\left[(3m_{1}+m_{2})\frac{G_{1}}{m_{1}}+(m_{1}-m_{2})G_{2}\right]\frac{m_{2}}{3},\nonumber\\
G_{E}=(m_{1}-m_{2})\left[\frac{G_{1}}{m_{1}}+G_{2}\right]\frac{m_{2}}{3}.
\end{eqnarray}

 In order to obtain the explicit expressions
 of the correlation function from the phenomenological side, we also perform summation over spins of the spin 3/2 particles using
\begin{equation}\label{raritabela}
\sum_{s}u_{\mu}(p,s)\bar u_{\nu}(p,s)=\frac{(\not\!p+m_)}{2m}\{-g_{\mu\nu}
+\frac{1}{3}\gamma_{\mu}\gamma_{\nu}-\frac{2p_{\mu}p_{\nu}}
{3m^{2}}-\frac{p_{\mu}\gamma_{\nu}-p_{\nu}\gamma_{\mu}}{3m}\}.
\end{equation}
  In principle,  the phenomenological part of the correlator can be obtained with the help of the Eqs.
   (\ref{T2}-\ref{raritabela}). As noted in \cite{kazem1,onbes}, at this point two problems appear:
   1) all Lorentz structures are not independent, b) not only spin 3/2, but spin 1/2 states  also give contributions
    to the correlation function. In other words the matrix element of the current $\eta_{\mu}$ between vacuum and spin 1/2 states is nonzero.
     In general, this matrix element can be written in the following way
\begin{equation}\label{spin12}
\langle0\mid \eta_{\mu}(0)\mid B(p,s=1/2)\rangle= (A  p_{\mu}+B\gamma_{\mu})u(p,s=1/2).
\end{equation}
Imposing the condition $\gamma_\mu \eta^\mu = 0$, one can
immediately obtain that $B=-\frac{A}{4}m$.

In order   to remove the contribution of  the spin 1/2 states and
deal with only independent structures in the correlation function, we
will follow \cite{kazem1,onbes} and remove those contributions by
ordering the Dirac matrices  in a specific way. For this aim, we
choose the  ordering for Dirac matrices as
$\not\!\varepsilon\not\!q\not\!p\gamma_{\mu}$. Using this ordering
for the correlator, we get
\begin{eqnarray}\label{final phenpart}
T_\mu &=& e \lambda_{1} \lambda_{2} \frac{1}{p^2-m_{2}^2} \frac{1}{(p+q)^2-m_{1}^2} \left[ \vphantom{\int_0^{x_2}}\right.
\nonumber \\ &&
\left[ \varepsilon_\mu (pq) - (\varepsilon p) q_\mu \right] \left\{
-2 G_1 m_{1} - G_2 m_{1} m_{2} + G_2 (p+q)^2 \right.
\nonumber \\ &&
+ \left. \left[ 2 G_1 - G_2(m_{1}-m_{2}) \right] \not\!p + m_{2} G_2 \not\!q - G_2 \not\!q\not\!p \right\} \gamma_5
\nonumber \\ &&
+ \left[ q_\mu \not\!\varepsilon - \varepsilon_\mu \not\!q \right] \left\{ G_1 (p^2 + m_{1} m_{2}) -
G_1 (m_{1}+m_{2}) \not\!p \right\} \gamma_5
\nonumber \\ &&
+ 2 G_1 \left[ \not\!\varepsilon (pq) - \not\!q (\varepsilon p) \right] q_\mu \gamma_5
\nonumber \\ &&
- G_1 \not\!\varepsilon \not\!q \left\{ m_{2} + \not\!p \right\} q_\mu \gamma_5
\nonumber \\ &&
\left. \mbox{other structures with $\gamma_\mu$ at the end or which are proportional to $(p+q)_\mu$} \vphantom{\int_0^{x_2}}\right].\nonumber\\
\end{eqnarray}
For determination of the form factors $G_{1}$ and $G_{2}$, we need
two  invariant structures. We obtained that among all structures,
the best convergence comes from the structures $\not\!\varepsilon
\not\!p\gamma_{5}q_{\mu}$ and $\not\!q\not\!p\gamma_{5}(\varepsilon
p)q_{\mu}$ for $G_{1}$ and $G_{2}$, respectively. To get the sum
rules expression for the form factors $G_{1}$ and $G_{2}$, we will
choose the same structures also from the QCD side and match the
corresponding coefficients. We also would like to note that, the correlation function receives contributions from contact terms. But, the contact terms do not give contributions to the chosen structures (for a detailed discussion see e. g. \cite{onbes,khodbey}  ).

On QCD side, the correlation function  can be calculated using
 the operator product expansion.  For this aim, we need the explicit
expressions of the interpolating currents of the heavy baryons with
the angular momentums $J=3/2$ and  $J=1/2$. The interpolating
currents for the spin $J=3/2$ baryons are written in such a way that
the light quarks should  enter the expression of currents in
symmetric way and the condition $\gamma^{\mu}\eta_{\mu}=0$ should be
satisfied. The general form of the currents for  spin $J=3/2$
baryons satisfying both aforementioned conditions can be written as
\cite{kazem1}
\begin{eqnarray}\label{currentguy}
\eta_{\mu}=A\epsilon_{abc}\left\{\vphantom{\int_0^{x_2}}(q_{1}^{aT}C\gamma_{\mu}q_{2}^{b})Q^{c}+(q_{2}^{aT}C\gamma_{\mu}Q^{b})q_{1}^{c}+
(Q^{aT}C\gamma_{\mu}q_{1}^{b})q_{2}^{c}\right\},
\end{eqnarray}
where C is the charge conjugation operator and  a, b and c are color
indices. The value of normalization factor A and quark fields $q_{1}$ and $q_{2}$  for
corresponding  heavy baryons is given in Table 1 (see \cite{kazem1}).
\begin{table}[h]
\centering
\begin{tabular}{|c||c|c|c|}\hline
  Heavy spin $\frac{3}{2}$ baryons&A & $q_{1}$& $q_{2}$\\\cline{1-4}
\hline\hline
 $\Sigma_{b(c)}^{*+(++)}$&$1/\sqrt{3}$ &u&u\\\cline{1-4}
 $\Sigma_{b(c)}^{*0(+)}$&$\sqrt{2/3}$&u&d\\\cline{1-4}
 $\Sigma_{b(c)}^{*-(0)}$&$1/\sqrt{3}$&d&d\\\cline{1-4}
 $\Xi_{b(c)}^{*0(+)}$&$\sqrt{2/3}$&s&u\\\cline{1-4}
 $\Xi_{b(c)}^{*-(0)}$&$\sqrt{2/3}$&s&d\\\cline{1-4}
$\Omega_{b(c)}^{*-(0)}$ &$1/\sqrt{3}$&s &s\\\cline{1-4}
 \end{tabular}
 \vspace{0.8cm}
\caption{The value of normalization factor A and quark fields $q_{1}$ and $q_{2}$  for
the corresponding heavy spin 3/2 baryons. }\label{tab:2}
\end{table}

The general form of the interpolating currents for the heavy spin $1/2$ baryons can be written in  the following  form:
\begin{eqnarray}\label{currentguy}
\eta_{\Sigma_{Q}}&=&-\frac{1}{\sqrt{2}}\epsilon_{abc}\left\{\vphantom{\int_0^{x_2}}(q_{1}^{aT}CQ^{b})\gamma_{5}q_{2}^{c
}+\beta(q_{1}^{aT}C\gamma_{5}Q^{b})q_{2}^{c} \right.\nonumber\\&-& \left. [(Q_{}^{aT}Cq_{2}^{b})\gamma_{5}q_{1}^{c}+\beta(Q_{}^{aT}C\gamma_{5}q_{2}^{b})q_{1}^{c}]\right\},\nonumber\\
\eta_{\Xi_{Q},\Lambda_{Q}}&=&\frac{1}{\sqrt{6}}\epsilon_{abc}\left\{\vphantom{\int_0^{x_2}}2(q_{1}^{aT}Cq_{2}^{b})\gamma_{5}Q^{c}
+\beta(q_{1}^{aT}C\gamma_{5}q_{2}^{b})Q^{c}+(q_{1}^{aT}CQ^{b})\gamma_{5}q_{2}^{c}\right.\nonumber\\
&+&\left. \beta(q_{1}^{aT}C\gamma_{5}Q^{b})q_{2}^{c}+(Q_{}^{aT}Cq_{2}^{b})\gamma_{5}q_{1}^{c}+\beta(Q_{}^{aT}C\gamma_{5}q_{2}^{b})q_{1}^{c}\right\},\nonumber\\
\end{eqnarray}
where $\beta$ is an arbitrary parameter and $\beta=-1$ corresponds to the Ioffe current. The quark fields $q_{1}$ and $q_{2}$
 for the corresponding heavy spin 1/2 baryons are as presented in Table 2.
\begin{table}[h]
\centering
\begin{tabular}{|c||c|c|}\hline
 Heavy spin $\frac{1}{2}$ baryons& $q_{1}$& $q_{2}$\\\cline{1-3}
\hline\hline
 $\Sigma_{b(c)}^{+(++)}$&u&u\\\cline{1-3}
 $\Sigma_{b(c)}^{0(+)}$&u&d\\\cline{1-3}
 $\Sigma_{b(c)}^{-(0)}$&d&d\\\cline{1-3}
 $\Xi_{b(c)}^{0(+)}$&u&s\\\cline{1-3}
 $\Xi_{b(c)}^{-(0)}$&d&s\\\cline{1-3}
 $\Lambda_{b(c)}^{0(+)}$ &u &d\\\cline{1-3}
 \end{tabular}
 \vspace{0.8cm}
\caption{The  quark fields $q_{1}$ and $q_{2}$  for
the corresponding heavy spin 1/2 baryons. }\label{tab:2}
\end{table}

  After performing all contractions of the quark fields in  Eq. (\ref{T}), we get the following expression for the correlation
functions responsible for the $\Sigma^{*0}_{b}\rightarrow
\Sigma^{0}_{b}\gamma$ and $\Xi^{*0}_{b}\rightarrow
\Xi^{0}_{b}\gamma$ transitions in terms of the light and heavy quark
propagators
\begin{eqnarray}\label{tree expresion.m}
T^{\Sigma^{*0}_{b}\rightarrow \Sigma^{0}_{b}}_{\mu}&=&\frac{i}{\sqrt{3}}\epsilon_{abc}\epsilon_{a'b'c'}\int
d^{4}xe^{ipx}\langle0[\gamma(q)]\mid\{-\gamma_{5}S^{c'c}_{d}Tr[S^{b'b}_{b}\gamma_{\mu}S'^{a'a}_{u}]\nonumber\\&+&
\gamma_{5}S^{c'c}_{u}Tr[S^{b'b}_{d}\gamma_{\mu}S'^{a'a}_{b}]+\beta Tr[\gamma_{5}S^{aa'}_{u}\gamma_{\mu}S'^{bb'}_{b}]S^{cc'}_{d}-\beta Tr[\gamma_{5}S^{aa'}_{b}\gamma_{\mu}S'^{bb'}_{d}]S^{cc'}_{u}
\nonumber\\&-&\gamma_{5}S^{c'a}_{d}\gamma_{\mu}S'^{a'b}_{u}S^{b'c}_{b}
+\gamma_{5}S^{c'b}_{d}\gamma_{\mu}S'^{b'a}_{u}S^{a'c}_{u}-\gamma_{5}S^{c'b}_{u}\gamma_{\mu}S'^{b'a}_{d}S^{a'c}_{b}
\nonumber\\&+&\gamma_{5}S^{c'a}_{u}\gamma_{\mu}S'^{a'b}_{b}S^{b'c}_{d}-\beta S^{ca'}_{d}\gamma_{\mu}S'^{ab'}_{u}\gamma_{5}S^{bc'}_{b}+\beta S^{c'b}_{d}\gamma_{\mu}S'^{b'a}_{b}\gamma_{5}S^{a'c}_{u}
\nonumber\\&-&\beta S^{c'b}_{u}\gamma_{\mu}S'^{b'a}_{d}\gamma_{5}S^{a'c}_{b}+\beta S^{c'a}_{u}\gamma_{\mu}S'^{a'b}_{b}\gamma_{5}S^{b'c}_{d}\mid 0\rangle,
\end{eqnarray}
\begin{eqnarray}\label{tree expresion.m2}
T^{\Xi^{*0}_{b}\rightarrow \Xi^{0}_{b}}_{\mu}&=&\frac{i}{3}\epsilon_{abc}\epsilon_{a'b'c'}\int
d^{4}xe^{ipx}\langle0[\gamma(q)]\mid\{\gamma_{5}S^{c'c}_{s}Tr[S^{b'b}_{b}\gamma_{\mu}S'^{a'a}_{u}]\nonumber\\&-&
2\gamma_{5}S^{c'c}_{b}Tr[S^{b'a}_{s}\gamma_{\mu}S'^{a'b}_{u}]+2\beta Tr[\gamma_{5}S^{a'b}_{u}\gamma_{\mu}S'^{b'a}_{s}]S^{c'c}_{b}-\beta Tr[\gamma_{5}S^{a'a}_{u}\gamma_{\mu}S'^{b'b}_{b}]S^{c'c}_{s}\nonumber\\&+&
\gamma_{5}S^{c'c}_{u}Tr[S^{b'b}_{s}\gamma_{\mu}S'^{a'a}_{b}]-\beta Tr[\gamma_{5}S^{a'a}_{b}\gamma_{\mu}S'^{b'b}_{s}]S^{c'c}_{u}-\gamma_{5}S^{c'b}_{u}\gamma_{\mu}S'^{b'a}_{s}S^{a'c}_{b}
\nonumber\\&+&2\gamma_{5}S^{c'a}_{b}\gamma_{\mu}S'^{b'b}_{s}S^{a'c}_{u}
-2\gamma_{5}S^{c'b}_{b}\gamma_{\mu}S'^{a'a}_{u}S^{b'c}_{s}+\gamma_{5}S^{c'a}_{s}\gamma_{\mu}S'^{a'b}_{u}S^{b'c}_{b}
\nonumber\\&-&\gamma_{5}S^{c'b}_{s}\gamma_{\mu}S'^{b'a}_{b}S^{a'c}_{u}+2\beta S^{c'a}_{b}\gamma_{\mu}S'^{b'b}_{s}\gamma_{5}S^{a'c}_{u}-2\beta S^{c'b}_{b}\gamma_{\mu}S'^{a'a}_{u}\gamma_{5}S^{b'c}_{s}
\nonumber\\&+&\beta S^{c'a}_{s}\gamma_{\mu}S'^{a'b}_{u}\gamma_{5}S^{b'c}_{b}-\beta S^{c'b}_{s}\gamma_{\mu}S'^{b'a}_{b}\gamma_{5}S^{a'c}_{u}+\gamma_{5}S^{c'a}_{u}\gamma_{\mu}S'^{a'b}_{b}S^{b'c}_{s}\nonumber\\
&-&\beta S^{c'b}_{u}\gamma_{\mu}S'^{b'a}_{s}\gamma_{5}S^{a'c}_{b}+\beta S^{c'a}_{u}\gamma_{\mu}S'^{a'b}_{b}\gamma_{5}S^{b'c}_{s}\beta\mid 0\rangle,
\end{eqnarray}
where $S'=CS^TC$. 

The correlation functions for all other possible transitions can be obtained by the following replacements
\begin{eqnarray}\label{replacements}
T^{\Sigma^{*-}_{b}\rightarrow \Sigma^{-}_{b}}_{\mu}&=&T^{\Sigma^{*0}_{b}\rightarrow \Sigma^{0}_{b}}_{\mu}
(u\rightarrow d),\nonumber\\
T^{\Sigma^{*+}_{b}\rightarrow \Sigma^{+}_{b}}_{\mu}&=&T^{\Sigma^{*0}_{b}\rightarrow \Sigma^{0}_{b}}_{\mu}
(d\rightarrow u),\nonumber\\
T^{\Sigma^{*+}_{c}\rightarrow \Sigma^{+}_{c}}_{\mu}&=&T^{\Sigma^{*0}_{b}\rightarrow \Sigma^{0}_{b}}_{\mu}
(b\rightarrow c),\nonumber\\
T^{\Sigma^{*0}_{c}\rightarrow \Sigma^{0}_{c}}_{\mu}&=&T^{\Sigma^{*-}_{b}\rightarrow \Sigma^{-}_{b}}_{\mu}
(b\rightarrow c),\nonumber\\
T^{\Sigma^{*++}_{c}\rightarrow \Sigma^{++}_{c}}_{\mu}&=&T^{\Sigma^{*+}_{b}\rightarrow \Sigma^{+}_{b}}_{\mu}
(b\rightarrow c),\nonumber\\
T^{\Xi^{*-}_{b}\rightarrow \Xi^{-}_{b}}_{\mu}&=&T^{\Xi^{*0}_{b}\rightarrow \Xi^{0}_{b}}_{\mu}(u\rightarrow d),\nonumber\\
T^{\Xi^{*0}_{c}\rightarrow \Xi^{0}_{c}}_{\mu}&=&T^{\Xi^{*-}_{b}\rightarrow \Xi^{-}_{b}}_{\mu}(b\rightarrow c),\nonumber\\
T^{\Xi^{*+}_{c}\rightarrow \Xi^{+}_{c}}_{\mu}&=&T^{\Xi^{*0}_{b}\rightarrow \Xi^{0}_{b}}_{\mu}(b\rightarrow c),\nonumber\\
T^{\Sigma^{*0}_{b}\rightarrow \Lambda^{0}_{b}}_{\mu}&=&T^{\Xi^{*0}_{b}\rightarrow \Xi^{0}_{b}}_{\mu}(s\rightarrow d),\nonumber\\
T^{\Sigma^{*+}_{c}\rightarrow \Lambda^{+}_{c}}_{\mu}&=&T^{\Sigma^{*0}_{b}\rightarrow \Lambda^{0}_{b}}_{\mu}(b\rightarrow c).
\end{eqnarray}

The correlators in Eqs. (\ref{tree expresion.m}, \ref{tree expresion.m2}) get three different contributions: 1) Perturbative
contributions, 2) Mixed contributions, i.e., the photon is radiated
from  short distance and at least one of the quarks forms a
condensate. 3) Non-perturbative contributions, i.e., when photon is
radiated at long distances. This contribution is described by the
matrix element $ \langle\gamma(q)\mid\bar q(x_{1})
 \Gamma q(x_{2})\mid0\rangle$ which is parameterized in terms of photon distribution amplitudes with definite twists.

The results of the contributions when the photon interacts with the quarks perturbatively
is obtained by  replacing the propagator of the quark that emits  the photon by
\begin{equation}\label{rep1guy}
S^{ab}_{\alpha \beta} \Rightarrow  \left\{ \int d^4 y
S^{free} (x-y) \not\!A
S^{free}(y)\right\}^{ab}_{\alpha \beta}.
\end{equation}
 The   free light and heavy quark propagators are defined as:
\begin{eqnarray}\label{free1guy}
S^{free}_{q} &=&\frac{i\not\!x}{2\pi^{2}x^{4}}-\frac{m_{q}}{4\pi^{2}x^{2}},\nonumber\\
S^{free}_{Q}
&=&\frac{m_{Q}^{2}}{4\pi^{2}}\frac{K_{1}(m_{Q}\sqrt{-x^2})}{\sqrt{-x^2}}-i
\frac{m_{Q}^{2}\not\!x}{4\pi^{2}x^2}K_{2}(m_{Q}\sqrt{-x^2}),\nonumber\\
\end{eqnarray}
where $K_{i}$ are Bessel functions.
The non-perturbative contributions to the
 correlation function can be easily obtained from Eq.
(\ref{tree expresion.m})  replacing one of the light quark
propagators that emits a photon by
\begin{equation}\label{rep2guy}
\label{rep} S^{ab}_{\alpha \beta} \rightarrow - \frac{1}{4} \bar q^a \Gamma_j q^b ( \Gamma_j )_{\alpha \beta}~,
\end{equation}
 where $\Gamma$ is the full set of Dirac matrices $\Gamma_j = \Big\{
1,~\gamma_5,~\gamma_\alpha,~i\gamma_5 \gamma_\alpha, ~\sigma_{\alpha
\beta} /\sqrt{2}\Big\}$ and sum over index $j$ is implied. Remaining
quark propagators are full propagators  involving the perturbative
as well as the  non-perturbative contributions.

The light cone expansion of the light and heavy quark propagators in
the presence of an  external field is done  in \cite{Balitsky}. The
operators $\bar q G q$, $\bar q G G q$ and $\bar q q \bar q q$,
where $G$ is the gluon field  strength tensor give contributions to
the propagators and in  \cite{Braun2}, it was shown that terms
with two gluons  as well as four quarks operators give negligible
small contributions, so we neglect them. Taking into account this
fact,  the expressions for the heavy  and light quark propagators
are written as:
\begin{eqnarray}\label{heavylightguy}
 S_Q (x)& =&  S_Q^{free} (x) - i g_s \int \frac{d^4 k}{(2\pi)^4}
e^{-ikx} \int_0^1 dv \Bigg[\frac{\not\!k + m_Q}{( m_Q^2-k^2)^2}
G^{\mu\nu}(vx)
\sigma_{\mu\nu} \nonumber \\
&+& \frac{1}{m_Q^2-k^2} v x_\mu G^{\mu\nu} \gamma_\nu \Bigg],
\nonumber \\
S_q(x) &=&  S_q^{free} (x) - \frac{m_q}{4 \pi^2 x^2} - \frac{\langle
\bar q q \rangle}{12} \left(1 - i \frac{m_q}{4} \not\!x \right) -
\frac{x^2}{192} m_0^2 \langle \bar q q \rangle \left( 1 - i
\frac{m_q}{6}\not\!x \right) \nonumber \\ &&
 - i g_s \int_0^1 du \left[\frac{\not\!x}{16 \pi^2 x^2} G_{\mu \nu} (ux) \sigma_{\mu \nu} - u x^\mu
G_{\mu \nu} (ux) \gamma^\nu \frac{i}{4 \pi^2 x^2} \right. \nonumber
\\ && \left. - i \frac{m_q}{32 \pi^2} G_{\mu \nu} \sigma^{\mu \nu}
\left( \ln \left( \frac{-x^2 \Lambda^2}{4} \right) + 2 \gamma_E
\right) \right],
 \end{eqnarray}
  where $\Lambda$ is the scale parameter and we choose it at factorization scale i.e., $\Lambda=(0.5~GeV-1~GeV)$ (see \cite{kazem1,revised2}).

 As we already noted,  for the calculation of the non-perturbative contributions,  the matrix
 elements  $ \langle\gamma(q)\mid\bar q
 \Gamma_{i}q\mid0\rangle$ are needed. These matrix  elements are calculated  in terms of
 the photon distribution amplitudes (DA's) as follows \cite{Ball}.
 \begin{eqnarray}
&&\langle \gamma(q) \vert  \bar q(x) \sigma_{\mu \nu} q(0) \vert  0
\rangle  = -i e_q \bar q q (\varepsilon_\mu q_\nu - \varepsilon_\nu
q_\mu) \int_0^1 du e^{i \bar u qx} \left(\chi \varphi_\gamma(u) +
\frac{x^2}{16} \mathbb{A}  (u) \right) \nonumber \\ &&
-\frac{i}{2(qx)}  e_q \qq \left[x_\nu \left(\varepsilon_\mu - q_\mu
\frac{\varepsilon x}{qx}\right) - x_\mu \left(\varepsilon_\nu -
q_\nu \frac{\varepsilon x}{q x}\right) \right] \int_0^1 du e^{i \bar
u q x} h_\gamma(u)
\nonumber \\
&&\langle \gamma(q) \vert  \bar q(x) \gamma_\mu q(0) \vert 0 \rangle
= e_q f_{3 \gamma} \left(\varepsilon_\mu - q_\mu \frac{\varepsilon
x}{q x} \right) \int_0^1 du e^{i \bar u q x} \psi^v(u)
\nonumber \\
&&\langle \gamma(q) \vert \bar q(x) \gamma_\mu \gamma_5 q(0) \vert 0
\rangle  = - \frac{1}{4} e_q f_{3 \gamma} \epsilon_{\mu \nu \alpha
\beta } \varepsilon^\nu q^\alpha x^\beta \int_0^1 du e^{i \bar u q
x} \psi^a(u)
\nonumber \\
&&\langle \gamma(q) | \bar q(x) g_s G_{\mu \nu} (v x) q(0) \vert 0
\rangle = -i e_q \qq \left(\varepsilon_\mu q_\nu - \varepsilon_\nu
q_\mu \right) \int {\cal D}\alpha_i e^{i (\alpha_{\bar q} + v
\alpha_g) q x} {\cal S}(\alpha_i)
\nonumber \\
&&\langle \gamma(q) | \bar q(x) g_s \tilde G_{\mu \nu} i \gamma_5 (v
x) q(0) \vert 0 \rangle = -i e_q \qq \left(\varepsilon_\mu q_\nu -
\varepsilon_\nu q_\mu \right) \int {\cal D}\alpha_i e^{i
(\alpha_{\bar q} + v \alpha_g) q x} \tilde {\cal S}(\alpha_i)
\nonumber \\
&&\langle \gamma(q) \vert \bar q(x) g_s \tilde G_{\mu \nu}(v x)
\gamma_\alpha \gamma_5 q(0) \vert 0 \rangle = e_q f_{3 \gamma}
q_\alpha (\varepsilon_\mu q_\nu - \varepsilon_\nu q_\mu) \int {\cal
D}\alpha_i e^{i (\alpha_{\bar q} + v \alpha_g) q x} {\cal
A}(\alpha_i)
\nonumber \\
&&\langle \gamma(q) \vert \bar q(x) g_s G_{\mu \nu}(v x) i
\gamma_\alpha q(0) \vert 0 \rangle = e_q f_{3 \gamma} q_\alpha
(\varepsilon_\mu q_\nu - \varepsilon_\nu q_\mu) \int {\cal
D}\alpha_i e^{i (\alpha_{\bar q} + v \alpha_g) q x} {\cal
V}(\alpha_i) \nonumber \\ && \langle \gamma(q) \vert \bar q(x)
\sigma_{\alpha \beta} g_s G_{\mu \nu}(v x) q(0) \vert 0 \rangle  =
e_q \qq \left\{
        \left[\left(\varepsilon_\mu - q_\mu \frac{\varepsilon x}{q x}\right)\left(g_{\alpha \nu} -
        \frac{1}{qx} (q_\alpha x_\nu + q_\nu x_\alpha)\right) \right. \right. q_\beta
\nonumber \\ && -
         \left(\varepsilon_\mu - q_\mu \frac{\varepsilon x}{q x}\right)\left(g_{\beta \nu} -
        \frac{1}{qx} (q_\beta x_\nu + q_\nu x_\beta)\right) q_\alpha
\nonumber \\ && -
         \left(\varepsilon_\nu - q_\nu \frac{\varepsilon x}{q x}\right)\left(g_{\alpha \mu} -
        \frac{1}{qx} (q_\alpha x_\mu + q_\mu x_\alpha)\right) q_\beta
\nonumber \\ &&+
         \left. \left(\varepsilon_\nu - q_\nu \frac{\varepsilon x}{q.x}\right)\left( g_{\beta \mu} -
        \frac{1}{qx} (q_\beta x_\mu + q_\mu x_\beta)\right) q_\alpha \right]
   \int {\cal D}\alpha_i e^{i (\alpha_{\bar q} + v \alpha_g) qx} {\cal T}_1(\alpha_i)
\nonumber \\ &&+
        \left[\left(\varepsilon_\alpha - q_\alpha \frac{\varepsilon x}{qx}\right)
        \left(g_{\mu \beta} - \frac{1}{qx}(q_\mu x_\beta + q_\beta x_\mu)\right) \right. q_\nu
\nonumber \\ &&-
         \left(\varepsilon_\alpha - q_\alpha \frac{\varepsilon x}{qx}\right)
        \left(g_{\nu \beta} - \frac{1}{qx}(q_\nu x_\beta + q_\beta x_\nu)\right)  q_\mu
\nonumber \\ && -
         \left(\varepsilon_\beta - q_\beta \frac{\varepsilon x}{qx}\right)
        \left(g_{\mu \alpha} - \frac{1}{qx}(q_\mu x_\alpha + q_\alpha x_\mu)\right) q_\nu
\nonumber \\ &&+
         \left. \left(\varepsilon_\beta - q_\beta \frac{\varepsilon x}{qx}\right)
        \left(g_{\nu \alpha} - \frac{1}{qx}(q_\nu x_\alpha + q_\alpha x_\nu) \right) q_\mu
        \right]
    \int {\cal D} \alpha_i e^{i (\alpha_{\bar q} + v \alpha_g) qx} {\cal T}_2(\alpha_i)
\nonumber \\ &&+
        \frac{1}{qx} (q_\mu x_\nu - q_\nu x_\mu)
        (\varepsilon_\alpha q_\beta - \varepsilon_\beta q_\alpha)
    \int {\cal D} \alpha_i e^{i (\alpha_{\bar q} + v \alpha_g) qx} {\cal T}_3(\alpha_i)
\nonumber \\ &&+
        \left. \frac{1}{qx} (q_\alpha x_\beta - q_\beta x_\alpha)
        (\varepsilon_\mu q_\nu - \varepsilon_\nu q_\mu)
    \int {\cal D} \alpha_i e^{i (\alpha_{\bar q} + v \alpha_g) qx} {\cal T}_4(\alpha_i)
                        \right\},
\end{eqnarray}
where
$\varphi_\gamma(u)$ is the leading twist 2, $\psi^v(u)$,
$\psi^a(u)$, ${\cal A}$ and ${\cal V}$ are the twist 3 and
$h_\gamma(u)$, $\mathbb{A}$, ${\cal T}_i$ ($i=1,~2,~3,~4$) are the
twist 4 photon DA's, respectively and  $\chi$ is the magnetic susceptibility of the quarks. The photon DA's is calculated in \cite{Ball}. The measure ${\cal D} \alpha_i$ is defined as
\begin{equation}
\int {\cal D} \alpha_i = \int_0^1 d \alpha_{\bar q} \int_0^1 d
\alpha_q \int_0^1 d \alpha_g \delta(1-\alpha_{\bar
q}-\alpha_q-\alpha_g).\nonumber \\
\end{equation}

The  coefficient of any structure in the expression of the correlation function  can be written in the form
\begin{eqnarray}\label{relationscorr}
T(q_{1},q_{2},Q)=e_{q_{1}}T_{1}(q_{1},q_{2},Q)+e_{q_{2}}T'_{1}(q_{1},q_{2},Q)+e_{Q}T_{2}(q_{1},q_{2},Q),
\end{eqnarray}
where $T_{1}$, $T'_{1}$ and  $T_{2}$ in the right side   correspond
to the radiation of the photon from the quarks $q_1$,  $q_2$ and
$Q$, respectively. Starting from the expressions of the
interpolating currents, one can easily obtain that the functions
$T_{1}$ and $T'_{1}$ differ only by $q_{1}\longleftrightarrow
q_{2}$ exchange for $\Sigma^{*}_{Q}\rightarrow \Sigma_{Q}$
transition. Using Eq. (\ref{replacements}),  in SU(2) symmetry
limit, we obtain
\begin{eqnarray}
T^{\Sigma^{*+(++)}_{b(c)}\rightarrow \Sigma^{+(++)}_{b(c)}} + T^{\Sigma^{*-(0)}_{b(c)}\rightarrow \Sigma^{-(0)}_{b(c)}}=2T^{\Sigma^{*0(+)}_{b(c)}\rightarrow \Sigma^{0(+)}_{b(c)}}.
\end{eqnarray}
Therefore, in the numerical analysis section, we have not presented the
numerical results  for
$T^{\Sigma^{*-(0)}_{b(c)}\rightarrow \Sigma^{-(0)}_{b(c)}}$.

Now, using the above equations, one can get the the correlation
function from the QCD side.  Separating the coefficients of the
structures $\not\!\varepsilon \not\!p\gamma_{5}q_{\mu}$ and
$\not\!q\not\!p\gamma_{5}(\varepsilon p)q_{\mu}$ respectively for
the form factors $G_{1}$ and $G_{2}$ from both QCD and
phenomenological representations and equating them, we  get  sum
rules for the form factors $G_{1}$ and $G_{2}$.  To suppress the
contributions of the higher states and continuum, Borel
transformations with respect to the variables $p^2$ and $(p+q)^2$
 are applied. The explicit forms of the  sum rules for the form factors $G_{1}$  and $G_{2}$ can be as follows. For $\Sigma^{*0}_{b}\rightarrow \Sigma^{0}_{b}$, we obtain
\begin{eqnarray}\label{magneticmoment1}
G_{1}&=&-\frac{1}{\lambda_{1}\lambda_{2}(m_{1}+m_{2})}e^{\frac{m_{1}^{2}}{M_{1}^{2}}}e^{\frac{m_{2}^{2}}{M_{2}^{2}}}
\left[\vphantom{\int_0^{x_2}}e_{q_{1}}\Pi_{1}+e_{q_{2}}\Pi_{1}(q_{1}\leftrightarrow q_{2})+e_{b}\Pi'_{1}\right]
\nonumber\\
G_{2}&=&\frac{1}{\lambda_{1}\lambda_{2}}e^{\frac{m_{1}^{2}}{M_{1}^{2}}}e^{\frac{m_{2}^{2}}{M_{2}^{2}}}
\left[\vphantom{\int_0^{x_2}}e_{q_{1}}\Pi_{2}+e_{q_{2}}\Pi_{2}(q_{1}\leftrightarrow q_{2})+e_{b}\Pi'_{2}\right],
\end{eqnarray}

where  $q_{1}=u$,  $q_{2}=d$. The functions $\Pi_i[\Pi'_{i}]$ can be written as:
\begin{eqnarray}\label{magneticmoment2}
\Pi_{i}[\Pi'_{i}]&=&\int_{m_{Q}^{2}}^{s_{0}}e^{\frac{-s}{M^{2}}}\rho_{i}(s)[\rho'_{i}(s)]ds+e^{\frac{-m_Q^2}{M^{2}}}
\Gamma_{i}[\Gamma'_{i}],
\end{eqnarray}
 where,
\begin{eqnarray}\label{rho1}
\sqrt{3}\rho_{1}(s)&=&\langle q_{1}q_{1}\rangle \langle q_{2}q_{2}\rangle\left[-\frac{\beta}{3}\chi \varphi_{\gamma}(u_{0})\right]
\nonumber\\&+&m_{0}^{2}\langle q_{2}q_{2}\rangle \left[-\frac{(1+\beta)}{144m_Q\pi^{2}}(1+\psi_{11})\vphantom{\int_0^{x_2}}\right]
\nonumber\\&+&\langle q_{2}q_{2}\rangle \left[\vphantom{\int_0^{x_2}}
\frac{\beta m_{Q}}{4\pi^{2}}(\psi_{10}-\psi_{21})\vphantom{\int_0^{x_2}}\right]
\nonumber\\&+&\langle q_{1}q_{1}\rangle \frac{m_Q}{96\pi^{2}}\left[\vphantom{\int_0^{x_2}}\right.
-3(1+\beta)(\psi_{10}-\psi_{21})\mathbb{A}(u_{0})\nonumber\\&+&2\left(\vphantom{\int_0^{x_2}}\right.
-2(\psi_{10}-\psi_{21})\{(1+5\beta)\eta_{1}-\eta_{2}-4\eta_{3}-\eta_{4}+6\eta_{5}\nonumber\\&+&
2\eta_{6}+3\eta_{7}-2\eta_{8}-\beta(\eta_{2}+\eta_{4}-2\eta_{5}-2\eta_{6}-7\eta_{7}+2\eta_{8})\}
\nonumber\\&-&3(3+\beta)(-1+\psi_{02}+2\psi_{10}-2\psi_{21})(u_{0}-1)\zeta_{1}+2\{(1+3\beta)\eta_{1}
\nonumber\\&-&(3+\beta)(\eta_{2}-\eta_{4})+(1+3\beta)\eta_{7}\}ln(\frac{s}{m_Q^{2}})
\nonumber\\&+&3(1+\beta)m_Q^2\chi\{2\psi_{10}-\psi_{20}+\psi_{31}-2ln(\frac{s}{m_Q^{2}})\}\varphi_{\gamma}(u_{0})\left.\vphantom{\int_0^{x_2}}\right)\left.\vphantom{\int_0^{x_2}}\right]+
\nonumber\\&+&\frac{m_Q^{4}(\beta-1)}{512\pi^{4}}\left[\vphantom{\int_0^{x_2}}\right.4\psi_{30}
-5\psi_{42}-4\psi_{52}+(4\psi_{30}-8\psi_{41}+5\psi_{42}+4\psi_{52})u_{0}
\nonumber\\&+&6\{2\psi_{10}-\psi_{20}-2ln(\frac{s}{m_Q^{2}})\}(1+u_{0})\left.\vphantom{\int_0^{x_2}}\right]\nonumber\\&+&
\frac{f_{3\gamma}m_Q^{2}(\beta-1)}{192\pi^{2}}\left[\vphantom{\int_0^{x_2}}8\{-(\psi_{20}-\psi_{31})(\eta'_{1}-\eta'_{2})+
[\psi_{10}-ln(\frac{s}{m_Q^{2}})](\eta'_{3}+\eta'_{1}-\eta'_{2})\}\right.\nonumber\\&+&6(\psi_{20}-\psi_{31})\psi^a(u_{0})
+(3\psi_{31}+4\psi_{32}+3\psi_{42})(u_{0}-1)[4\psi^v(u_{0})-\frac{d\psi^{a}(u_{0})}{du_{0}}]\left.\vphantom{\int_0^{x_2}}\right],\nonumber
\end{eqnarray}
\begin{eqnarray}\label{rho1'}
\sqrt{3}\rho^{'}_{1}(s)&=&m_{0}^{2}(\langle q_{1}q_{1}\rangle+\langle q_{2}q_{2}\rangle)\left[\vphantom{\int_0^{x_2}}
\frac{5(1+3\beta)(1+\psi_{11})}{144\pi^{2}m_Q}
\vphantom{\int_0^{x_2}}\right]\nonumber\\&+&
(\langle q_{1}q_{1}\rangle+\langle q_{2}q_{2}\rangle)\frac{m_Q(1+3\beta)}{8\pi^{2}}\left[\vphantom{\int_0^{x_2}}\psi_{10}-\psi_{21}-ln(\frac{s}{m_Q^{2}})\vphantom{\int_0^{x_2}}\right]\nonumber\\&+&
\frac{m_Q^{4}(1-\beta)}{768\pi^{4}}
\left[\vphantom{\int_0^{x_2}}\right.2(-3\psi_{20}-12\psi_{30}+18\psi_{31}+8\psi_{32}+18\psi_{41})
-3(\psi_{42}+4\psi_{52})
\nonumber\\&+&\{2(3\psi_{20}+6\psi_{30}-18\psi_{31}-8\psi_{32})+3(-8\psi_{41}+\psi_{42}+4\psi_{52})\}u_{0}
\nonumber\\&-&12\psi_{10}(7+5u_{0})-72(\psi_{10}+u_{0})ln(\frac{m_Q^{2}}{s})
\nonumber\\&+&12\{7+2\psi_{10}(-1+u_{0})-u_{0}\}ln(\frac{s}{m_Q^{2}})\left.\vphantom{\int_0^{x_2}}\right],
\end{eqnarray}
\begin{eqnarray}\label{rho2}
\sqrt{3}\rho_{2}(s)&=&\langle q_{1}q_{1}\rangle \frac{1}{12m_Q\pi^{2}}\left[\vphantom{\int_0^{x_2}}\right.
3(3+\beta)(-1+\psi_{03}+2\psi_{12}+\psi_{22})(u_{0}-1)\zeta_{2}\nonumber\\&+&2\left(\vphantom{\int_0^{x_2}}\right.
-2(1+\beta)+(\beta-1)\psi_{11}+2(1+3\beta)(\psi_{12}+\psi_{22})\nonumber\\&-&\frac{3+\beta}{2}(-1+2\psi_{12}+\psi_{22})(\zeta_{9}-\zeta_{5})\left.\vphantom{\int_0^{x_2}}\right)\left.\vphantom{\int_0^{x_2}}\right]+
\nonumber\\&+&\frac{m_Q^{2}(\beta-1)}{128\pi^{4}}\left[\vphantom{\int_0^{x_2}}\right.(12\psi_{31}
+19\psi_{32}+\psi_{33}+12\psi_{42})(u_{0}-1)u_{0}\left.\vphantom{\int_0^{x_2}}\right]\nonumber\\&+&
\frac{f_{3\gamma}(\beta-1)}{96\pi^{2}}\left[\vphantom{\int_0^{x_2}}4\{(1-\psi_{02})\eta_{9}+(-3+3\psi_{02}+2\psi_{10}-2\psi_{21})\eta_{11}\right.\nonumber\\&-&(-1+\psi_{02}+2\psi_{10}-2\psi_{21})\eta_{10}+(-1+3\psi_{02}-2\psi_{03})(u_{0}-1)\zeta_{11}\}\nonumber\\&-&(-1+3\psi_{02}-2\psi_{03})(u_{0}-1)\psi^a(u_{0})\left.\vphantom{\int_0^{x_2}}\right],\nonumber
\end{eqnarray}
\begin{eqnarray}\label{rho2'}
\sqrt{3}\rho^{'}_{2}(s)&=&
\frac{m_Q^{2}(\beta-1)}{192\pi^{4}}
\left[\vphantom{\int_0^{x_2}}\right.(-1+u_{0})u_{0}\{12\psi_{10}-6\psi_{20}\nonumber\\&+&24\psi_{31}+27\psi_{32}+\psi_{33}
+18\psi_{42}-12ln(\frac{s}{m_Q^{2}})\}\left.\vphantom{\int_0^{x_2}}\right],
\end{eqnarray}

\begin{eqnarray}\label{Gamma1}
\sqrt{3}\Gamma_{1}&=&m_{0}^{2}<q_{1}q_{1}><q_{2}q_{2}>\left[\frac{\beta m_Q^4\mathbb{A}(u_{0})}{48M^{6}}+\frac{m_Q^{2}}{432M^{4}}\{12[\eta_{2}-2\eta_{3}\right.\nonumber\\&-&\eta_{4}+2\eta_{5}+\beta(2\eta_{1}
+\eta_{2}-2\eta_{3}+\eta_{4}-2\eta_{6}+2\eta_{8})+3(u_{0}-1)\zeta_{1}]-\beta\mathbb{A}(u_{0})\}
\nonumber\\&+&\frac{5\beta}{54}\chi\varphi_{\gamma}(u_{0})+\frac{1}{108M^{2}}\{(1-u_{0})(\beta-5)\zeta_{1}+9\beta
m_Q^{2}\chi\varphi_{\gamma}(u_{0}\}\left.\vphantom{\int_0^{x_2}}\right]
\nonumber\\&+&<q_{1}q_{1}><q_{2}q_{2}>\left[\frac{\beta m_Q^2\mathbb{A}(u_{0})}{12M^{2}}+\frac{1}{36}\{-4[\eta_{2}-2\eta_{3}\right.\nonumber\\&-&\eta_{4}+2\eta_{5}+\beta(2\eta_{1}
+\eta_{2}-2\eta_{3}+\eta_{4}-2\eta_{6}+2\eta_{8})+3(u_{0}-1)\zeta_{1}]+3\beta\mathbb{A}(u_{0})\}
\left.\vphantom{\int_0^{x_2}}\right]\nonumber\\&+&m_{0}^{2}<q_{2}q_{2}>\left[\frac{-\beta m_Q}{16\pi^{2}}+\{\frac{\beta m_{Q}^{3}}{24M^{4}}+\frac{(1+\beta)m_{Q}}{216M^{2}}\}f_{3\gamma}\psi^a(u_{0})\right]
\nonumber\\&+&<q_{2}q_{2}>\left[\frac{-\beta}{6}f_{3\gamma}m_{Q}\psi^a(u_{0})\right],
\end{eqnarray}
\begin{eqnarray}\label{Gamma1'}
\sqrt{3}\Gamma^{'}_{1}&=&m_{0}^{2}<q_{1}q_{1}><q_{2}q_{2}>\left[\frac{-\beta m_Q^2}{3M^{4}}\vphantom{\int_0^{x_2}}\right]
+<q_{1}q_{1}><q_{2}q_{2}>\left[\frac{2\beta }{3}\vphantom{\int_0^{x_2}}\right],\nonumber
\end{eqnarray}
\begin{eqnarray}\label{Gamma2}
\sqrt{3}\Gamma_{2}&=&m_{0}^{2}<q_{1}q_{1}><q_{2}q_{2}>\left[-\frac{1}{54M^{4}}(u_{0}-1)(\beta-5)\zeta_{2}\right.
\nonumber\\&+&\left.\frac{m_Q^{2}}{18M^{6}}\{3(u_{0}-1)\zeta_{2}-\beta(\zeta_{5}-\zeta_{9})\}\vphantom{\int_0^{x_2}}\right]\nonumber\\
&+&<q_{1}q_{1}><q_{2}q_{2}>\left[\frac{2}{9M^2}\{-3(u_{0}-1)\zeta_{2}+\beta(\zeta_{5}-\zeta_{9})\}\vphantom{\int_0^{x_2}}\right],\nonumber\\
\end{eqnarray}
\begin{eqnarray}\label{Gamma2'}
\Gamma^{'}_{2}=0.
\end{eqnarray}
For $\Xi^{*0}_{b}\rightarrow \Xi^{0}_{b}$,  we have
\begin{eqnarray}\label{magneticmoment11}
G_{1}&=&-\frac{1}{\lambda_{1}\lambda_{2}(m_{1}+m_{2})}e^{\frac{m_{1}^{2}}{M_{1}^{2}}}e^{\frac{m_{2}^{2}}{M_{2}^{2}}}
\left[\vphantom{\int_0^{x_2}}e_{q_{1}}\Theta_{1}-e_{q_{2}}\Theta_{1}(q_{1}\leftrightarrow q_{2})+e_{b}\Theta^{'}_{1}\right]
\nonumber\\
G_{2}&=&\frac{1}{\lambda_{1}\lambda_{2}}e^{\frac{m_{1}^{2}}{M_{1}^{2}}}e^{\frac{m_{2}^{2}}{M_{2}^{2}}}
\left[\vphantom{\int_0^{x_2}}e_{q_{1}}\Theta_{2}-e_{q_{2}}\Theta_{2}(q_{1}\leftrightarrow q_{2})+e_{b}\Theta^{'}_{2}\right],
\end{eqnarray}

where  $q_{1}=u$,  $q_{2}=s$. The functions $\Theta_i[\Theta^{'}_i]$ are also defined  as:
\begin{eqnarray}\label{magneticmoment22}
\Theta_i[\Theta^{'}_i]&=&\int_{m_{Q}^{2}}^{s_{0}}e^{\frac{-s}{M^{2}}}\varrho_{i}(s)[\varrho^{'}_{i}(s)]ds+e^{\frac{-m_Q^2}{M^{2}}}
\Delta_{i}[\Delta^{'}_{i}],
\end{eqnarray}
and
\begin{eqnarray}\label{rhho1}
\varrho_{1}(s)&=&\langle q_{1}q_{1}\rangle \langle
q_{2}q_{2}\rangle\left[\frac{1+2\beta}{9}\chi
\varphi_{\gamma}(u_{0})\right] \nonumber\\&+&m_{0}^{2}\langle
q_{2}q_{2}\rangle
\left[\frac{(1+\beta)}{144m_Q\pi^{2}}(1+\psi_{11})\vphantom{\int_0^{x_2}}\right]
\nonumber\\&+&\langle q_{2}q_{2}\rangle
\left[\vphantom{\int_0^{x_2}} \frac{\beta
m_{Q}}{4\pi^{2}}(\psi_{21}-\psi_{10})\vphantom{\int_0^{x_2}}\right]
\nonumber\\&+&\langle q_{1}q_{1}\rangle
\frac{m_Q}{288\pi^{2}}\left[\vphantom{\int_0^{x_2}}\right.
3(1+5\beta)(\psi_{10}-\psi_{21})\mathbb{A}(u_{0})\nonumber\\&+&2\left(\vphantom{\int_0^{x_2}}\right.
-2(\psi_{10}-\psi_{21})\{(-1+7\beta)\eta_{1}+\eta_{2}-4\eta_{3}+\eta_{4}+2\eta_{5}\nonumber\\&-&
2\eta_{6}-3\eta_{7}+2\eta_{8}+\beta(5\eta_{2}-8\eta_{3}+5\eta_{4}-2\eta_{5}-10\eta_{6}-3\eta_{7}+10\eta_{8})\}
\nonumber\\&+&3(-1+\beta)(-1+\psi_{02}+2\psi_{10}-2\psi_{21})(u_{0}-1)\zeta_{1}\nonumber\\&+&2(-1+\beta)(\eta_{1}
+\eta_{2}-\eta_{4} +\eta_{7})ln(\frac{s}{m_Q^{2}})
\nonumber\\&-&3(1+5\beta)m_Q^2\chi\{2\psi_{10}-\psi_{20}+\psi_{31}-2ln(\frac{s}{m_Q^{2}})\}\varphi_{\gamma}(u_{0})\left.\vphantom{\int_0^{x_2}}\right)\left.\vphantom{\int_0^{x_2}}\right]+
\nonumber\\&+&\frac{m_Q^{4}(\beta-1)}{512\pi^{4}}\left[\vphantom{\int_0^{x_2}}\right.6\psi_{20}-4\psi_{30}
+5\psi_{42}+4\psi_{52}+(6\psi_{20}-4\psi_{30}\nonumber\\&+&8\psi_{41}-5\psi_{42}-4\psi_{52})u_{0}
-12\{\psi_{10}-ln(\frac{s}{m_Q^{2}})\}(1+u_{0})\left.\vphantom{\int_0^{x_2}}\right]\nonumber\\&-&
\frac{f_{3\gamma}m_Q^{2}(\beta-1)}{576\pi^{2}}\left[\vphantom{\int_0^{x_2}}8\{(\psi_{20}-\psi_{31})(\eta'_{1}-\eta'_{2})+
[\psi_{10}-ln(\frac{s}{m_Q^{2}})](3\eta'_{3}-\eta'_{1}+\eta'_{2})\}\right.\nonumber\\&+&18(\psi_{20}-\psi_{31})\psi^a(u_{0})
+3(3\psi_{31}+4\psi_{32}+3\psi_{42})(u_{0}-1)[4\psi^v(u_{0})-\frac{d\psi^{a}(u_{0})}{du_{0}}]\left.\vphantom{\int_0^{x_2}}\right],\nonumber\\
\end{eqnarray}
\begin{eqnarray}\label{rhho1'}
\varrho^{'}_{1}(s)&=&m_{0}^{2}(\langle q_{1}q_{1}\rangle-\langle q_{2}q_{2}\rangle)\left[\vphantom{\int_0^{x_2}}
\frac{5(\beta-1)(1+\psi_{11})}{432\pi^{2}m_Q}
\vphantom{\int_0^{x_2}}\right]\nonumber\\&+&
(\langle q_{1}q_{1}\rangle-\langle q_{2}q_{2}\rangle)\frac{m_Q(\beta-1)}{24\pi^{2}}\left[\vphantom{\int_0^{x_2}}\psi_{10}-\psi_{21}-ln(\frac{s}{m_Q^{2}})\vphantom{\int_0^{x_2}}\right]\nonumber\\&+&
\frac{m_Q^{3}m_{q_{2}}(\beta-1)}{128\pi^{4}}
\left[\vphantom{\int_0^{x_2}}\right.6\psi_{10}-\psi_{20}+\psi_{31}
-2\{3+\psi_{10}\}ln(\frac{s}{m_Q^{2}})\left.\vphantom{\int_0^{x_2}}\right],\nonumber\\
\end{eqnarray}
\begin{eqnarray}\label{rhho2}
\varrho_{2}(s)&=&\langle q_{1}q_{1}\rangle
\frac{1-\beta}{36m_Q\pi^{2}}\left[\vphantom{\int_0^{x_2}}\right.
3(-1+\psi_{03}+2\psi_{12}+\psi_{22})(u_{0}-1)\zeta_{2}\nonumber\\&+&2
\psi_{11}(2\zeta_{3}+\zeta_{5}-2\zeta_{7}-\zeta_{9})+2\{\frac{1}{2}(\zeta_{9}-\zeta_{5})
\nonumber\\&+&(2\psi_{12}+\psi_{22})[2\zeta_{3}+\frac{3}{2}(\zeta_{5}-\zeta_{9})-2\zeta_{7}]\}\left.\vphantom{\int_0^{x_2}}\right]+
\nonumber\\&-&\frac{m_Q^{2}(\beta-1)}{128\pi^{4}}\left[\vphantom{\int_0^{x_2}}\right.(12\psi_{31}
+19\psi_{32}+\psi_{33}+12\psi_{42})(u_{0}-1)u_{0}\left.\vphantom{\int_0^{x_2}}\right]\nonumber\\&-&
\frac{f_{3\gamma}(\beta-1)}{288\pi^{2}}\left[\vphantom{\int_0^{x_2}}4\{(-1+\psi_{02}+4\psi_{10}-4\psi_{21})\eta_{9}
-3(-1+\psi_{02}+2\psi_{10}-2\psi_{21})\eta_{11}\right.\nonumber\\&+&(-1+\psi_{02}-2\psi_{10}+2\psi_{21})\eta_{10}
+3(-1+3\psi_{02}-2\psi_{03})(u_{0}-1)\zeta_{11}\}\nonumber\\&-&3(-1+3\psi_{02}-2\psi_{03})(u_{0}-1)\psi^a(u_{0})
\left.\vphantom{\int_0^{x_2}}\right],\nonumber
\end{eqnarray}
\begin{eqnarray}\label{rhho2'}
\varrho^{'}_{2}(s)&=&0,
\end{eqnarray}
\begin{eqnarray}\label{delta1}
\Delta_{1}&=&m_{0}^{2}<q_{1}q_{1}><q_{2}q_{2}>\left[\frac{(1+2\beta)
m_Q^4\mathbb{A}(u_{0})}{144M^{6}}+\frac{m_Q^{2}}{1296M^{4}}\{12[(1-\beta)(\eta_{2}-2\eta_{3})\right.
\nonumber\\&-&3(1+\beta)\eta_{4}+2(2+\beta)\eta_{5}+2(1+2\beta)(\eta_{6}
+\eta_{7}-\eta_{8})+3(2+\beta)(u_{0}-1)\zeta_{1}]\nonumber\\&+&(1+2\beta)\mathbb{A}(u_{0})\}
-\frac{5(1+2\beta)}{162}\chi\varphi_{\gamma}(u_{0})\nonumber\\&+&\frac{1}{108M^{2}}\{(-1+u_{0})(5+3\beta)\zeta_{1}-3(1+2\beta)
m_Q^{2}\chi\varphi_{\gamma}(u_{0}\}\left.\vphantom{\int_0^{x_2}}\right]
\nonumber\\&+&<q_{1}q_{1}><q_{2}q_{2}>\left[-\frac{(1+2\beta)
m_Q^2\mathbb{A}(u_{0})}{36M^{2}}+\frac{1}{108}\{4[(\beta-1)\eta_{2}+2\eta_{3}\right.\nonumber\\&+&3\eta_{4}
-2(2\eta_{5}+\eta_{6}+\eta_{7}-\eta_{8}-3\zeta_{1})-6u_{0}\zeta_{1}\nonumber\\&-&\beta(2\eta_{3}
-3\eta_{4}+2\eta_{5}+4(\eta_{6}+\eta_{7}-\eta_{8})+3(u_{0}-1)\zeta_{1})]-3(1+2\beta)\mathbb{A}(u_{0})\}
\left.\vphantom{\int_0^{x_2}}\right]\nonumber\\&+&m_{0}^{2}<q_{2}q_{2}>\left[\frac{\beta
m_Q}{16\pi^{2}}-\{\frac{\beta
m_{Q}^{3}}{24M^{4}}+\frac{(1+\beta)m_{Q}}{216M^{2}}\}f_{3\gamma}\psi^a(u_{0})\right]
\nonumber\\&+&<q_{2}q_{2}>\left[\frac{\beta}{6}f_{3\gamma}m_{Q}\psi^a(u_{0})\right],
\end{eqnarray}
\begin{eqnarray}\label{delta1'}
\Delta^{'}_{1}&=&m_{0}^{2}<q_{1}q_{1}><q_{2}q_{2}>\frac{5(\beta-1)}{432}\left[\frac{ m_Q^3m_{q_{2}}}{M^{6}}-\frac{ m_Qm_{q_{2}}}{M^{4}}\vphantom{\int_0^{x_2}}\right]
\nonumber\\&-&<q_{1}q_{1}><q_{2}q_{2}>\left[\frac{(\beta-1) m_Qm_{q_{2}}}{36M^{2}}\vphantom{\int_0^{x_2}}\right],\nonumber
\end{eqnarray}
\begin{eqnarray}\label{delta2}
\Delta_{2}&=&m_{0}^{2}<q_{1}q_{1}><q_{2}q_{2}>\left[\frac{1}{54M^{4}}(u_{0}-1)(3\beta+5)\zeta_{2}\right.
\nonumber\\&+&\left.\frac{m_Q^{2}}{54M^{6}}\{3(2+\beta)(u_{0}-1)\zeta_{2}-(1+2\beta)(\zeta_{5}-\zeta_{9})\}\vphantom{\int_0^{x_2}}\right]\nonumber\\
&+&<q_{1}q_{1}><q_{2}q_{2}>\left[\frac{-2}{27M^2}\{3(2+\beta)(u_{0}-1)\zeta_{2}-(1+2\beta)(\zeta_{5}-\zeta_{9})\}\vphantom{\int_0^{x_2}}\right],\nonumber\\
\end{eqnarray}
\begin{eqnarray}\label{delta2'}
\Delta^{'}_{2}=0.
\end{eqnarray}
 Note that, in the above equations, the  terms proportional to $m_{s}$ are omitted  because of their lengthy
 expressions, but they have been  taken into account in numerical calculations. The contributions of the
terms $\sim <G^2>$ are also calculated, but their numerical values
are very small and therefore for customary in the expressions these
terms are also omitted. The functions  entering Eqs.
(\ref{rho1}-\ref{delta2'}) are given as
\begin{eqnarray}\label{etalar}
\eta_{i} &=& \int {\cal D}\alpha_i \int_0^1 dv f_{i}(\alpha_i)
\delta(\alpha_{ q} + v \alpha_g -  u_0),
\nonumber \\
\eta'_{i} &=& \int {\cal D}\alpha_i \int_0^1 dv g_{i}(\alpha_i)
\delta'(\alpha_{ q} + v \alpha_g -  u_0),
\nonumber \\
\zeta_{i}&=&\int_{u_{0}}^1 h_{i}(u)du~~~(i=1,2,11),\nonumber \\
\zeta_{i}&=& \int {\cal D}\alpha_i \int_0^1 d\bar v h_{i}(\alpha_i)
\theta(\alpha_{ q} + v \alpha_g -  u_0)~~~(i=3-10),\nonumber \\
\psi_{nm}&=&\frac{{( {s-m_{Q}}^2 )
}^n}{s^m{(m_{Q}^{2})}^{n-m}},\nonumber \\
\end{eqnarray}
 and  $f_{1}(\alpha_i)={\cal S}(\alpha_i)$, $f_{2}(\alpha_i)=\tilde{\cal S}(\alpha_i)$, $f_{3}(\alpha_i)=v \tilde{\cal S}(\alpha_i)$,  $f_{4}(\alpha_i)=h_{5}(\alpha_i)={\cal T}_{2}(\alpha_i)$,  $f_{5}(\alpha_i)=h_{6}(\alpha_i)=v{\cal T}_{2}(\alpha_i)$,  $f_{6}(\alpha_i)=h_{8}(\alpha_i)=v{\cal T}_{3}(\alpha_i)$,  $f_{7}(\alpha_i)=h_{9}(\alpha_i)={\cal T}_{4}(\alpha_i)$,  $f_{8}(\alpha_i)=h_{10}(\alpha_i)=v{\cal T}_{4}(\alpha_i)$, $f_{9}(\alpha_i)={\cal A}(\alpha_i)$, $f_{10}(\alpha_i)=g_{3}(\alpha_i)=v{\cal A}(\alpha_i)$, $f_{11}(\alpha_i)=g_{2}(\alpha_i)=v{\cal V}(\alpha_i)$, $g_{1}(\alpha_i)={\cal V}(\alpha_i)$,  $h_{1}(u)=h_{\gamma}(u)$,   $h_{2}(u)=(u-u_{0})h_{\gamma}(u)$, $h_{3}(\alpha_i)={\cal T}_{1}(\alpha_i)$, $h_{4}(\alpha_i)=v{\cal T}_{1}(\alpha_i)$, $h_{7}(\alpha_i)={\cal T}_{3}(\alpha_i)$ and  $h_{11}(u)=\psi^{v}(u)$ are functions in terms of the photon distribution amplitudes. Note that in the above equations, the Borel parameter $M^2$  is defined as $M^{2}=\frac{M_{1}^{2}M_{2}^{2}}{M_{1}^{2}+M_{2}^{2}}$ and
$u_{0}=\frac{M_{1}^{2}}{M_{1}^{2}+M_{2}^{2}}$.  Since the masses of
the initial and final baryons are very close to each other, we  set
$M_{1}^{2}=M_{2}^{2}=2M^{2}$ and $u_{0}=1/2$.

From Eqs. (\ref{magneticmoment1}) and (\ref{magneticmoment11}) it is clear that for the calculation of the transition  magnetic dipole and electric quadrupole moments,  the expressions for the
 residues $\lambda_{1}$ and $\lambda_{2}$ are needed.   These residues are
determined from two point sum rules. For the chosen interpolating currents, we get the following results for the residues $\lambda_{1}$ and $\lambda_{2}$ (see also \cite{kazem1,kazem2}):
\begin{eqnarray}\label{resediueguy1}
\lambda_{1}^{2}e^{\frac{-m_{1}^{2}}{M^{2}}}=A^{2}\left[\vphantom{\int_0^{x_2}}\Pi'+\Pi'(q_{1}\longleftrightarrow q_{2})\right],
\end{eqnarray}
where
\begin{eqnarray}\label{resediueguy2}
\Pi'&=&
\vphantom{\int_0^{x_2}}\int_{m_Q^2}^{s_0}ds~ e^{-s/M^2}\left\{m_{0}^{2}<q_{1}q_{1}>\left[\frac{(m_{q_{1}}-6m_{Q})(\psi_{22}
+2\psi_{12}-1)}{192m_{Q}^{2}\pi^{2}}\right]\right.\nonumber\\&-&<q_{1}q_{1}>\left[\frac{2(\psi_{02}+2\psi_{10}-2\psi_{21}-1)m_{Q}+(\psi_{02}-1)(3m_{q_{1}}-2m_{q_{2}})}{32\pi^{2}}\right]\nonumber\\&-&\frac{m_{Q}^{3}}{512\pi^{4}}\left[-8(3\psi_{31}
+2\psi_{32})m_{q_{1}}+3\{(2\psi_{30}-4\psi_{41}+\frac{5}{2}\psi_{42}+2\psi_{52})m_{Q}-4\psi_{42}m_{q_{1}}\}\right.\nonumber\\&+&\left.\left.6(2\psi_{10}-\psi_{20})(\frac{3}{2}m_{Q}-2m_{q_{1}})-6(4m_{q_{1}}-3m_{Q})ln(\frac{m_{Q}^{2}}{s})\right]\right\}\nonumber\\&+&e^{-m_{Q}^{2}/M^2}\left\{m_{0}^{2}<q_{1}q_{1}><q_{2}q_{2}>\left[\frac{-5m_{Q}^{3}m_{q_{1}}}{144M^{6}}+\frac{m_{Q}(m_{Q}+5m_{q_{1}})}{24M^{4}}-\frac{5}{24M^{2}}\right]\right.\nonumber\\&+&<q_{1}q_{1}><q_{2}q_{2}>\left[\frac{1}{12}-\frac{m_{Q}m_{q_{1}}}{12M^{2}}\right]+m_{0}^{2}<q_{1}q_{1}>\left[\frac{6m_{q_{2}}-7m_{q_{1}}}{192\pi^{2}}\right]\left.\vphantom{\int_0^{x_2}}\right\},
\end{eqnarray}
\begin{eqnarray}\label{residu2}
-\lambda_{2\Sigma_{Q}(\Xi_{Q})}^{2}e^{-m_{\Sigma_{Q}(\Xi_{Q})}^{2}/M^{2}}
&=&\int_{m_{Q}^{2}}^{s_{0}}e^{\frac{-s}{M^{2}}}\sigma_{\Sigma_{Q}(\Xi_{Q})}(s)ds+e^{\frac{-m_Q^2}{M^{2}}}\omega_{\Sigma_{Q}(\Xi_{Q})},
\end{eqnarray}
with
\begin{eqnarray}\label{residurho1}
\sigma_{\Sigma_{Q}}(s)&=&(<\overline{d}d>+<\overline{u}u>)\frac{(\beta^{2}-1)}{64 \pi^{2}}\Bigg\{\frac{m_{0}^{2}}{4 m_{Q}}(6\psi_{00}-13\psi_{02}-6\psi_{11})\nonumber\\&+&3m_{Q}(2\psi_{10}-\psi_{11}-\psi_{12}+2\psi_{21})\Bigg\}\nonumber\\&+&\frac{ m_{Q}^{4}}{2048 \pi^{4}} [5+\beta(2+5\beta)][12\psi_{10}-6\psi_{20}+2\psi_{30}-4\psi_{41}+\psi_{42}-12 ln(\frac{s}{m_{Q}^{2}})],\nonumber\\
\end{eqnarray}
\begin{eqnarray}\label{residurho2}
\sigma_{\Xi_{Q}}(s)&=&(<\overline{s}s>+<\overline{u}u>)\frac{(\beta-1)}{192
\pi^{2}}\Bigg\{\frac{m_{0}^{2}}{4 m_{Q}}[6(1+\beta)\psi_{00}
-(7+11\beta)\psi_{02}\nonumber\\&-&6(1+\beta)\psi_{11}]+(1+5\beta)m_{Q}(2\psi_{10}-\psi_{11}-\psi_{12}+2\psi_{21})\Bigg\}
\nonumber\\&+&\frac{ m_{Q}^{4}}{2048 \pi^{4}}
[5+\beta(2+5\beta)][12\psi_{10}-6\psi_{20}+2\psi_{30}-4\psi_{41}+\psi_{42}-12
ln(\frac{s}{m_{Q}^{2}})], \nonumber\\
\end{eqnarray}
\begin{eqnarray}\label{lamgamma1}
\omega_{\Sigma_{Q}}&=&\frac{ (\beta-1)^{2}}{24}<\overline{d}d><\overline{u}u>\left[\vphantom{\int_0^{x_2}}\right.\frac{m_{Q}^{2}m_{0}^{2}}{2 M^{4}}+\frac{m_{0}^{2}}{4 M^{2}}-1\Bigg],\nonumber\\
\omega_{\Xi_{Q}}&=&\frac{
(\beta-1)}{72}<\overline{s}s><\overline{u}u>\left[\vphantom{\int_0^{x_2}}\right.
\frac{m_{Q}^{2}m_{0}^{2}}{2
M^{4}}(13+11\beta)\nonumber\\&+&\frac{m_{0}^{2}}{4
M^{2}}(25+23\beta)-(13+11\beta)\Bigg],
\end{eqnarray}
\begin{eqnarray}\label{residubey}
\lambda_{2\Lambda_{Q}}&=&\lambda_{2\Xi_{Q}}(s\rightarrow d).
\end{eqnarray}
In the expression for  $\lambda_{2\Xi_{Q}}$  also  the terms
proportional to the strange quark mass are taken into account  in
our numerical calculations, but omitted in the above formulas.
%

\section{Numerical analysis}
This section is devoted to the numerical  analysis for the magnetic dipole $G_{M}$ and electric quadrupole  $G_{E}$ as well as the calculation of the decay rates for the  considered radiative transitions. The input parameters used in the analysis of the sum rules are taken to be: $\uu(1~GeV) = \dd(1~GeV)= -(0.243)^3~GeV^3$, $\ss(1~GeV) = 0.8
\uu(1~GeV)$, $m_0^2(1~GeV) = (0.8\pm0.2)~GeV^2$ \cite{Belyaev},  $\Lambda =
(0.5-1)~GeV$ and $f_{3 \gamma} = - 0.0039~GeV^2$ \cite{Ball}. The value
of the magnetic susceptibility  was
obtained  in different papers as $\chi(1~GeV)=-3.15\pm0.3~GeV^{-2}$  \cite{Ball},  $\chi(1~GeV)=-(2.85\pm0.5)~GeV^{-2}$
\cite{Rohrwild} and   $\chi(1~GeV)=-4.4~GeV^{-2}$\cite{Kogan}. From sum rules for the  magnetic dipole $G_{M}$ and electric quadrupole  $G_{E}$,  it is clear that we also need to know   the explicit form of the  photon DA's \cite{Ball}:
\begin{eqnarray}
\varphi_\gamma(u) &=& 6 u \bar u \left( 1 + \varphi_2(\mu)
C_2^{\frac{3}{2}}(u - \bar u) \right),
\nonumber \\
\psi^v(u) &=& 3 \left(3 (2 u - 1)^2 -1 \right)+\frac{3}{64} \left(15
w^V_\gamma - 5 w^A_\gamma\right)
                        \left(3 - 30 (2 u - 1)^2 + 35 (2 u -1)^4
                        \right),
\nonumber \\
\psi^a(u) &=& \left(1- (2 u -1)^2\right)\left(5 (2 u -1)^2 -1\right)
\frac{5}{2}
    \left(1 + \frac{9}{16} w^V_\gamma - \frac{3}{16} w^A_\gamma
    \right),
\nonumber \\
{\cal A}(\alpha_i) &=& 360 \alpha_q \alpha_{\bar q} \alpha_g^2
        \left(1 + w^A_\gamma \frac{1}{2} (7 \alpha_g - 3)\right),
\nonumber \\
{\cal V}(\alpha_i) &=& 540 w^V_\gamma (\alpha_q - \alpha_{\bar q})
\alpha_q \alpha_{\bar q}
                \alpha_g^2,
\nonumber \\
h_\gamma(u) &=& - 10 \left(1 + 2 \kappa^+\right) C_2^{\frac{1}{2}}(u
- \bar u),
\nonumber \\
\mathbb{A}(u) &=& 40 u^2 \bar u^2 \left(3 \kappa - \kappa^+
+1\right) \nonumber \\ && +
        8 (\zeta_2^+ - 3 \zeta_2) \left[u \bar u (2 + 13 u \bar u) \right.
\nonumber \\ && + \left.
                2 u^3 (10 -15 u + 6 u^2) \ln(u) + 2 \bar u^3 (10 - 15 \bar u + 6 \bar u^2)
        \ln(\bar u) \right],
\nonumber \\
{\cal T}_1(\alpha_i) &=& -120 (3 \zeta_2 + \zeta_2^+)(\alpha_{\bar
q} - \alpha_q)
        \alpha_{\bar q} \alpha_q \alpha_g,
\nonumber \\
{\cal T}_2(\alpha_i) &=& 30 \alpha_g^2 (\alpha_{\bar q} - \alpha_q)
    \left((\kappa - \kappa^+) + (\zeta_1 - \zeta_1^+)(1 - 2\alpha_g) +
    \zeta_2 (3 - 4 \alpha_g)\right),
\nonumber \\
{\cal T}_3(\alpha_i) &=& - 120 (3 \zeta_2 - \zeta_2^+)(\alpha_{\bar
q} -\alpha_q)
        \alpha_{\bar q} \alpha_q \alpha_g,
\nonumber \\
{\cal T}_4(\alpha_i) &=& 30 \alpha_g^2 (\alpha_{\bar q} - \alpha_q)
    \left((\kappa + \kappa^+) + (\zeta_1 + \zeta_1^+)(1 - 2\alpha_g) +
    \zeta_2 (3 - 4 \alpha_g)\right),\nonumber \\
{\cal S}(\alpha_i) &=& 30\alpha_g^2\{(\kappa +
\kappa^+)(1-\alpha_g)+(\zeta_1 + \zeta_1^+)(1 - \alpha_g)(1 -
2\alpha_g)\nonumber \\&+&\zeta_2
[3 (\alpha_{\bar q} - \alpha_q)^2-\alpha_g(1 - \alpha_g)]\},\nonumber \\
\tilde {\cal S}(\alpha_i) &=&-30\alpha_g^2\{(\kappa -
\kappa^+)(1-\alpha_g)+(\zeta_1 - \zeta_1^+)(1 - \alpha_g)(1 -
2\alpha_g)\nonumber \\&+&\zeta_2 [3 (\alpha_{\bar q} -
\alpha_q)^2-\alpha_g(1 - \alpha_g)]\}.
\end{eqnarray}
The constants entering  the above DA's are obtained as
\cite{Ball} $\varphi_2(1~GeV) = 0$, $w^V_\gamma = 3.8 \pm 1.8$,
$w^A_\gamma = -2.1 \pm 1.0$, $\kappa = 0.2$, $\kappa^+ = 0$,
$\zeta_1 = 0.4$, $\zeta_2 = 0.3$, $\zeta_1^+ = 0$ and $\zeta_2^+ =
0$.

\begin{figure}[h!]
\begin{center}
\includegraphics[width=9cm]{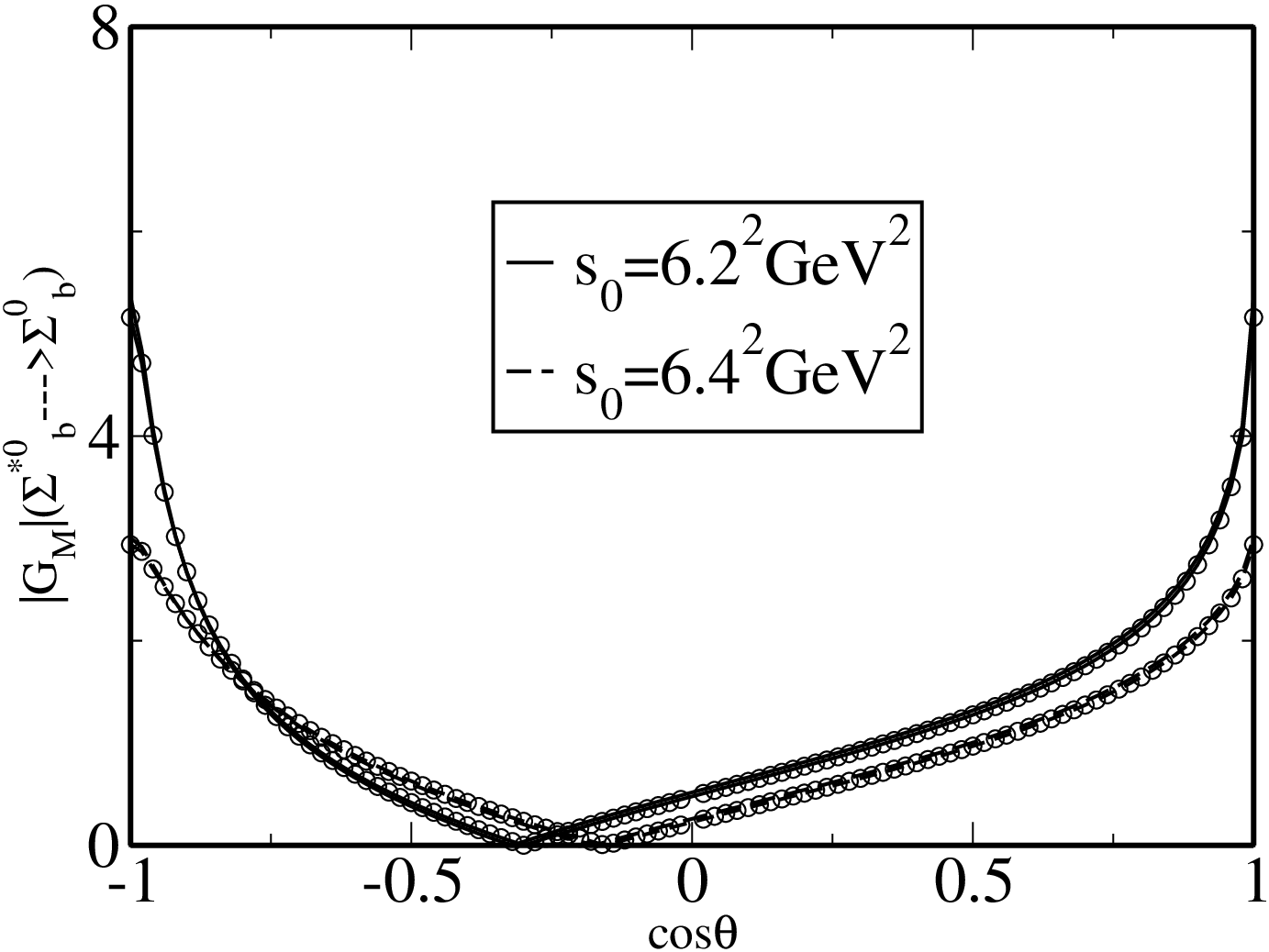}
\end{center}
\caption{The dependence of the magnetic dipole form factor  $G_{M}$ for $\Sigma^{*0}_{b}\rightarrow \Sigma^{0}_{b}$ on $cos\theta$ for two  fixed values of the continuum
threshold $s_{0}$. Bare lines and  lines with the circles  correspond to the $M^{2}=20~GeV^2$ and  $M^{2}=25~GeV^2$, respectively.} \label{fig25}
\end{figure}

\begin{figure}[h!]
\begin{center}
\includegraphics[width=9cm]{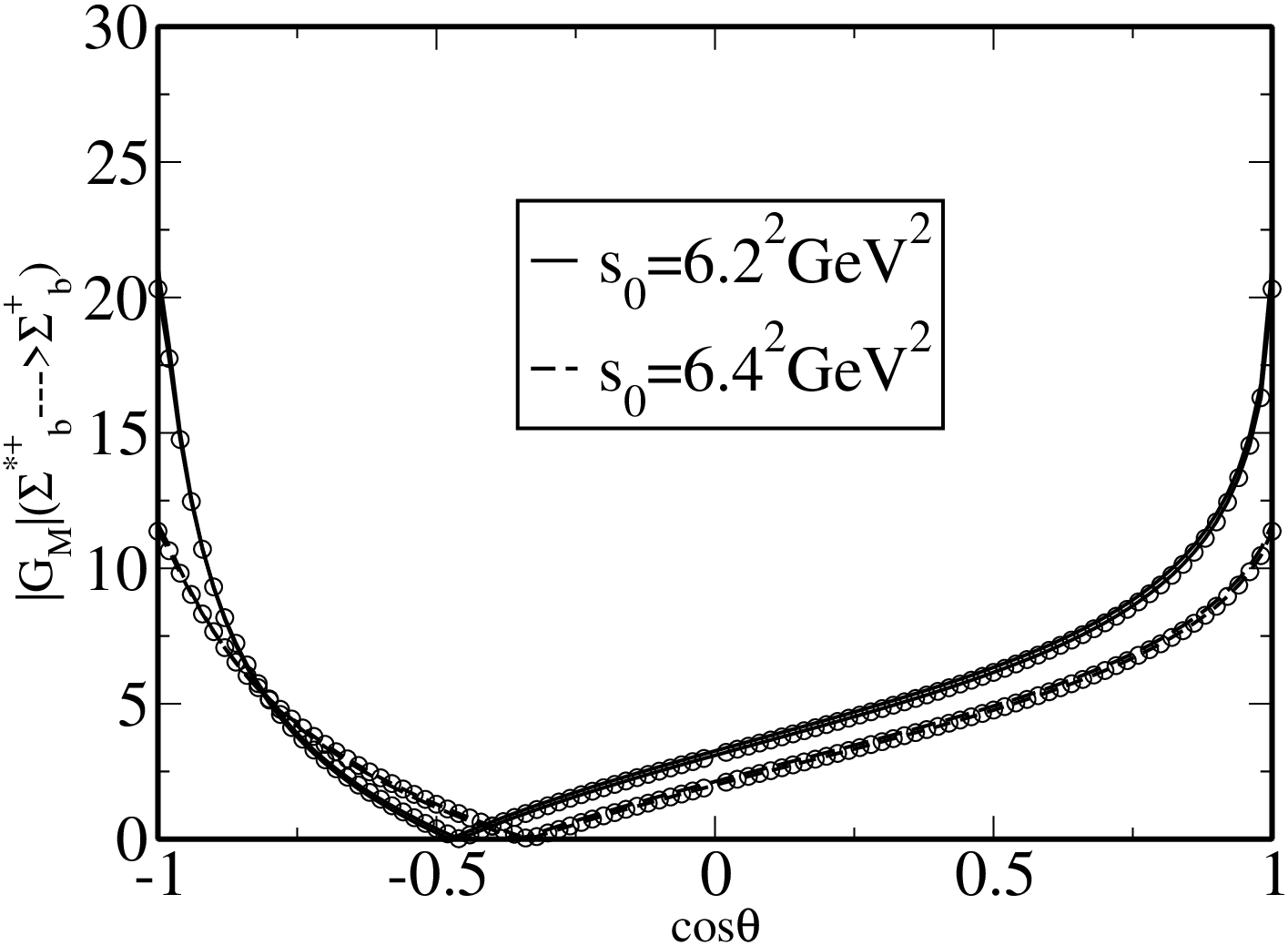}
\end{center}
\caption{The same as Fig. \ref{fig25}, but  for $\Sigma^{*+}_{b}\rightarrow \Sigma^{+}_{b}$.} \label{fig26}
\end{figure}

\begin{figure}[h!]
\begin{center}
\includegraphics[width=9cm]{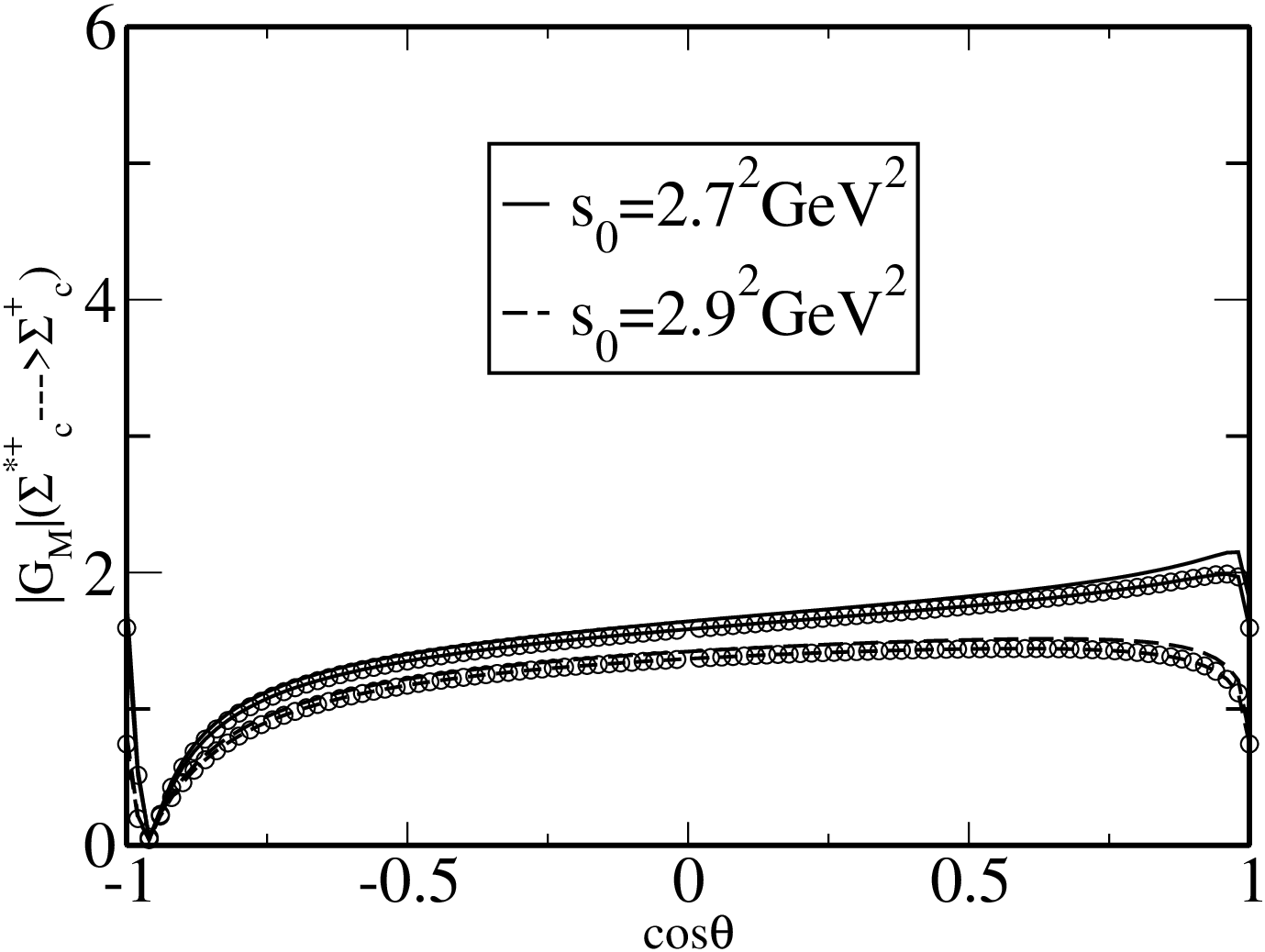}
\end{center}
\caption{The same as Fig. \ref{fig25}, but  for $\Sigma^{*+}_{c}\rightarrow \Sigma^{+}_{c}$ and   $M^{2}=6~GeV^2$ and  $M^{2}=9~GeV^2$.} \label{fig29}
\end{figure}

\begin{figure}[h!]
\begin{center}
\includegraphics[width=9cm]{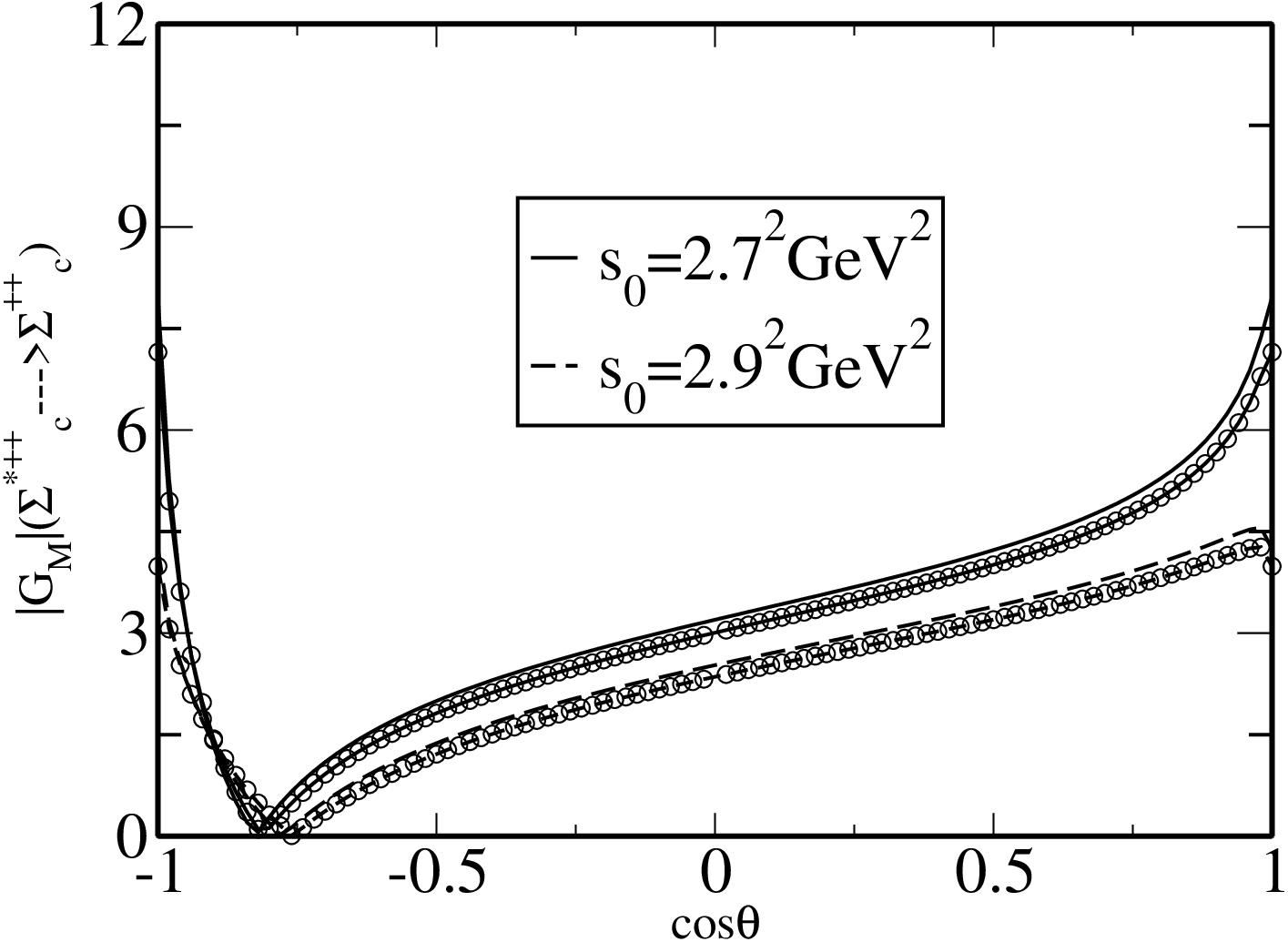}
\end{center}
\caption{The same as Fig. \ref{fig29}, but  for $\Sigma^{*++}_{c}\rightarrow \Sigma^{++}_{c}$.} \label{fig30}
\end{figure}

\begin{figure}[h!]
\begin{center}
\includegraphics[width=9cm]{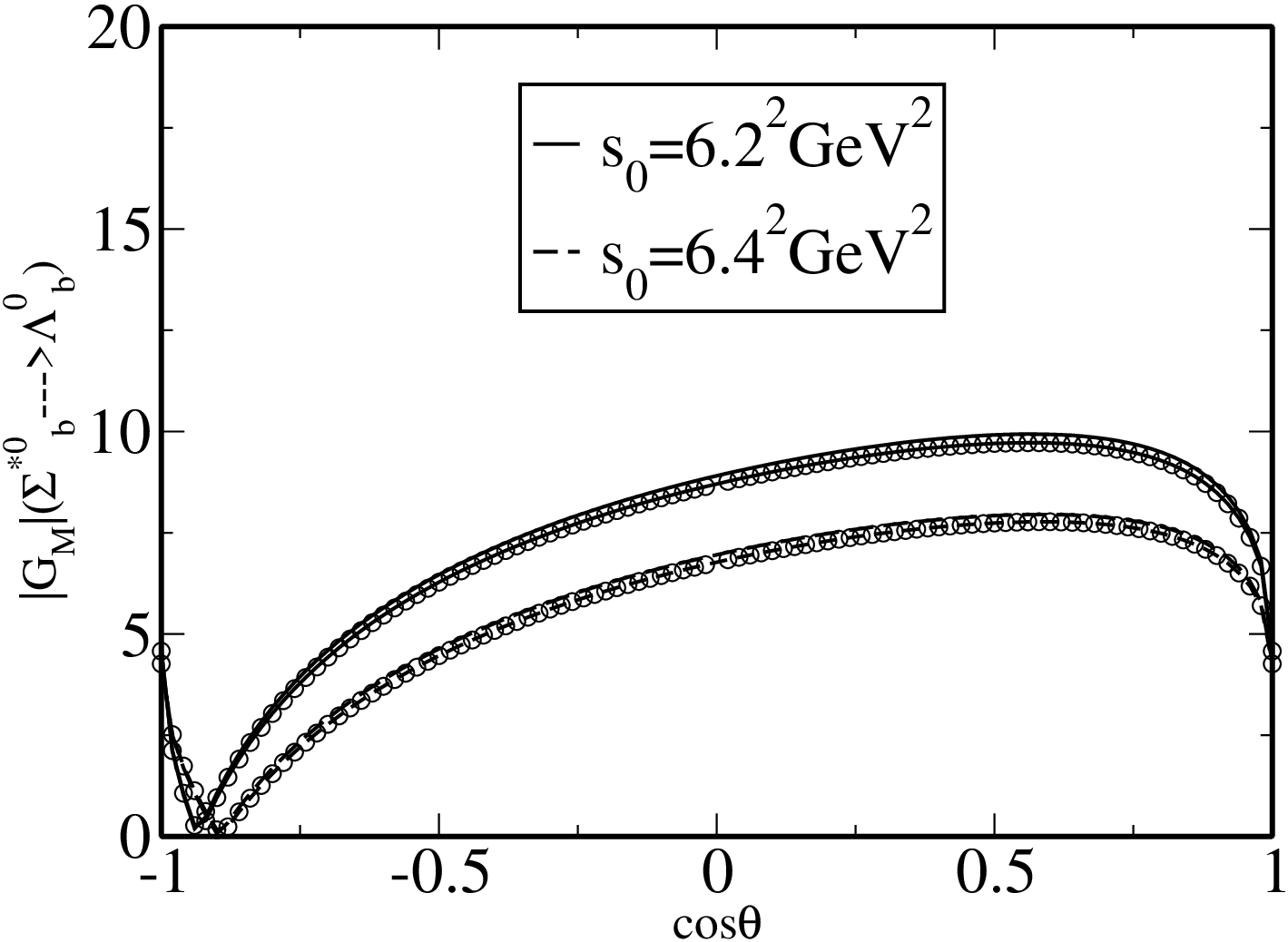}
\end{center}
\caption{The same as Fig. \ref{fig25}, but  for $\Sigma^{*0}_{b}\rightarrow \Lambda^{0}_{b}$.} \label{fig31}
\end{figure}

\begin{figure}[h!]
\begin{center}
\includegraphics[width=9cm]{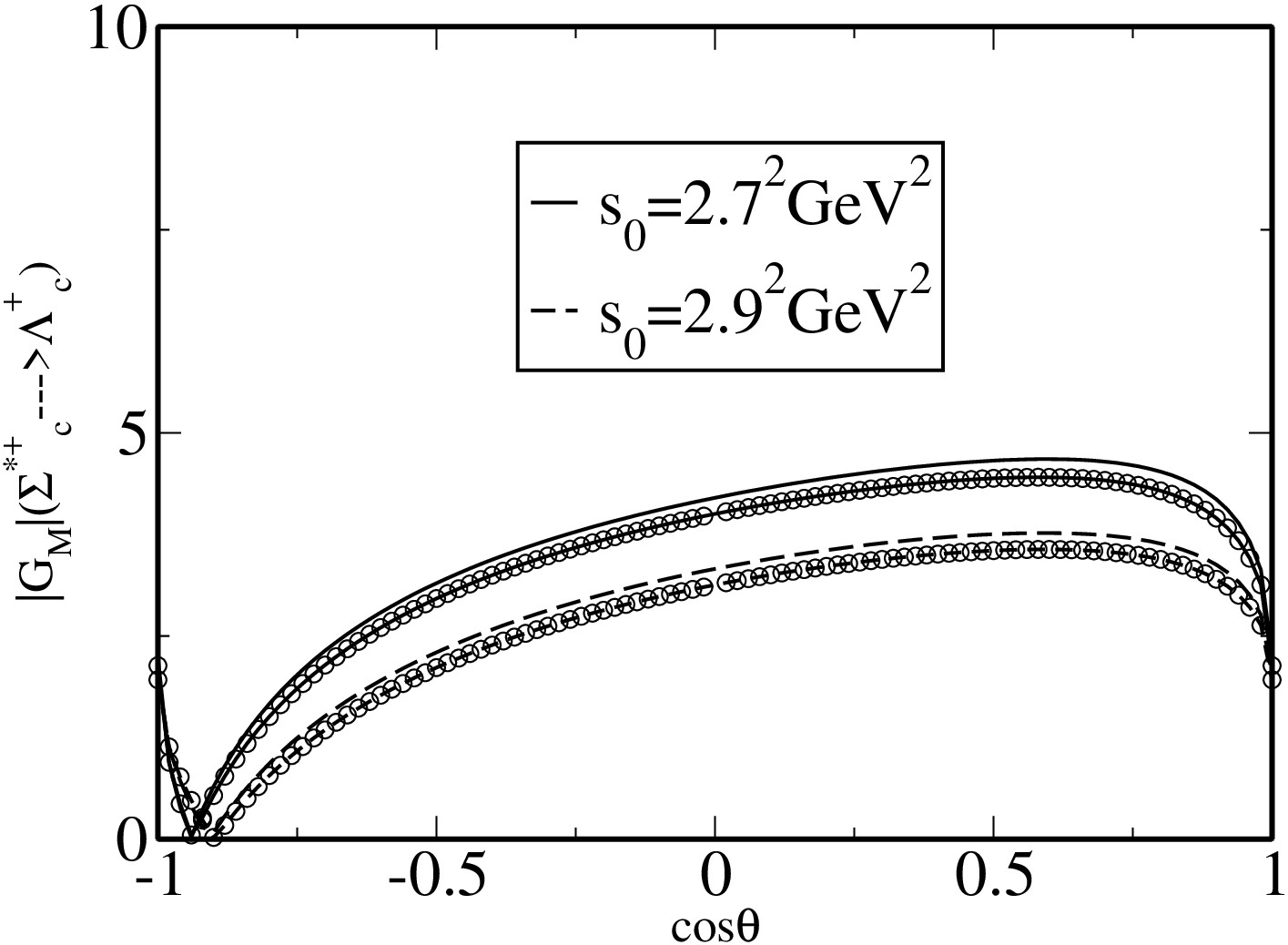}
\end{center}
\caption{The same as Fig. \ref{fig29}, but  for $\Sigma^{*+}_{c}\rightarrow \Lambda^{+}_{c}$.} \label{fig32}
\end{figure}

\begin{figure}[h!]
\begin{center}
\includegraphics[width=9cm]{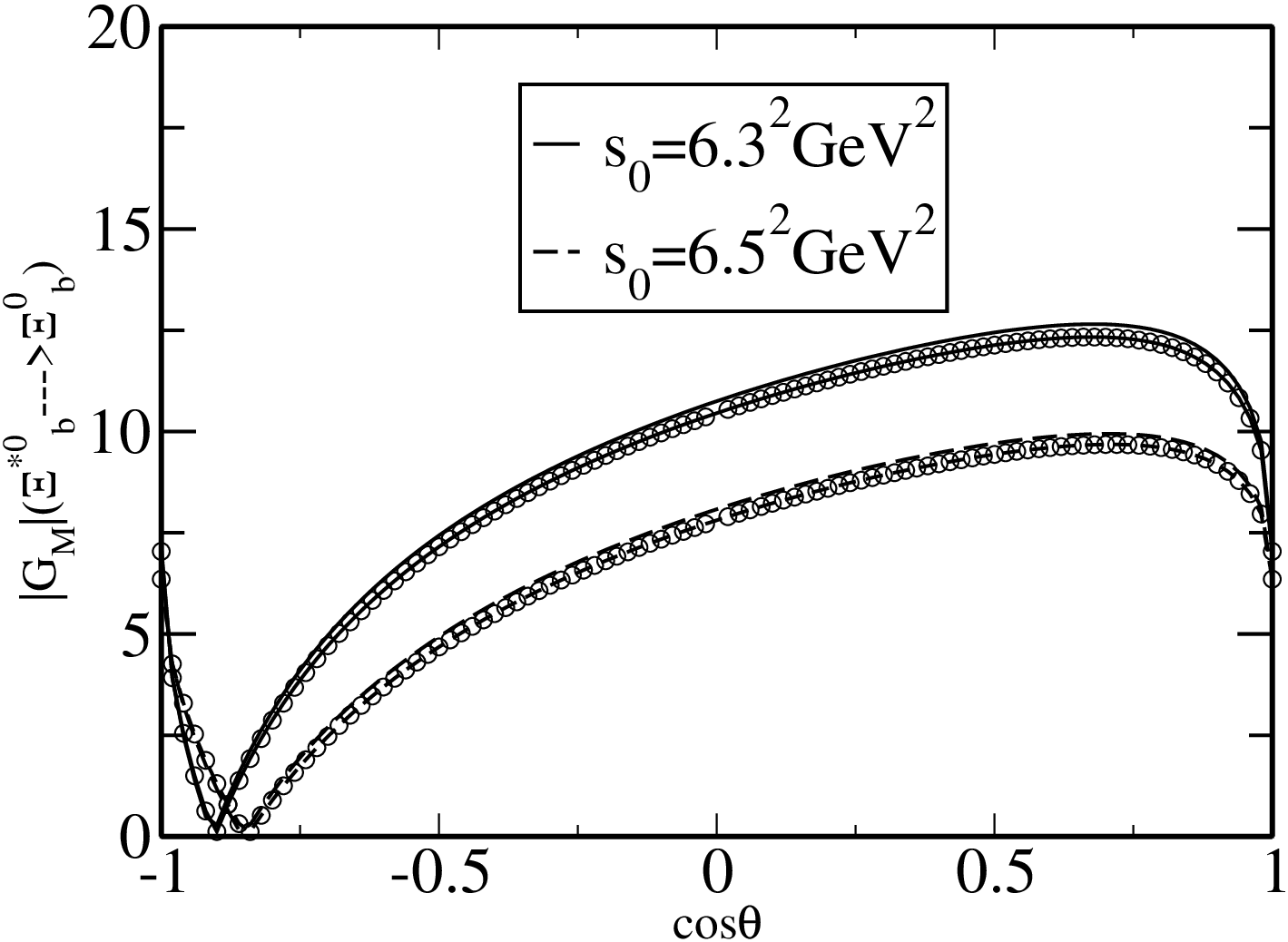}
\end{center}
\caption{The same as Fig. \ref{fig25}, but  for $\Xi^{*0}_{b}\rightarrow \Xi^{0}_{b}$.} \label{fig33}
\end{figure}

\begin{figure}[h!]
\begin{center}
\includegraphics[width=9cm]{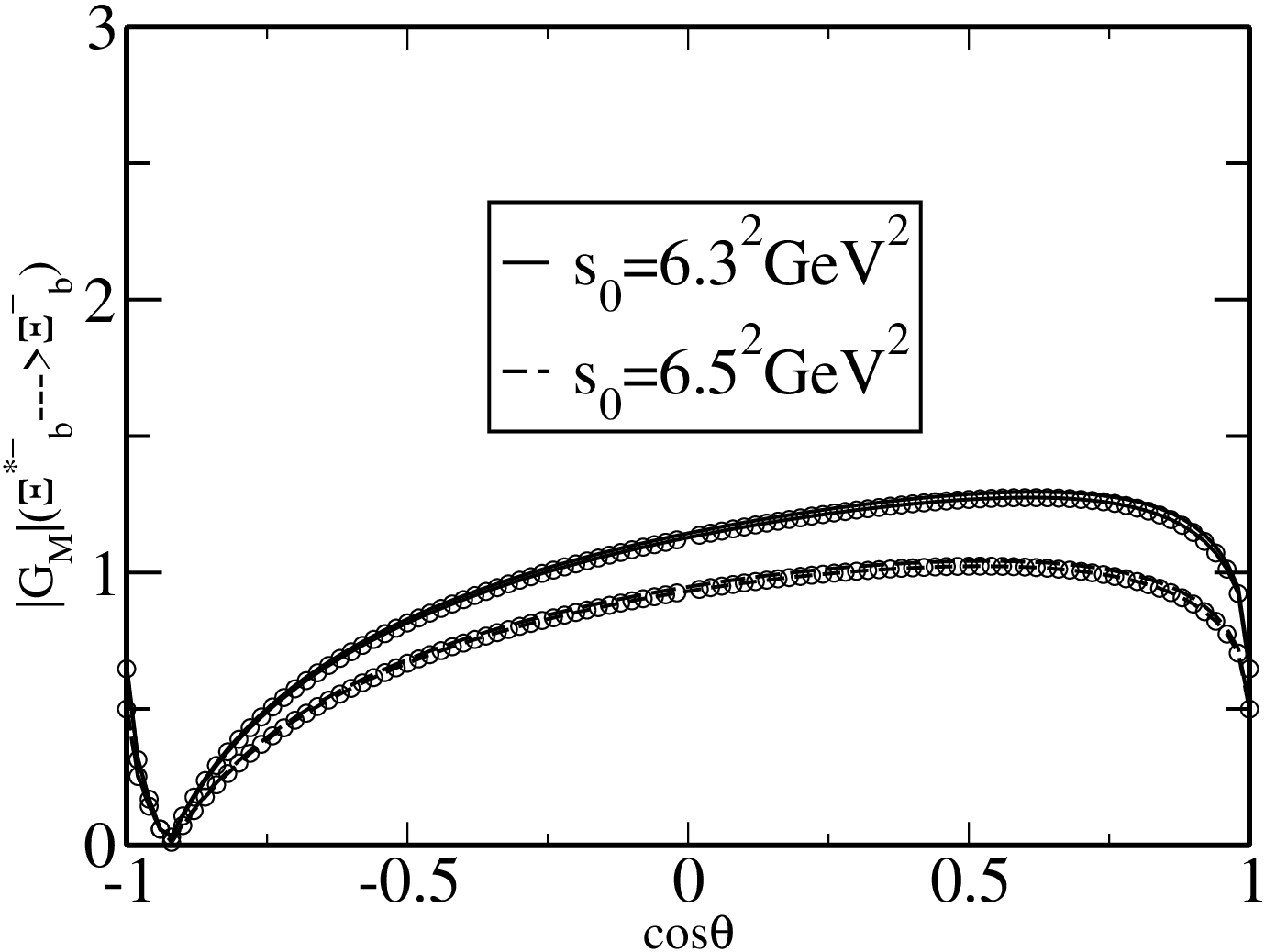}
\end{center}
\caption{The same as Fig. \ref{fig25}, but  for $\Xi^{*-}_{b}\rightarrow \Xi^{-}_{b}$.} \label{fig34}
\end{figure}

\begin{figure}[h!]
\begin{center}
\includegraphics[width=9cm]{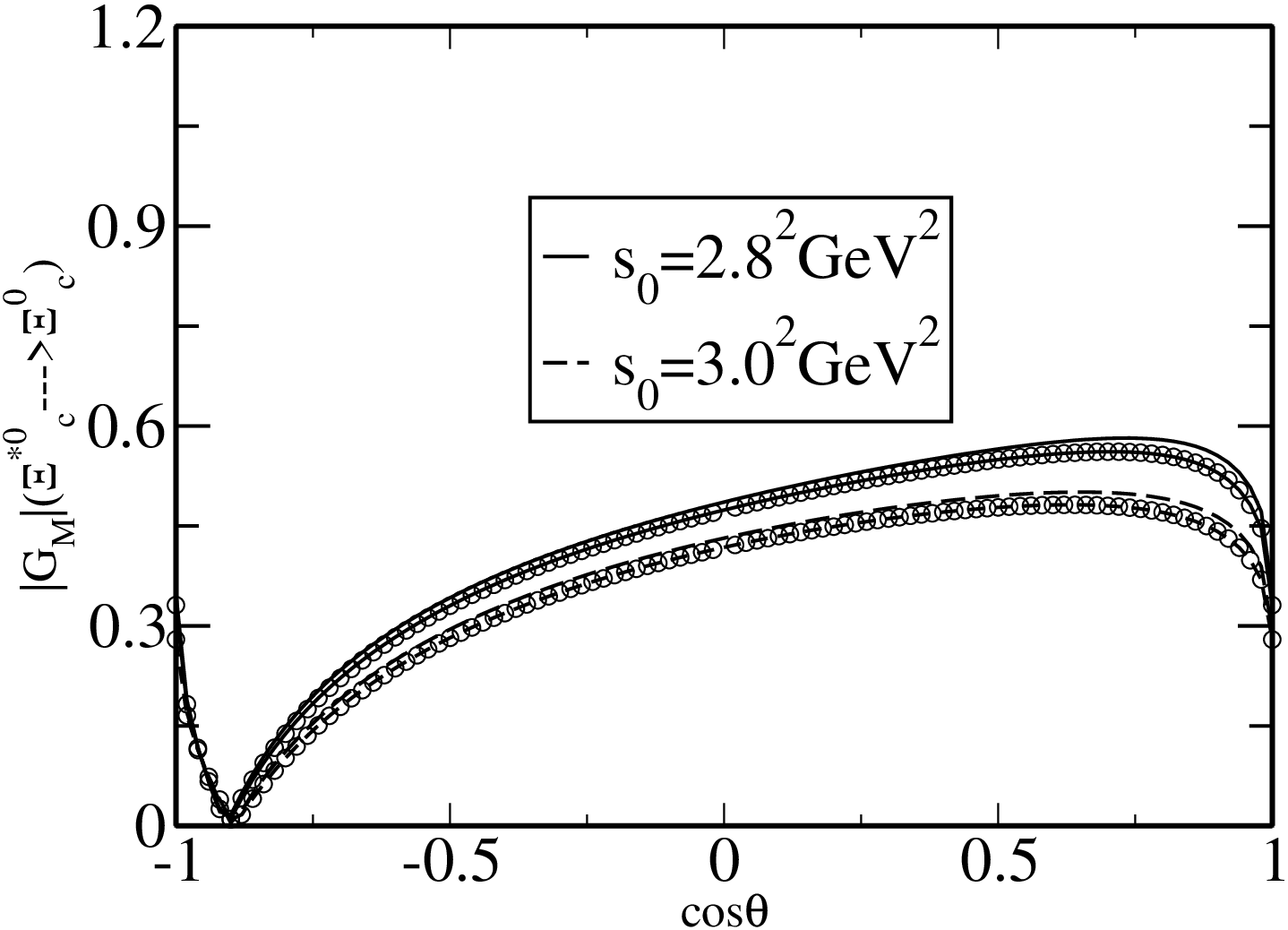}
\end{center}
\caption{The same as Fig. \ref{fig29}, but  for $\Xi^{*0}_{c}\rightarrow \Xi^{0}_{c}$.} \label{fig35}
\end{figure}

\begin{figure}[h!]
\begin{center}
\includegraphics[width=9cm]{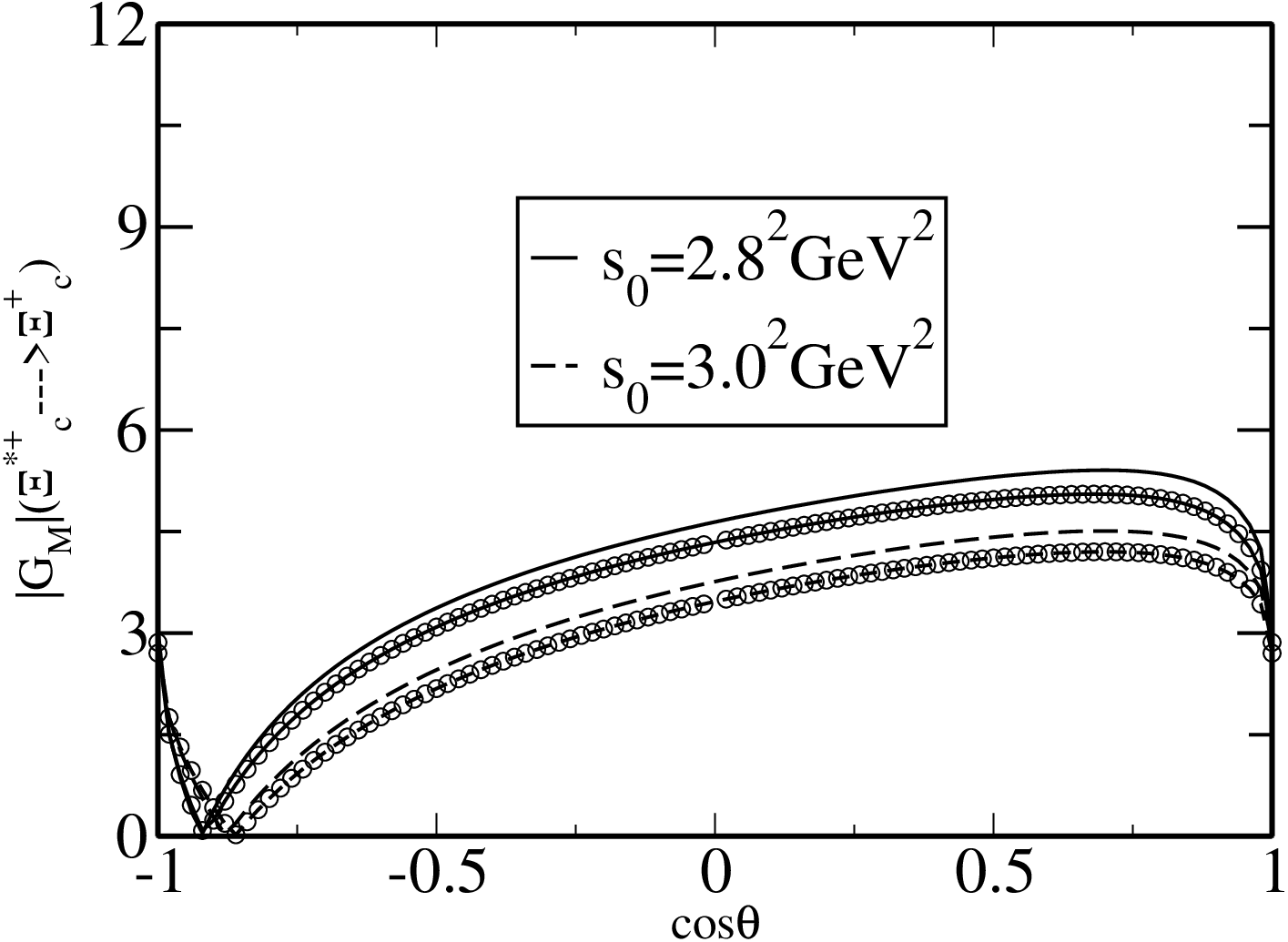}
\end{center}
\caption{The same as Fig. \ref{fig29}, but  for $\Xi^{*+}_{c}\rightarrow \Xi^{+}_{c}$.} \label{fig36}
\end{figure}

\begin{figure}[h!]
\begin{center}
\includegraphics[width=9cm]{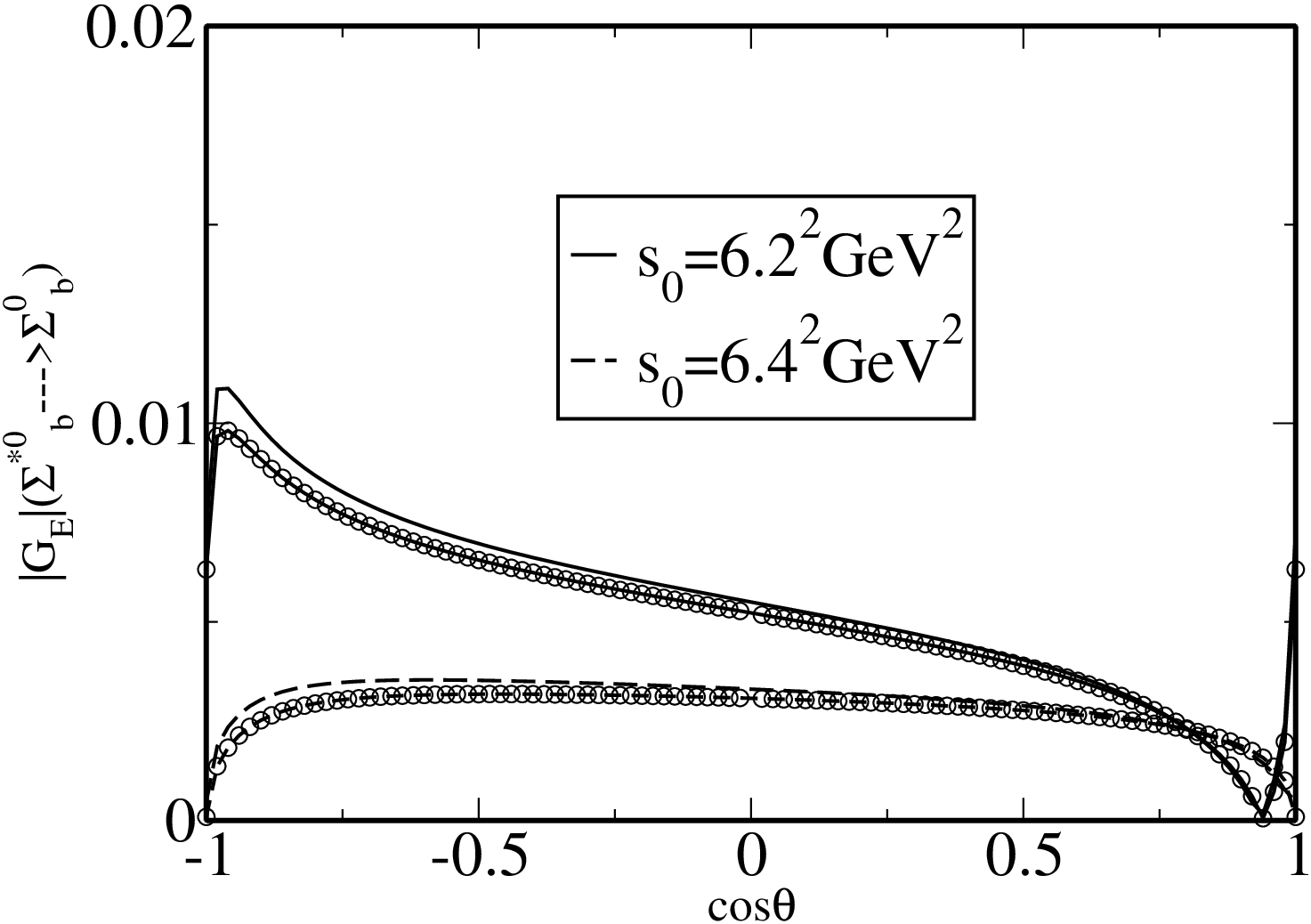}
\end{center}
\caption{The dependence of the electric dipole form factor  $G_{E}$ for $\Sigma^{*0}_{b}\rightarrow \Sigma^{0}_{b}$ on $cos\theta$ for two  fixed values of the continuum
threshold $s_{0}$. Bare lines and  lines with the circles  correspond to the $M^{2}=20~GeV^2$ and  $M^{2}=25~GeV^2$, respectively.} \label{fig37}
\end{figure}

\begin{figure}[h!]
\begin{center}
\includegraphics[width=9cm]{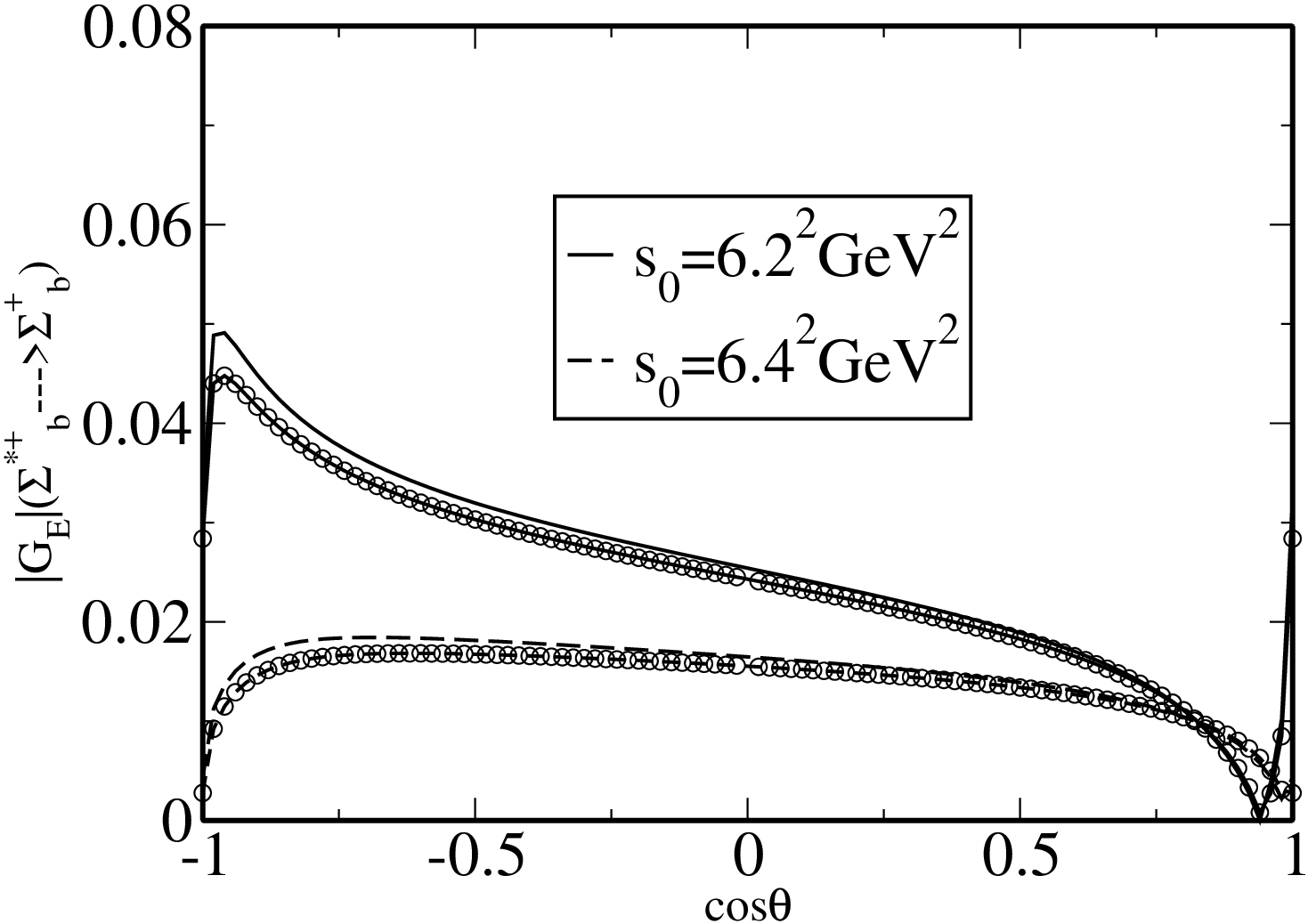}
\end{center}
\caption{The same as Fig. \ref{fig37}, but  for $\Sigma^{*+}_{b}\rightarrow \Sigma^{+}_{b}$.} \label{fig38}
\end{figure}

\begin{figure}[h!]
\begin{center}
\includegraphics[width=9cm]{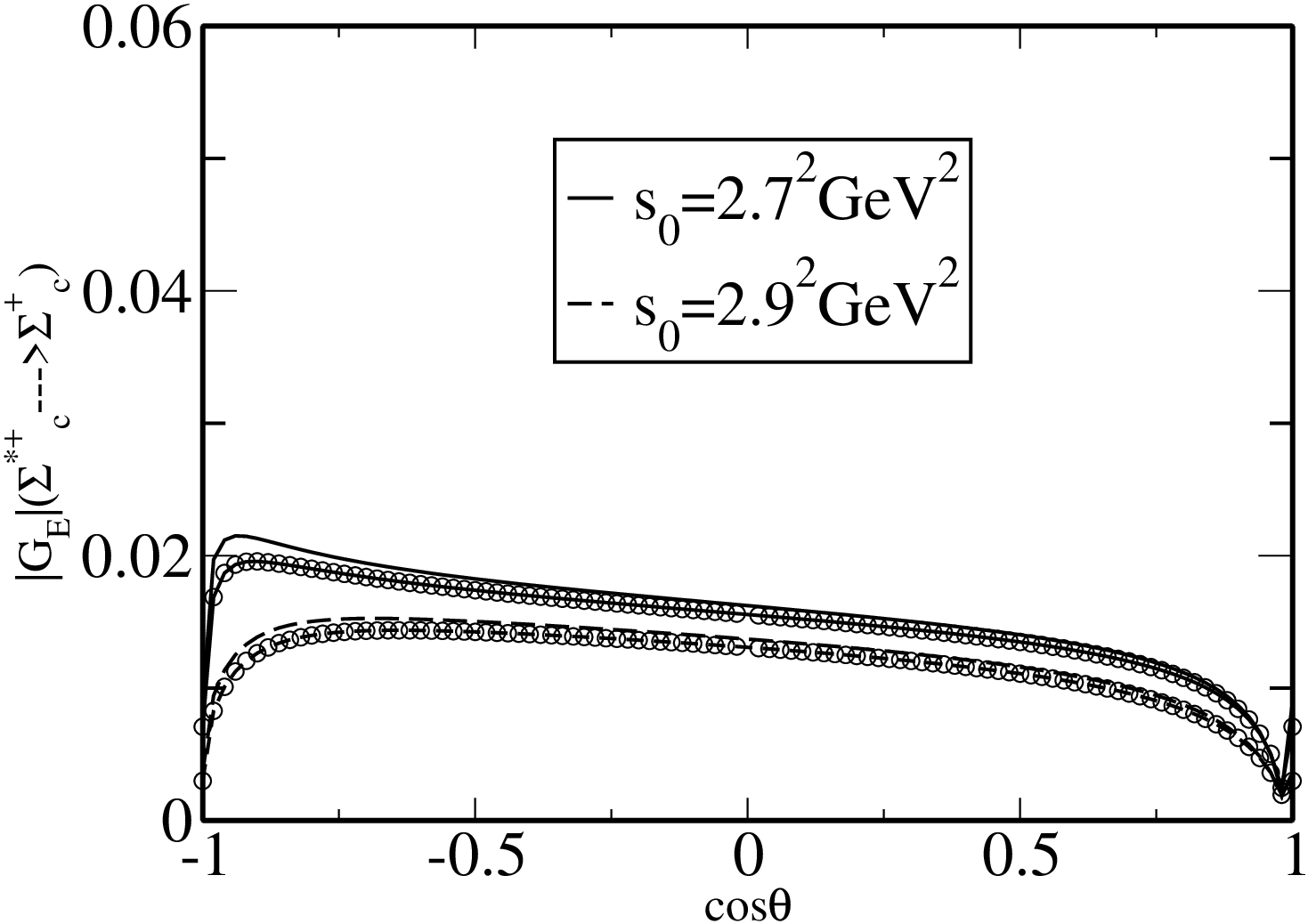}
\end{center}
\caption{The same as Fig. \ref{fig37}, but  for $\Sigma^{*+}_{c}\rightarrow \Sigma^{+}_{c}$ and   $M^{2}=6~GeV^2$ and  $M^{2}=9~GeV^2$.} \label{fig41}
\end{figure}

\begin{figure}[h!]
\begin{center}
\includegraphics[width=9cm]{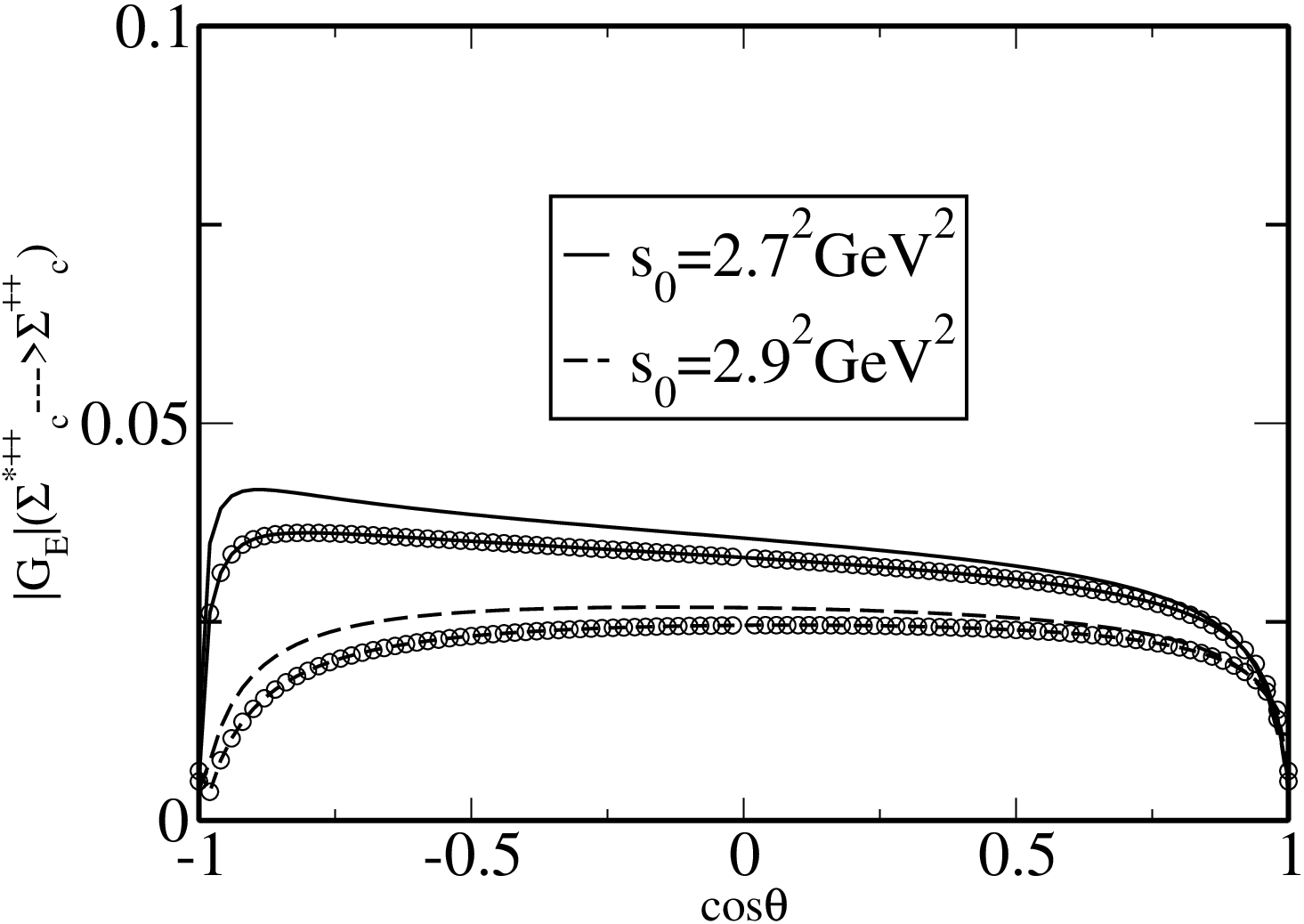}
\end{center}
\caption{The same as Fig. \ref{fig41}, but  for $\Sigma^{*++}_{c}\rightarrow \Sigma^{++}_{c}$.} \label{fig42}
\end{figure}

\begin{figure}[h!]
\begin{center}
\includegraphics[width=9cm]{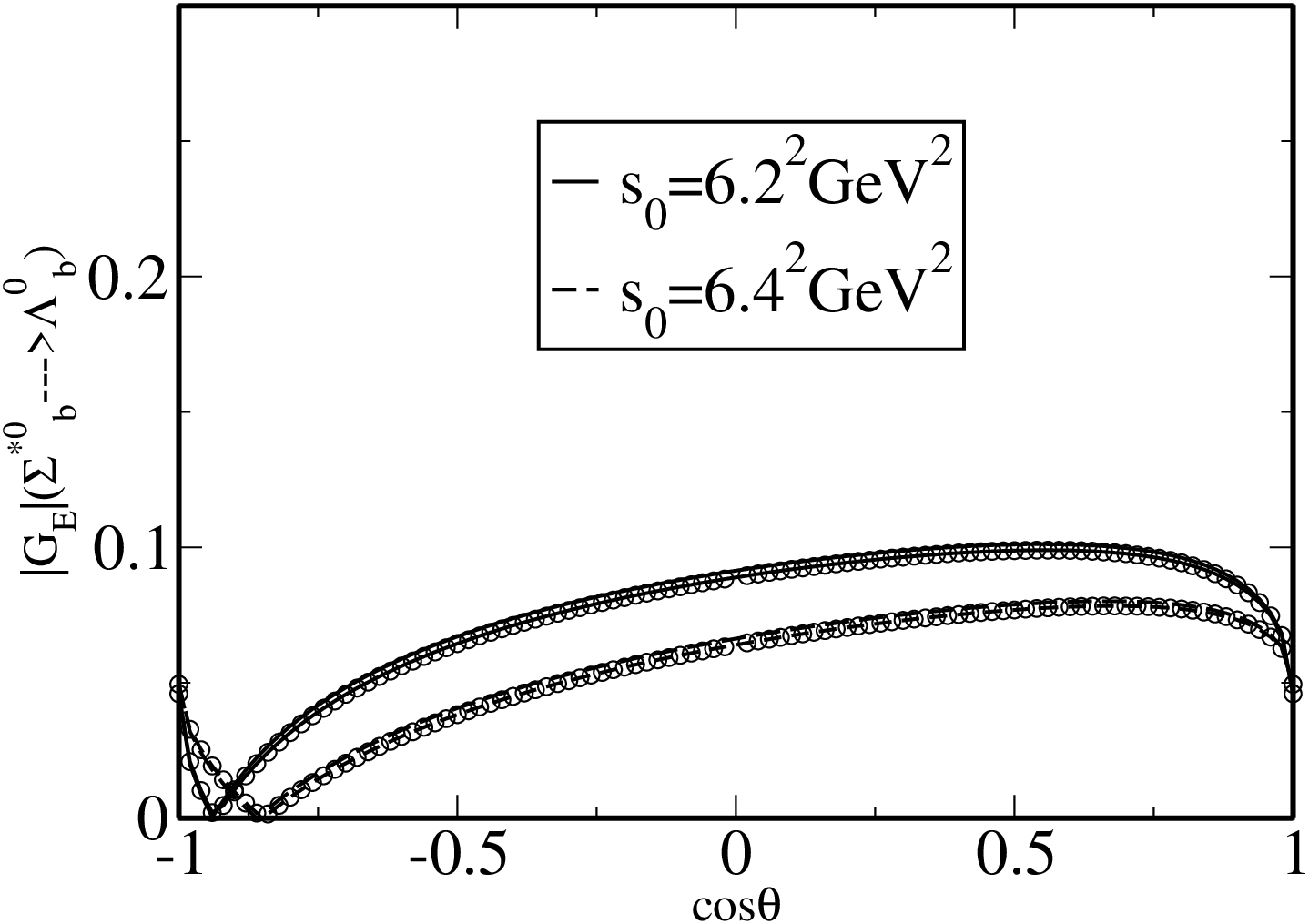}
\end{center}
\caption{The same as Fig. \ref{fig37}, but  for $\Sigma^{*0}_{b}\rightarrow \Lambda^{0}_{b}$.} \label{fig43}
\end{figure}

\begin{figure}[h!]
\begin{center}
\includegraphics[width=9cm]{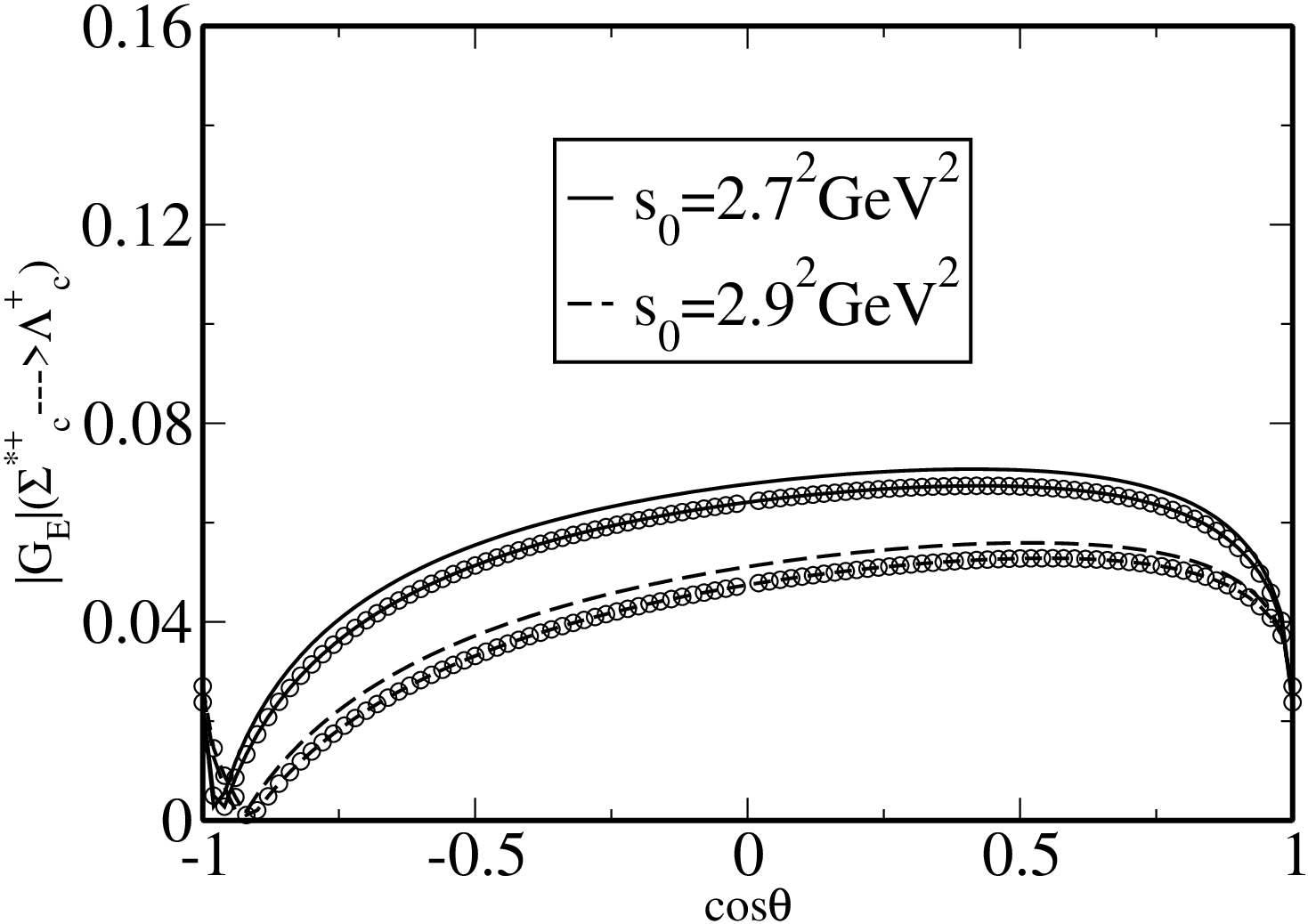}
\end{center}
\caption{The same as Fig. \ref{fig41}, but  for $\Sigma^{*+}_{c}\rightarrow \Lambda^{+}_{c}$.} \label{fig44}
\end{figure}

\begin{figure}[h!]
\begin{center}
\includegraphics[width=9cm]{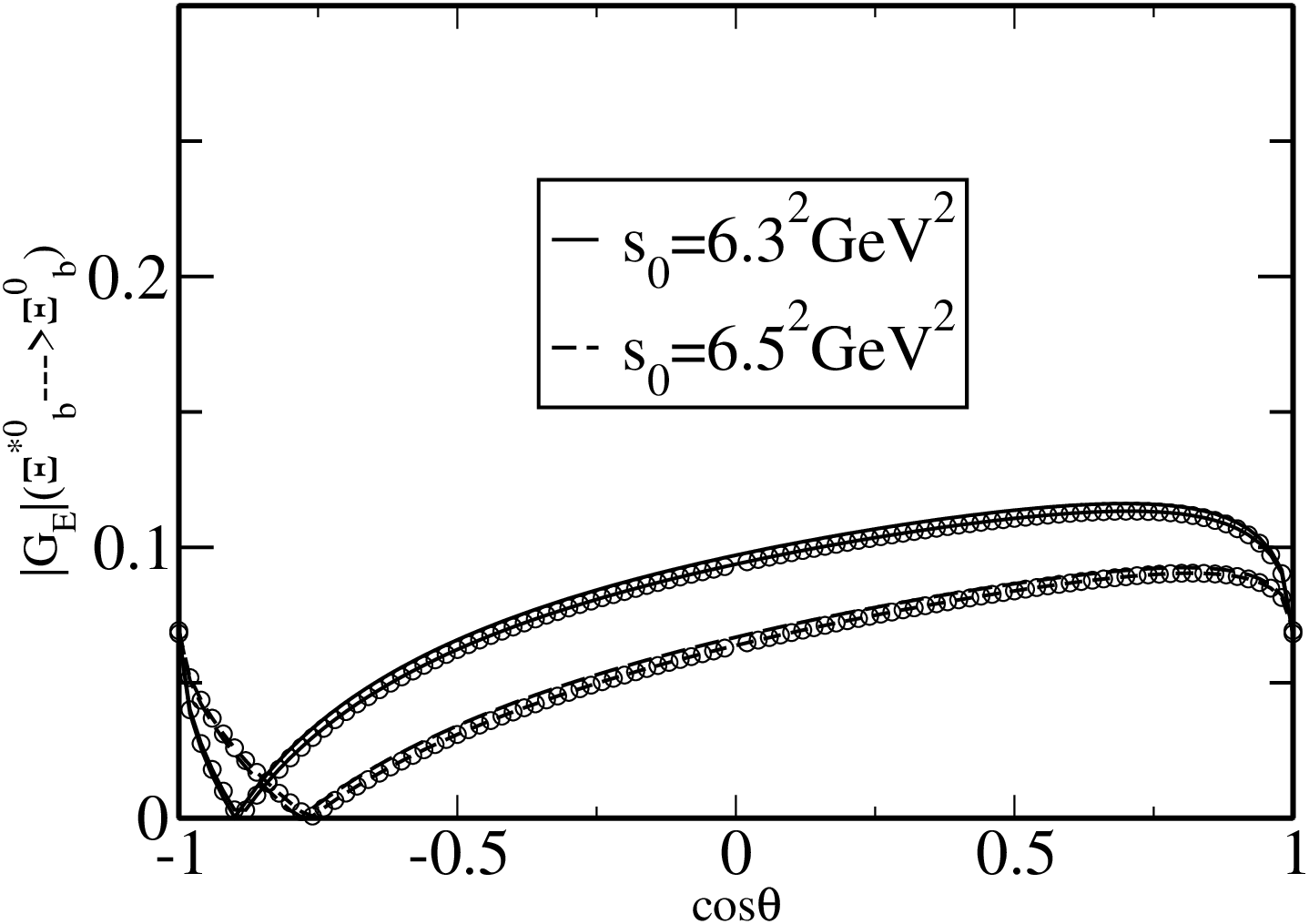}
\end{center}
\caption{The same as Fig. \ref{fig37}, but  for $\Xi^{*0}_{b}\rightarrow \Xi^{0}_{b}$.} \label{fig45}
\end{figure}

\begin{figure}[h!]
\begin{center}
\includegraphics[width=9cm]{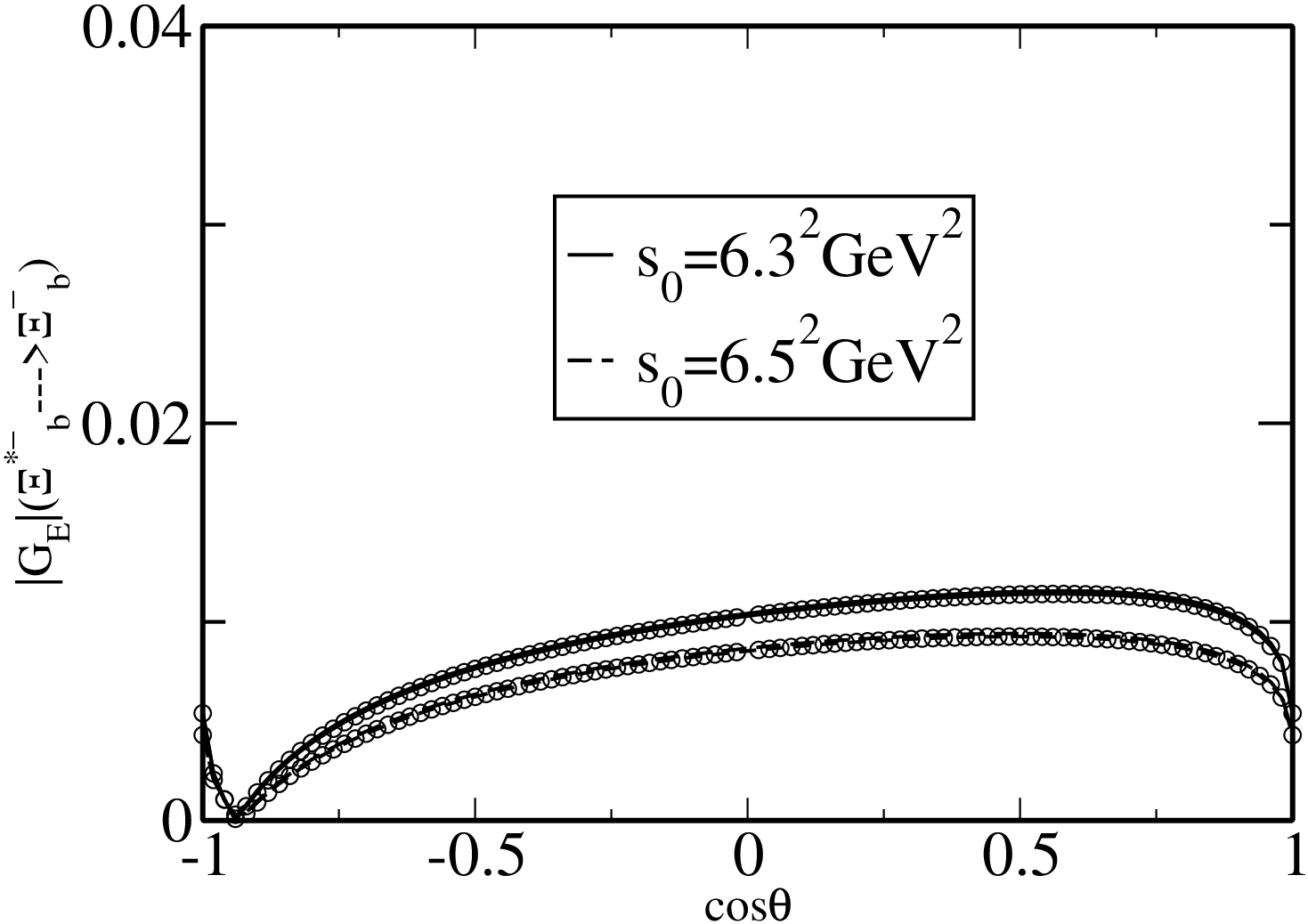}
\end{center}
\caption{The same as Fig. \ref{fig37}, but  for $\Xi^{*-}_{b}\rightarrow \Xi^{-}_{b}$.} \label{fig46}
\end{figure}

\begin{figure}[h!]
\begin{center}
\includegraphics[width=9cm]{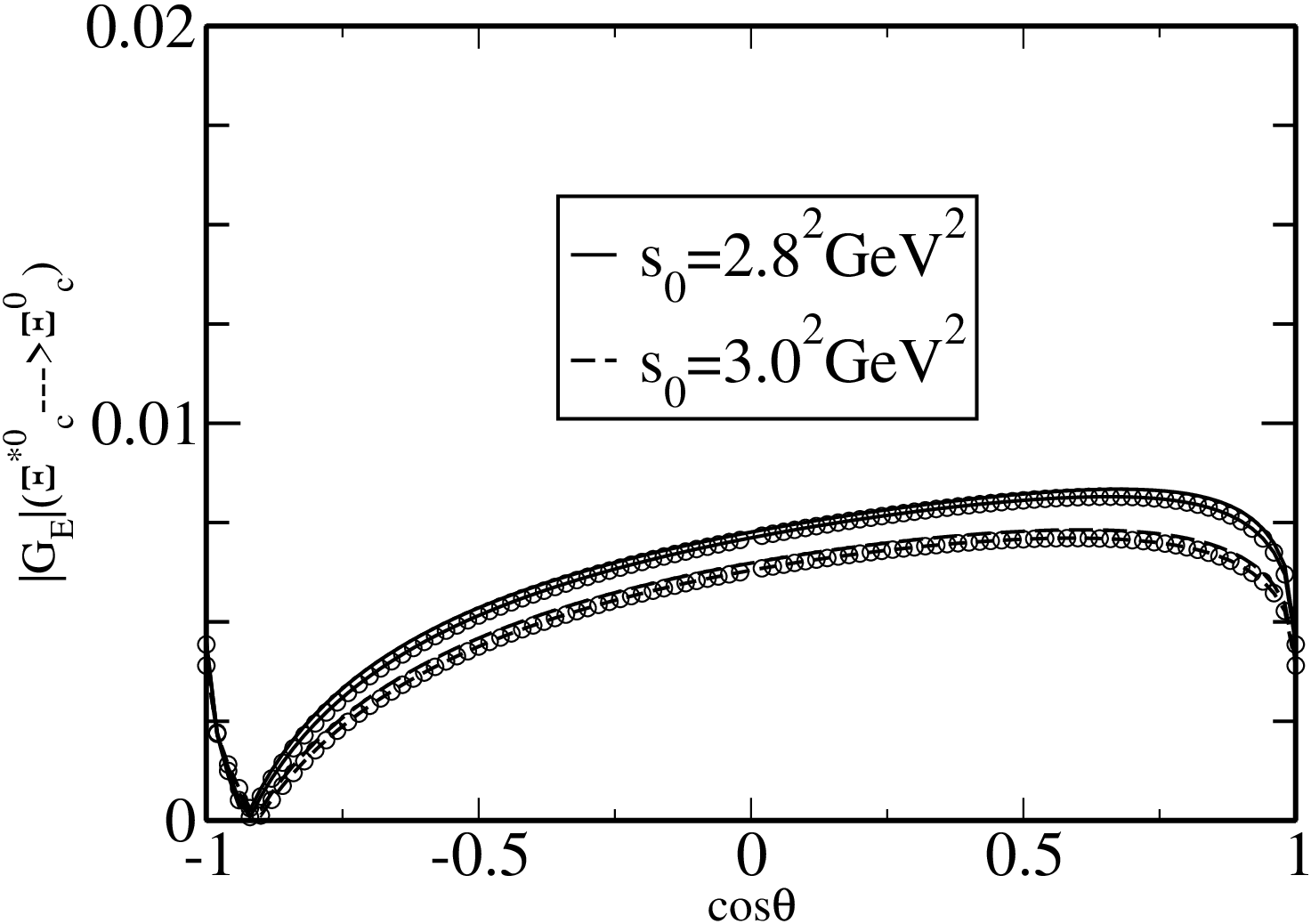}
\end{center}
\caption{The same as Fig. \ref{fig41}, but  for $\Xi^{*0}_{c}\rightarrow \Xi^{0}_{c}$.} \label{fig47}
\end{figure}

\begin{figure}[h!]
\begin{center}
\includegraphics[width=9cm]{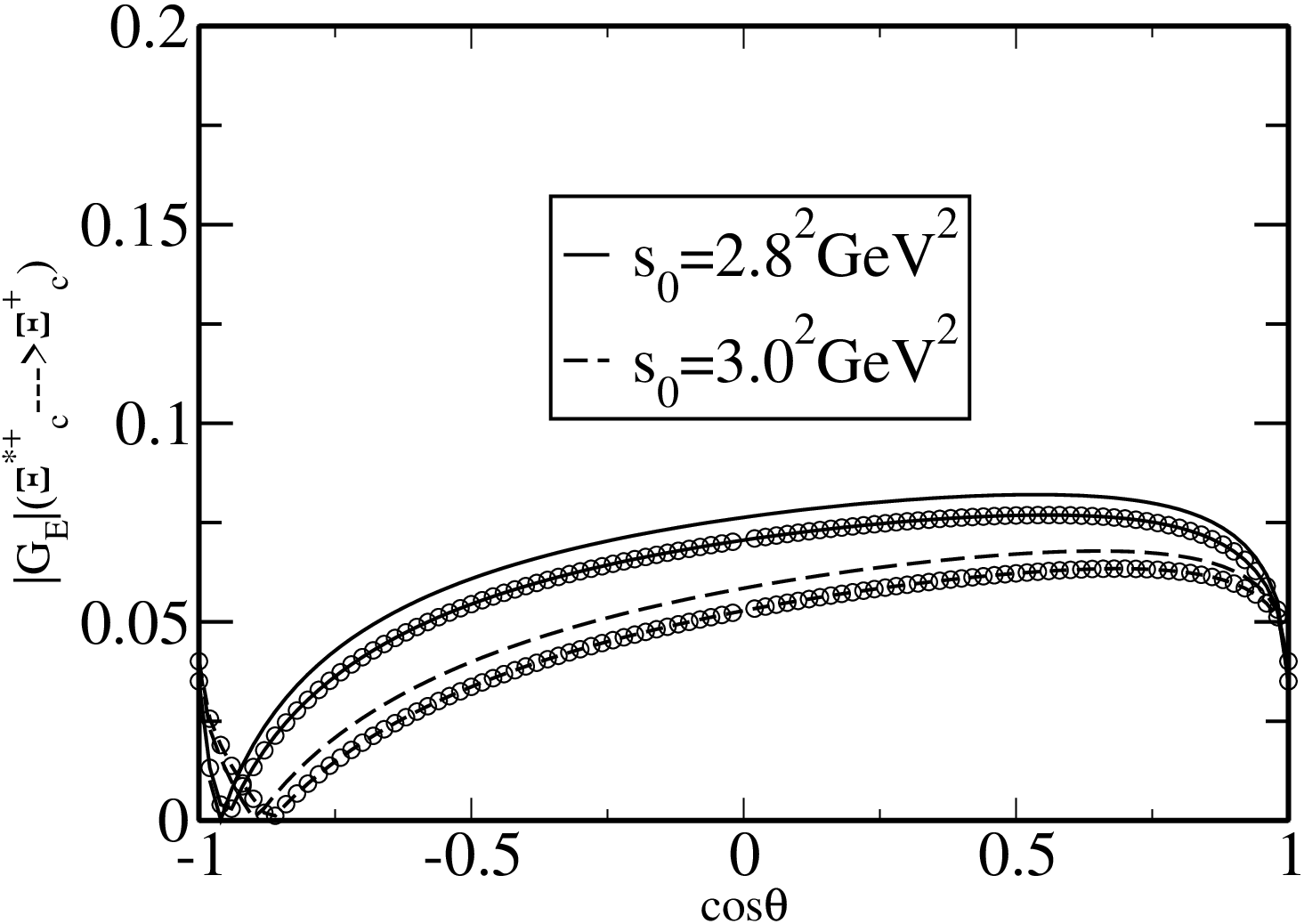}
\end{center}
\caption{The same as Fig. \ref{fig41}, but  for $\Xi^{*+}_{c}\rightarrow \Xi^{+}_{c}$.} \label{fig48}
\end{figure}

\begin{figure}[h!]
\begin{center}
\includegraphics[width=9cm]{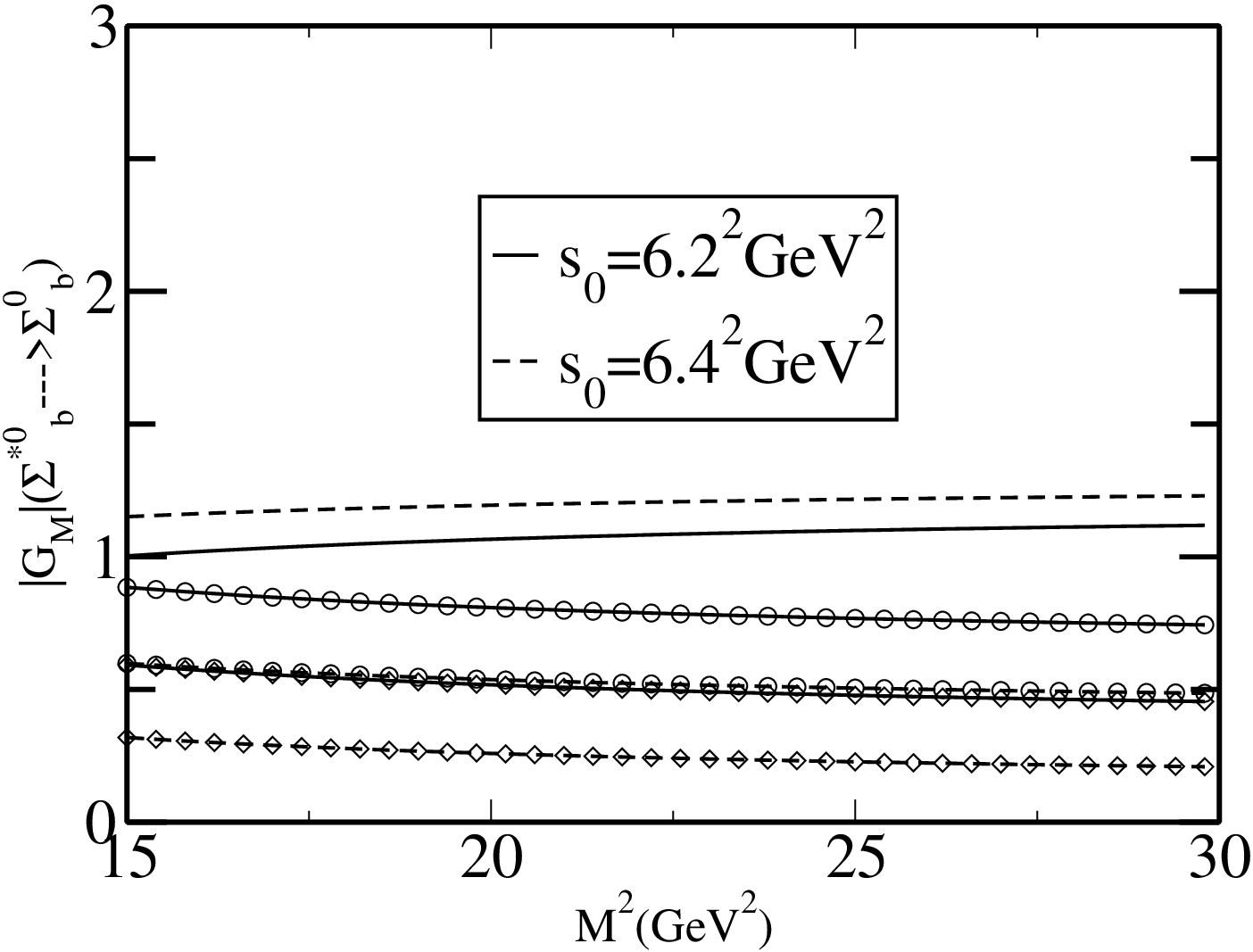}
\end{center}
\caption{The dependence of the magnetic dipole form factor  $G_{M}$ for $\Sigma^{*0}_{b}\rightarrow \Sigma^{0}_{b}$ on the
Borel mass parameter $M^{2}$ for two  fixed values of the continuum
threshold $s_{0}$. Bare lines, lines with the circles and lines with the dimonds correspond to the Ioffe currents ($\beta=-1$), $\beta=5$ and $\beta=\infty$, respectively.} \label{fig1}
\end{figure}

\begin{figure}[h!]
\begin{center}
\includegraphics[width=9cm]{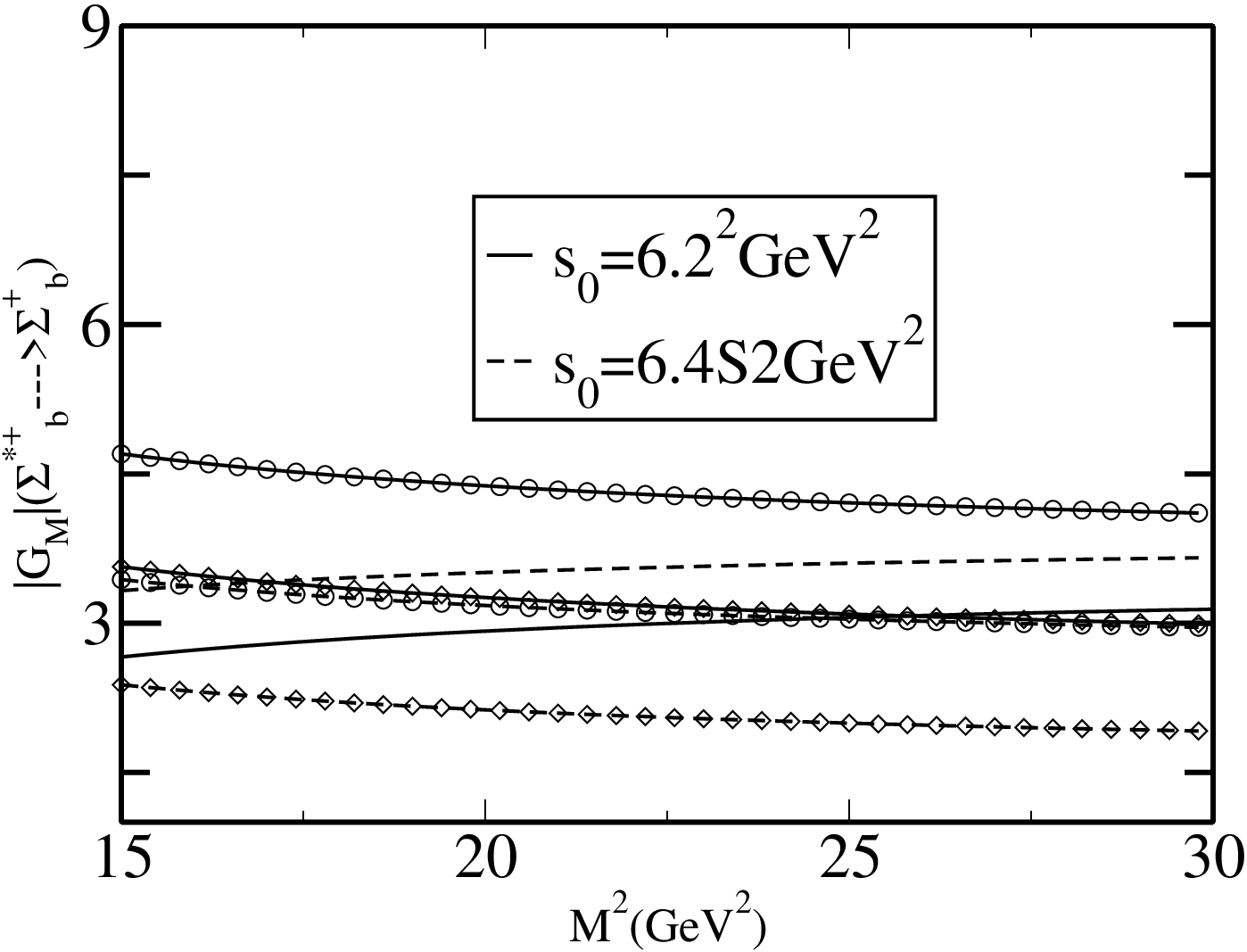}
\end{center}
\caption{The same as Fig. \ref{fig1}, but  for $\Sigma^{*+}_{b}\rightarrow \Sigma^{+}_{b}$.} \label{fig2}
\end{figure}

\begin{figure}[h!]
\begin{center}
\includegraphics[width=9cm]{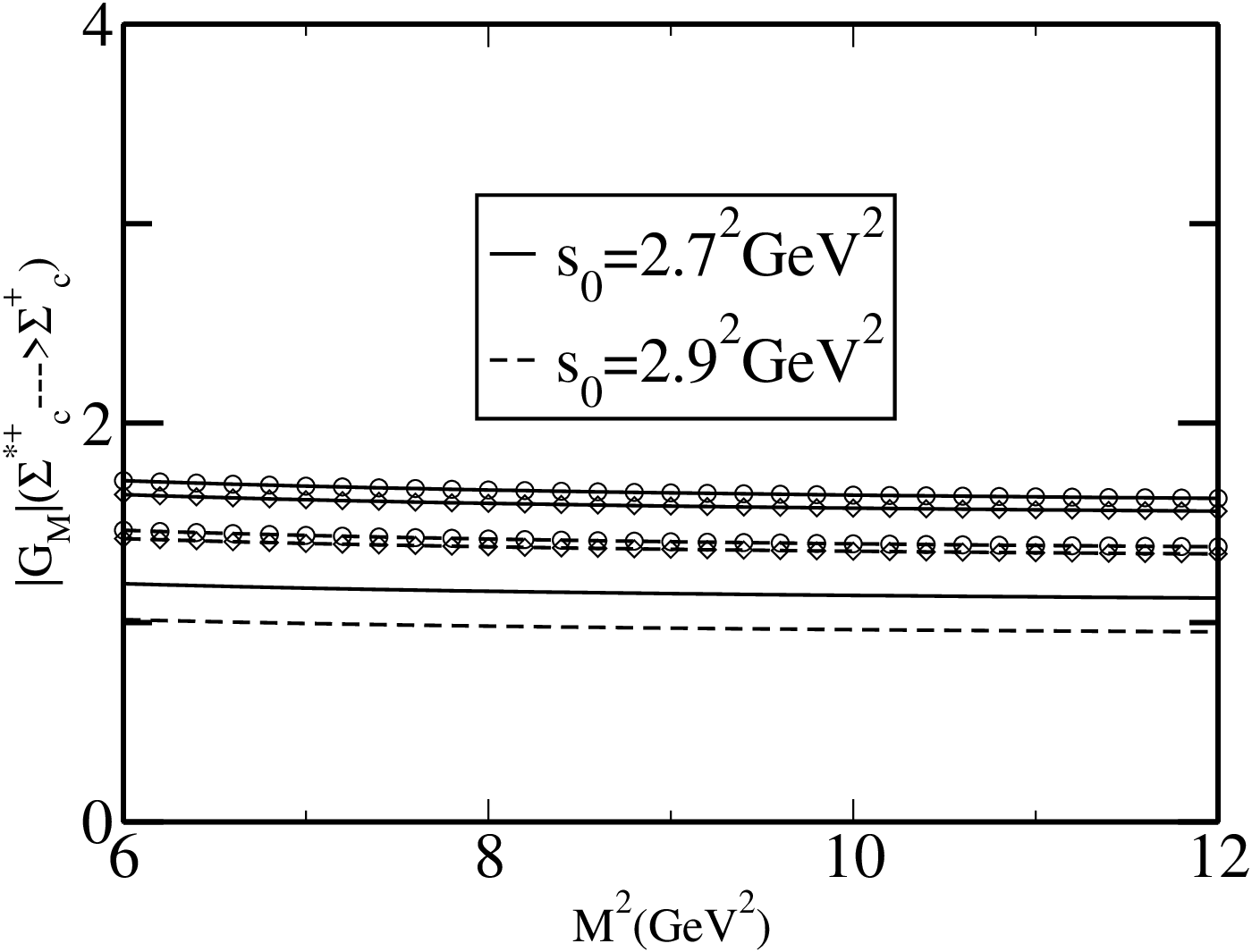}
\end{center}
\caption{The same as Fig. \ref{fig1}, but  for $\Sigma^{*+}_{c}\rightarrow \Sigma^{+}_{c}$ and   $M^{2}=6~GeV^2$ and  $M^{2}=9~GeV^2$.} \label{fig5}
\end{figure}

\begin{figure}[h!]
\begin{center}
\includegraphics[width=9cm]{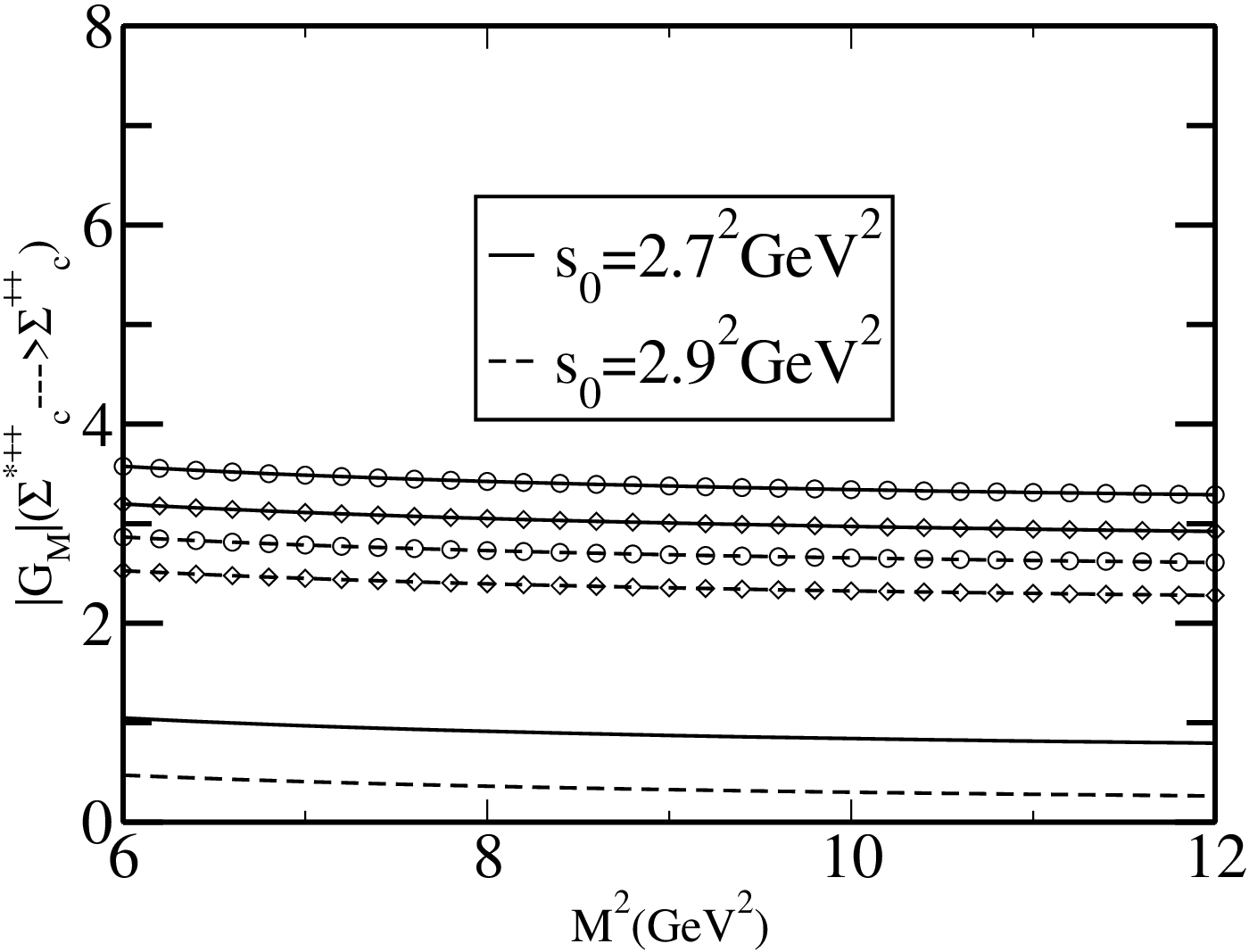}
\end{center}
\caption{The same as Fig. \ref{fig5}, but  for $\Sigma^{*++}_{c}\rightarrow \Sigma^{++}_{c}$.} \label{fig6}
\end{figure}

\begin{figure}[h!]
\begin{center}
\includegraphics[width=9cm]{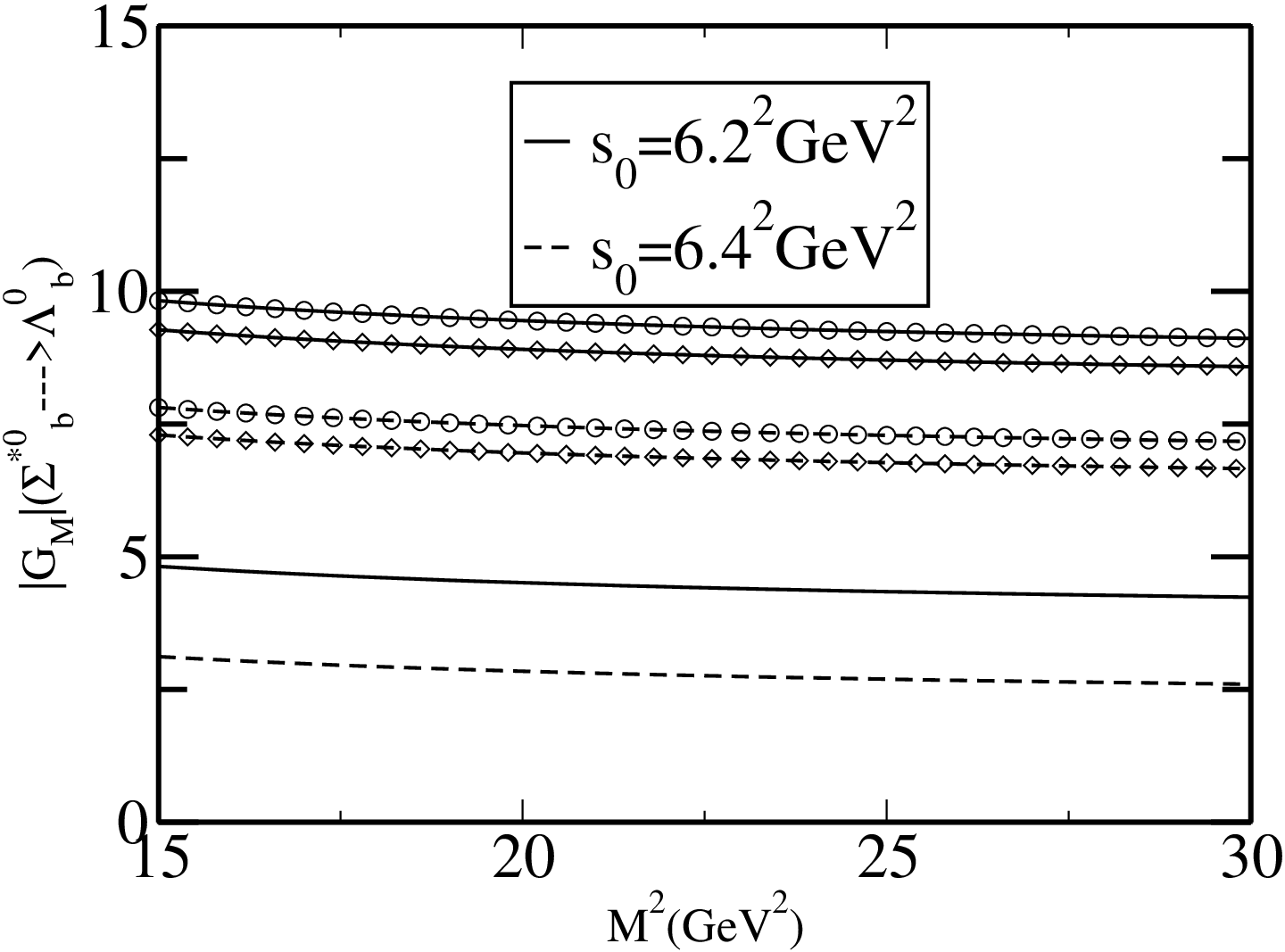}
\end{center}
\caption{The same as Fig. \ref{fig1}, but  for $\Sigma^{*0}_{b}\rightarrow \Lambda^{0}_{b}$.} \label{fig7}
\end{figure}

\begin{figure}[h!]
\begin{center}
\includegraphics[width=9cm]{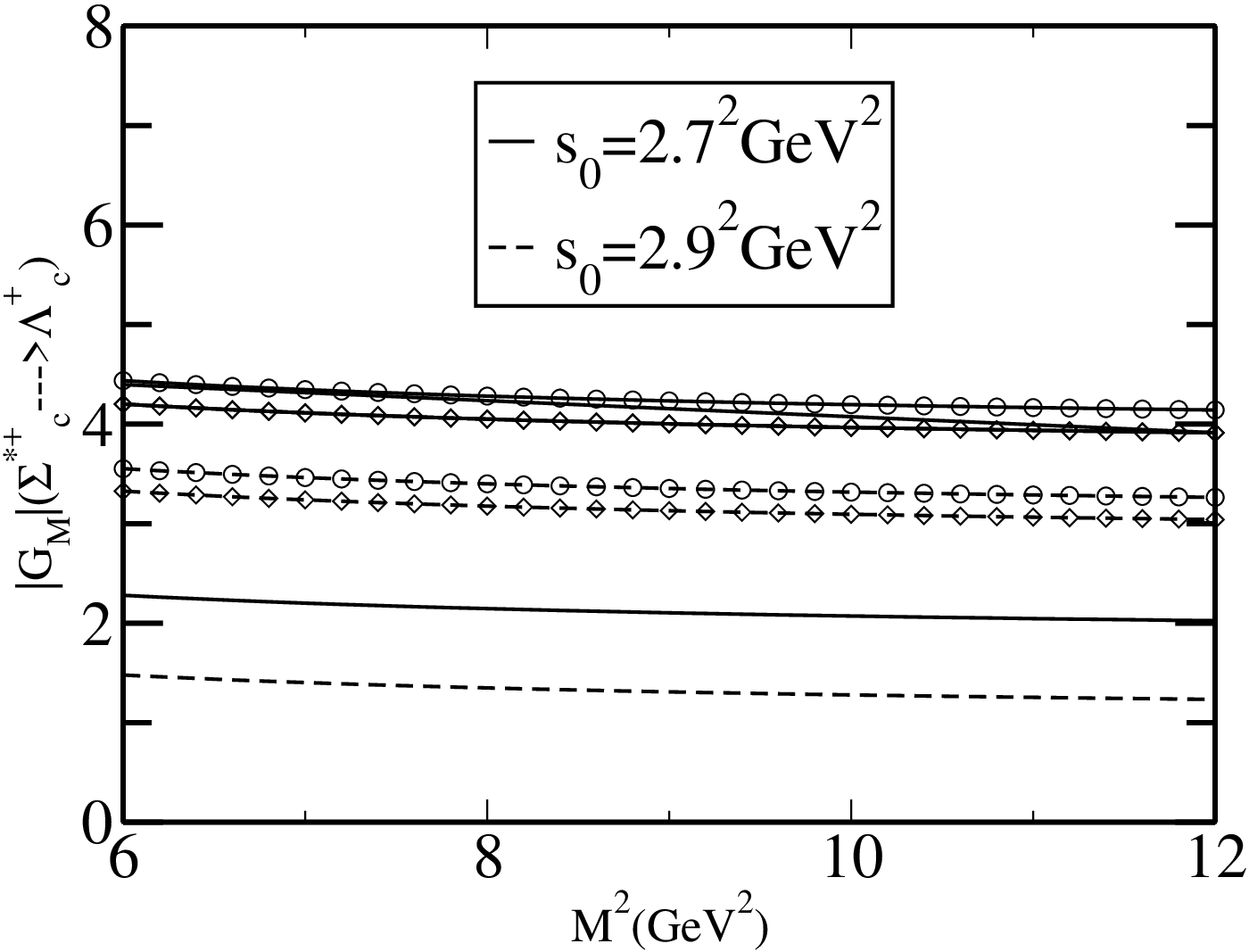}
\end{center}
\caption{The same as Fig. \ref{fig5}, but  for $\Sigma^{*+}_{c}\rightarrow \Lambda^{+}_{c}$.} \label{fig8}
\end{figure}

\begin{figure}[h!]
\begin{center}
\includegraphics[width=9cm]{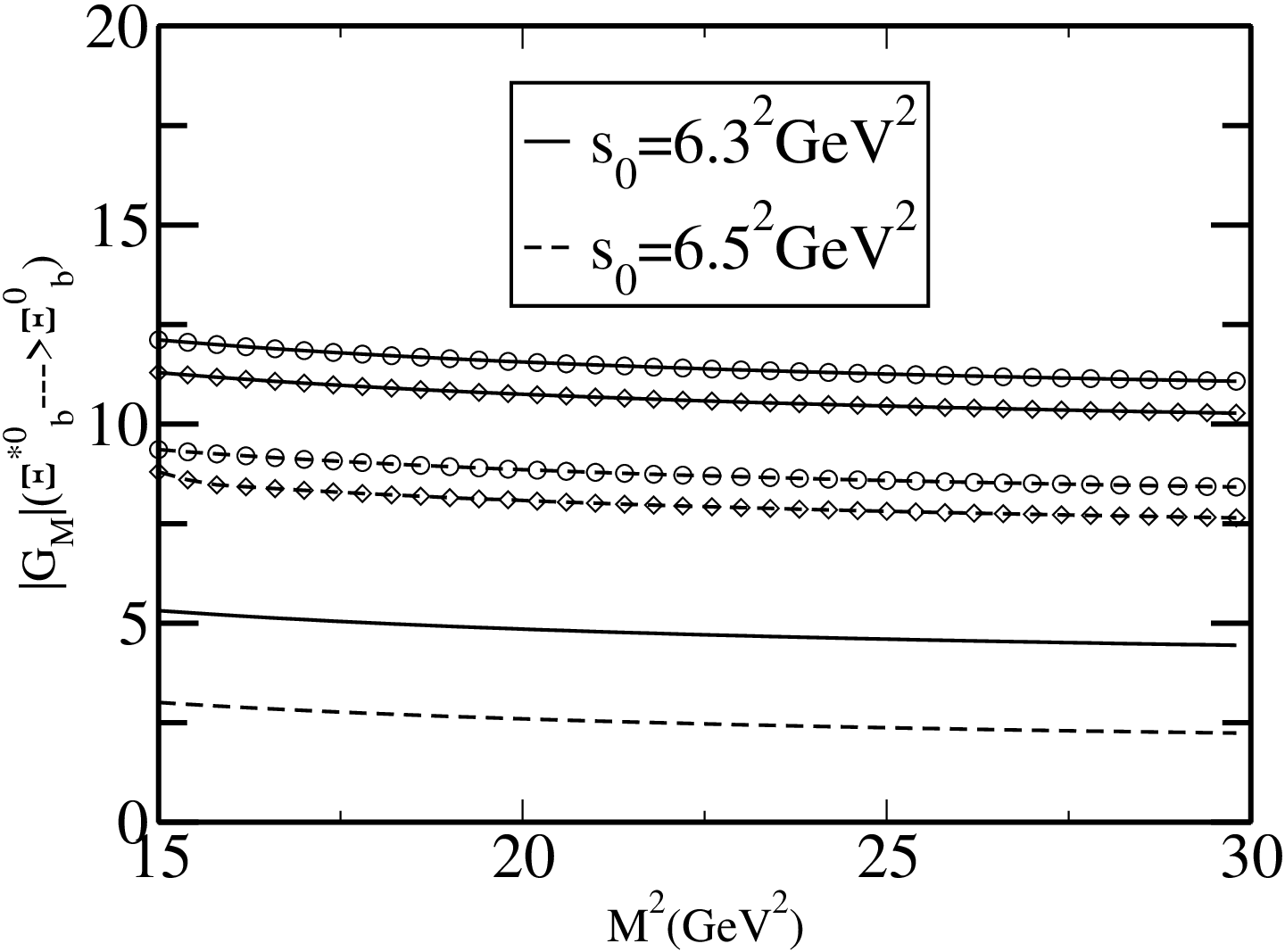}
\end{center}
\caption{The same as Fig. \ref{fig1}, but  for $\Xi^{*0}_{b}\rightarrow \Xi^{0}_{b}$.} \label{fig9}
\end{figure}

\begin{figure}[h!]
\begin{center}
\includegraphics[width=9cm]{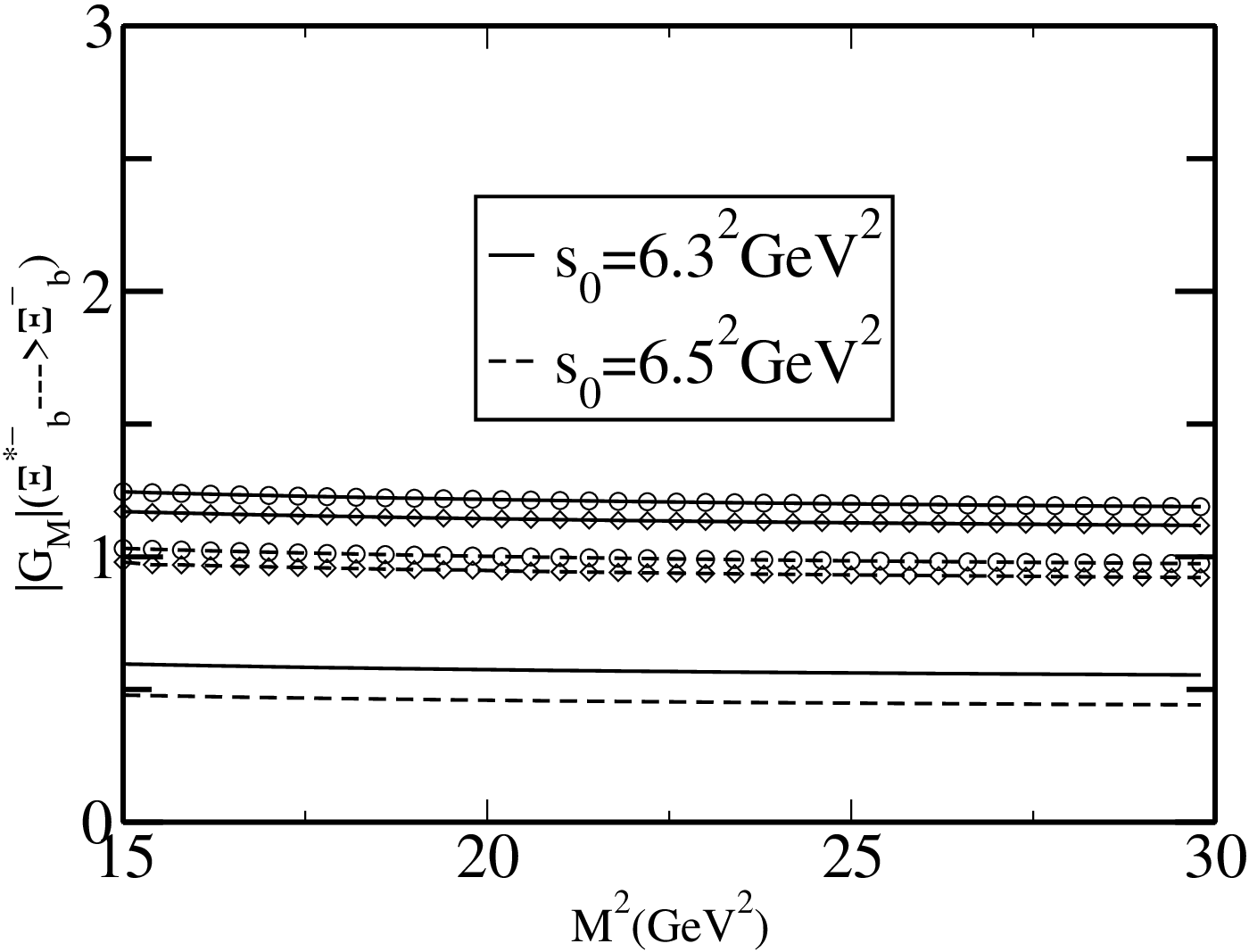}
\end{center}
\caption{The same as Fig. \ref{fig1}, but  for $\Xi^{*-}_{b}\rightarrow \Xi^{-}_{b}$.} \label{fig10}
\end{figure}

\begin{figure}[h!]
\begin{center}
\includegraphics[width=9cm]{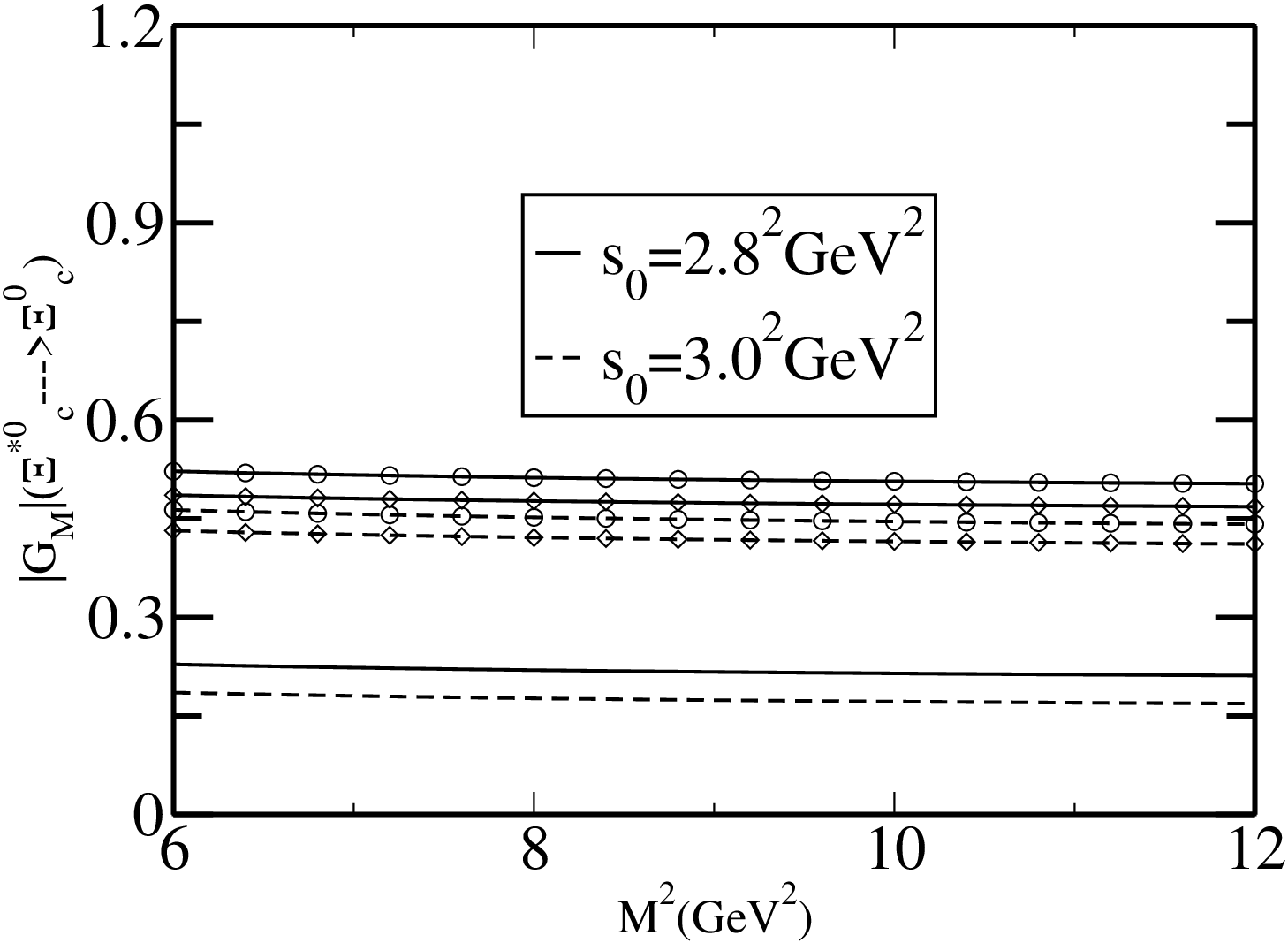}
\end{center}
\caption{The same as Fig. \ref{fig5}, but  for $\Xi^{*0}_{c}\rightarrow \Xi^{0}_{c}$.} \label{fig11}
\end{figure}

\begin{figure}[h!]
\begin{center}
\includegraphics[width=9cm]{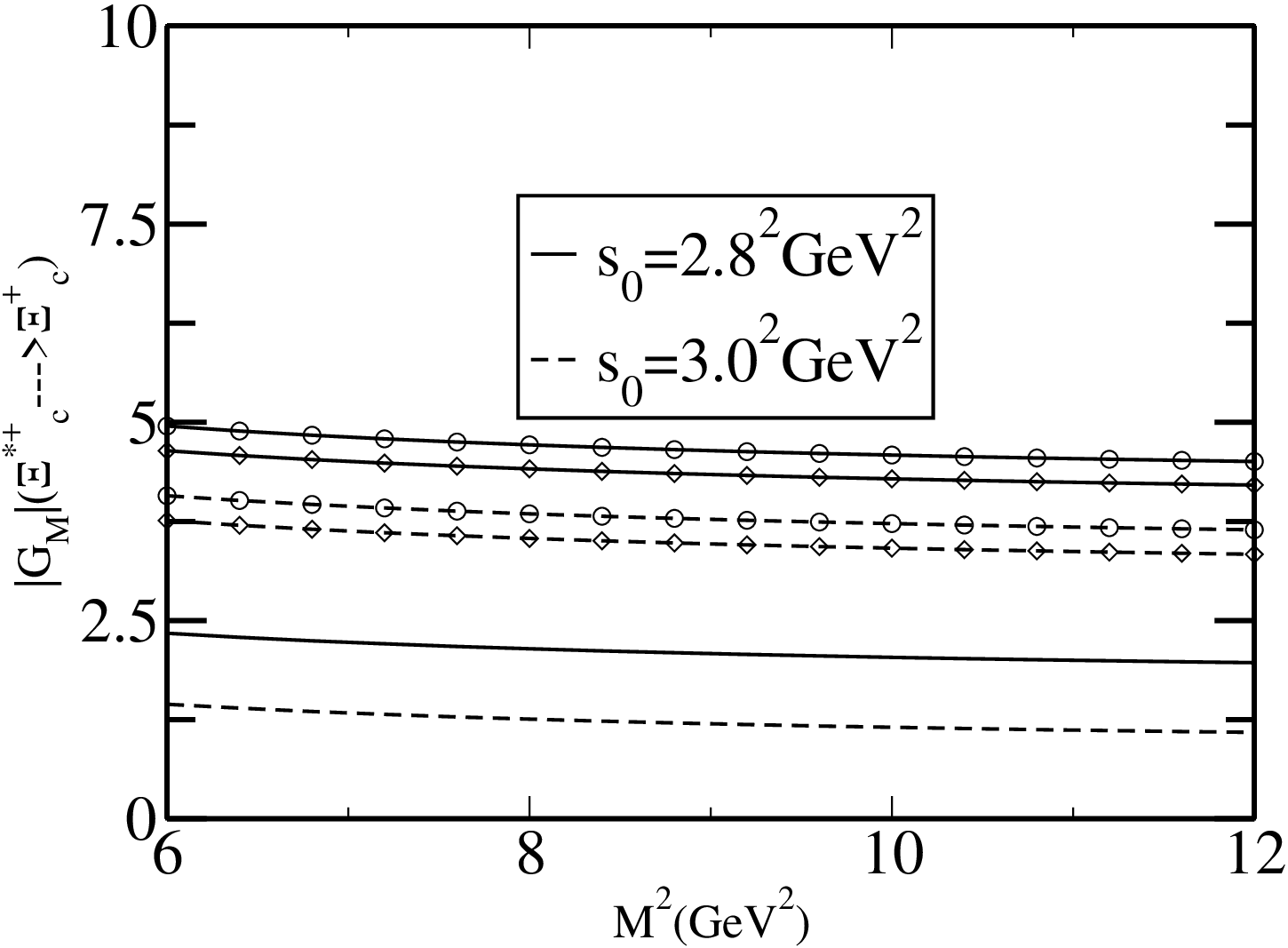}
\end{center}
\caption{The same as Fig. \ref{fig5}, but  for $\Xi^{*+}_{c}\rightarrow \Xi^{+}_{c}$.} \label{fig12}
\end{figure}

\begin{figure}[h!]
\begin{center}
\includegraphics[width=9cm]{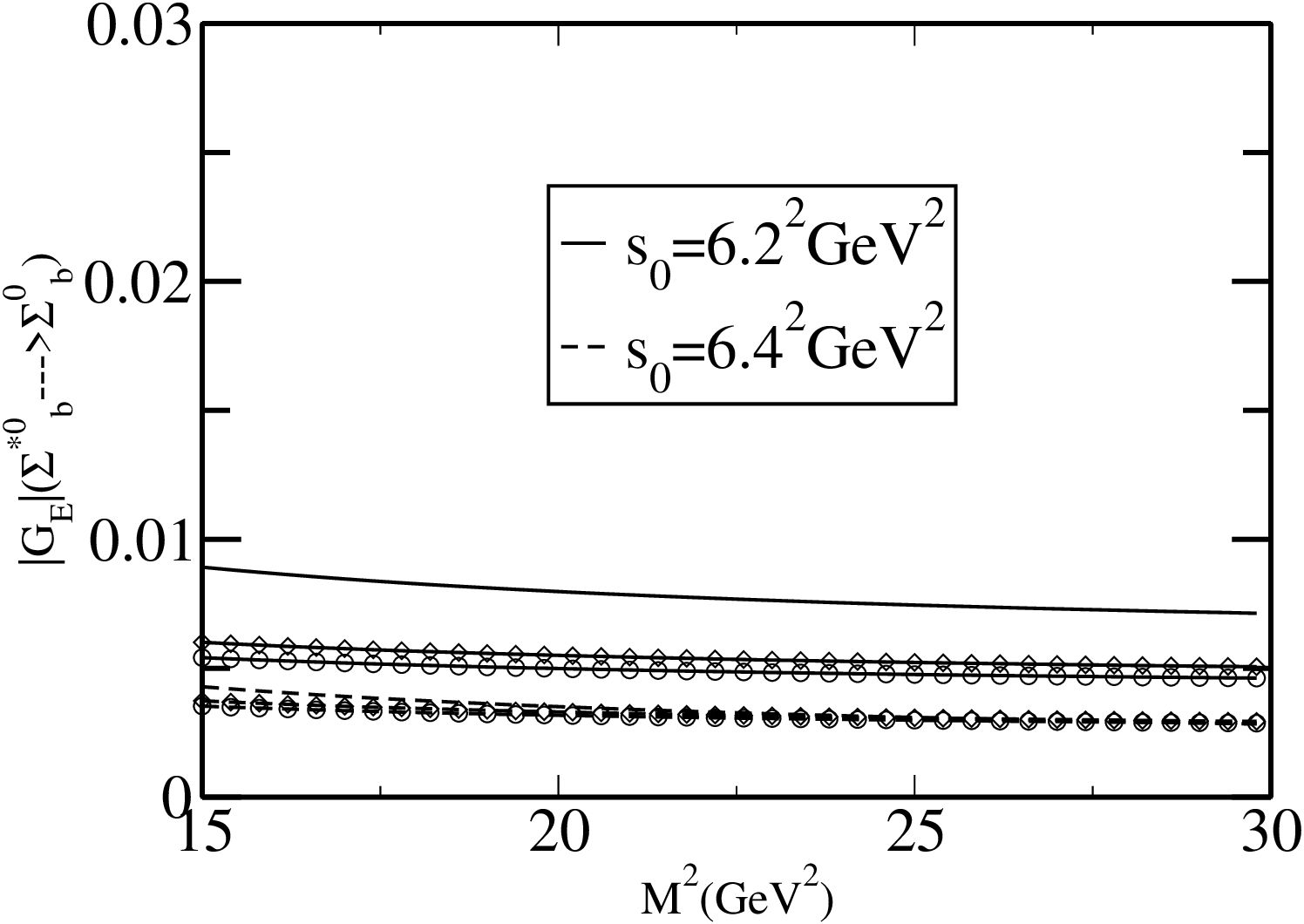}
\end{center}
\caption{The dependence of the electric dipole form factor  $G_{E}$ for $\Sigma^{*0}_{b}\rightarrow \Sigma^{0}_{b}$ on the
Borel mass parameter $M^{2}$ for two  fixed values of the continuum
threshold $s_{0}$. Bare lines, lines with the circles and lines with the dimonds correspond to the Ioffe currents ($\beta=-1$), $\beta=5$ and $\beta=\infty$, respectively.} \label{fig13}
\end{figure}

\begin{figure}[h!]
\begin{center}
\includegraphics[width=9cm]{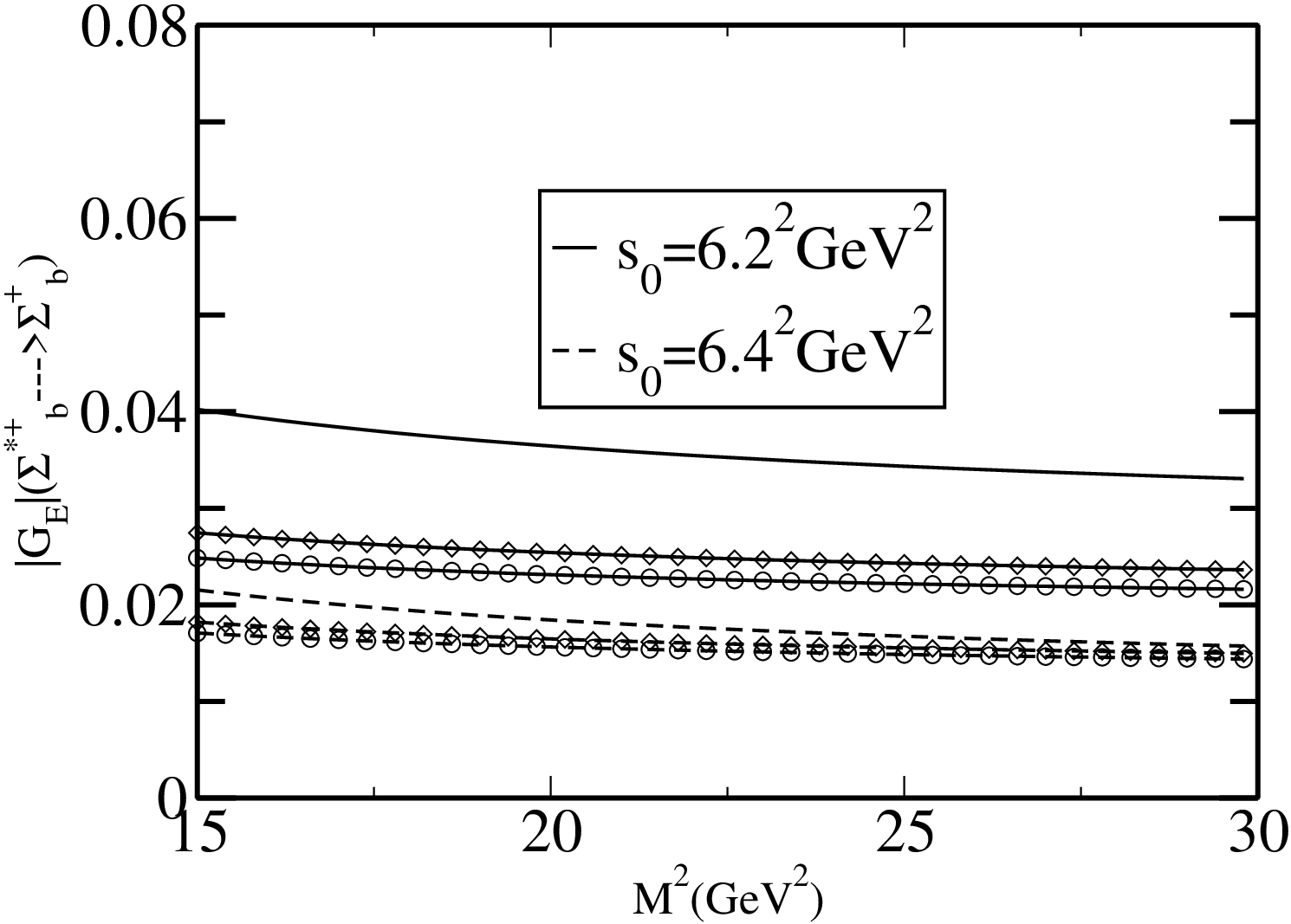}
\end{center}
\caption{The same as Fig. \ref{fig13}, but  for $\Sigma^{*+}_{b}\rightarrow \Sigma^{+}_{b}$.} \label{fig14}
\end{figure}

\begin{figure}[h!]
\begin{center}
\includegraphics[width=9cm]{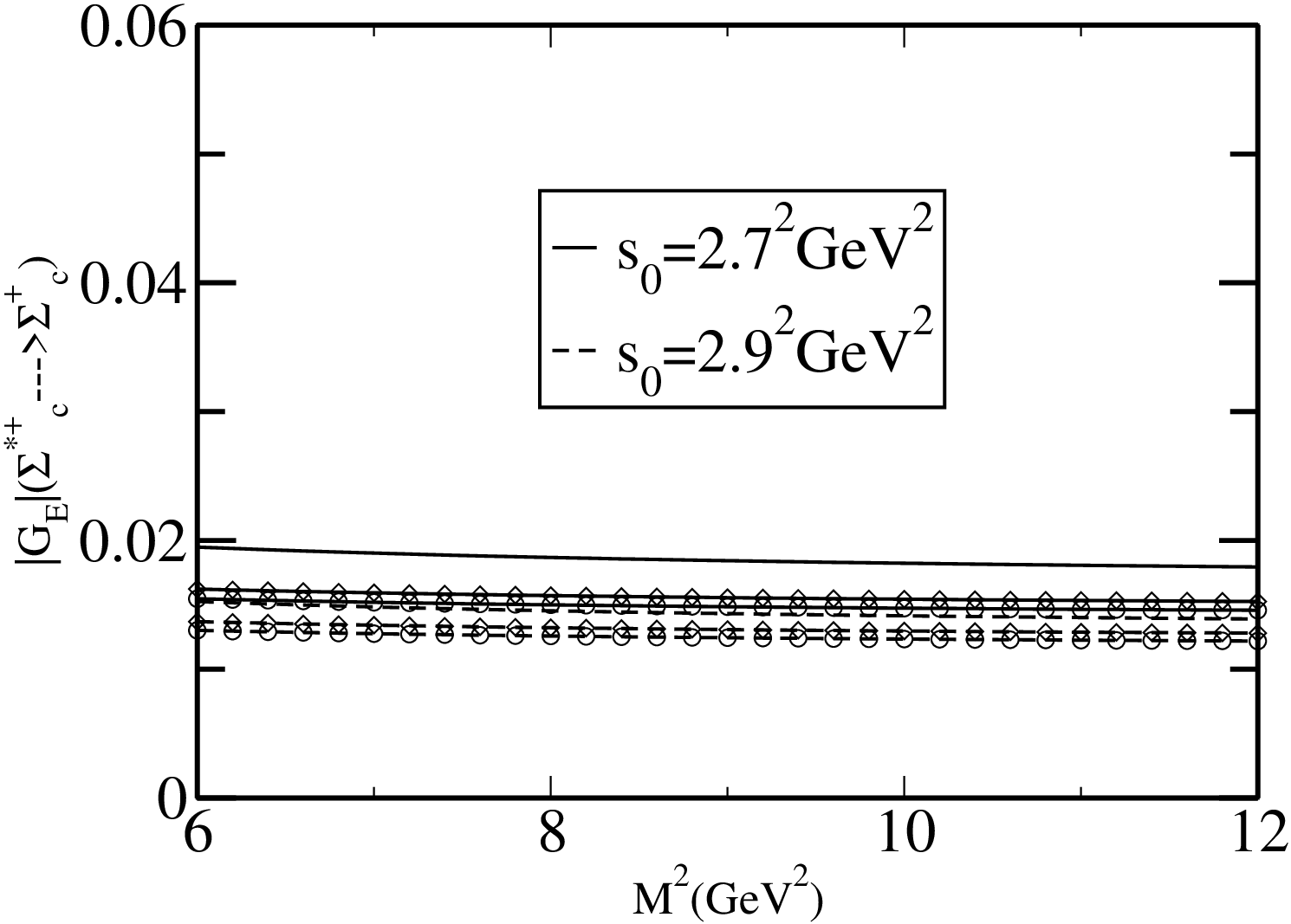}
\end{center}
\caption{The same as Fig. \ref{fig13}, but  for $\Sigma^{*+}_{c}\rightarrow \Sigma^{+}_{c}$ and   $M^{2}=6~GeV^2$ and  $M^{2}=9~GeV^2$.} \label{fig17}
\end{figure}

\begin{figure}[h!]
\begin{center}
\includegraphics[width=9cm]{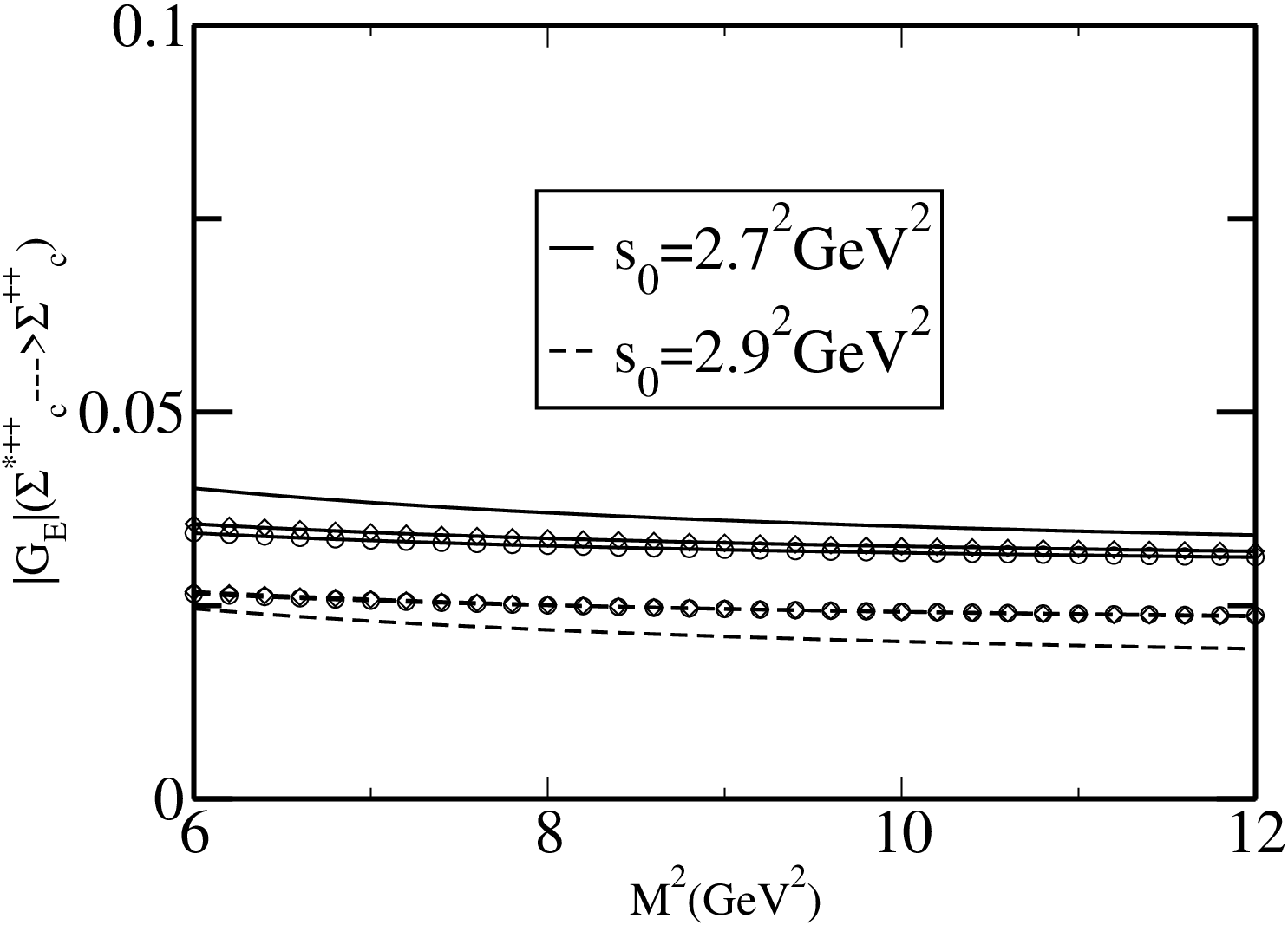}
\end{center}
\caption{The same as Fig. \ref{fig17}, but  for $\Sigma^{*++}_{c}\rightarrow \Sigma^{++}_{c}$.} \label{fig18}
\end{figure}

\begin{figure}[h!]
\begin{center}
\includegraphics[width=9cm]{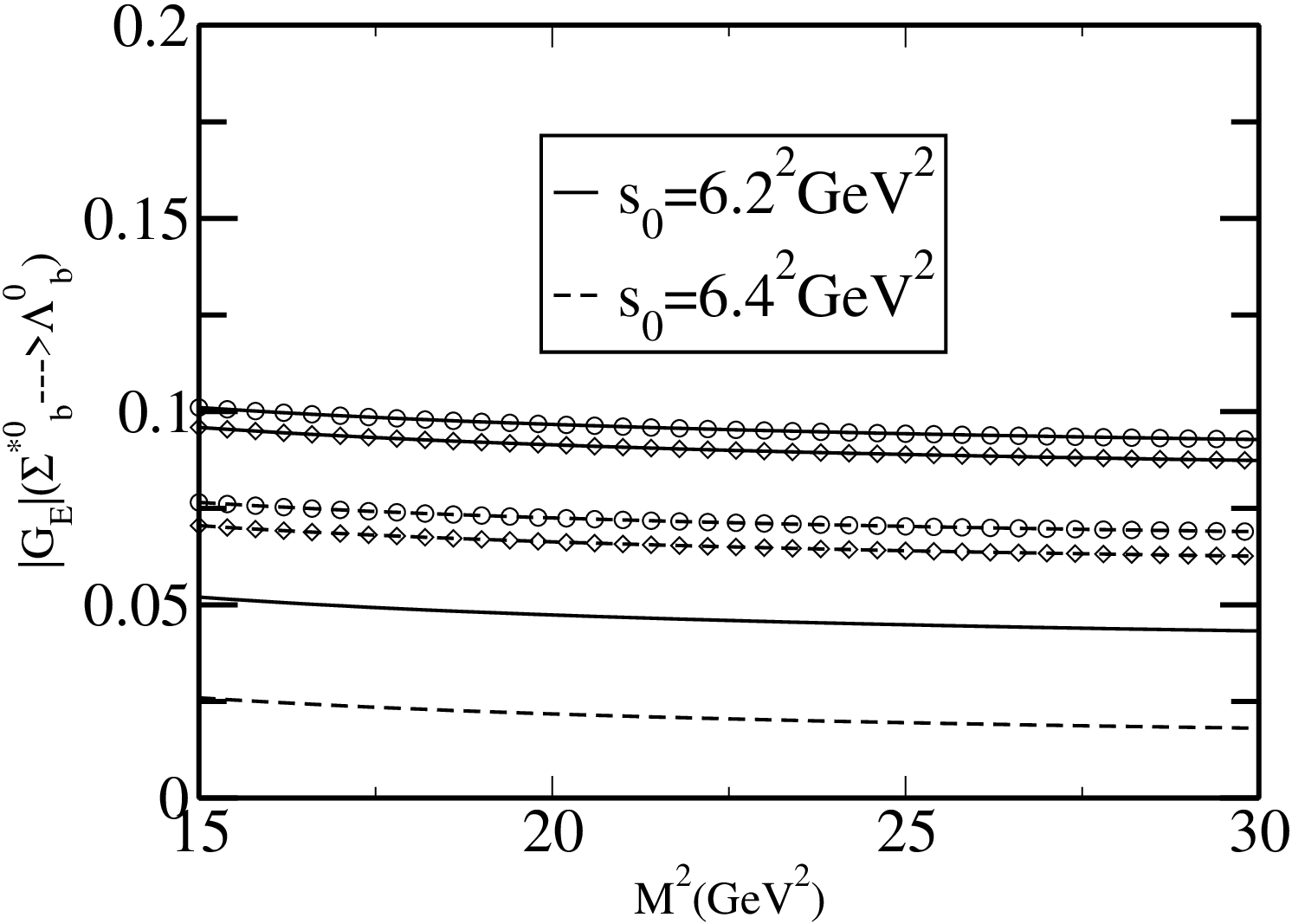}
\end{center}
\caption{The same as Fig. \ref{fig13}, but  for $\Sigma^{*0}_{b}\rightarrow \Lambda^{0}_{b}$.} \label{fig19}
\end{figure}

\begin{figure}[h!]
\begin{center}
\includegraphics[width=9cm]{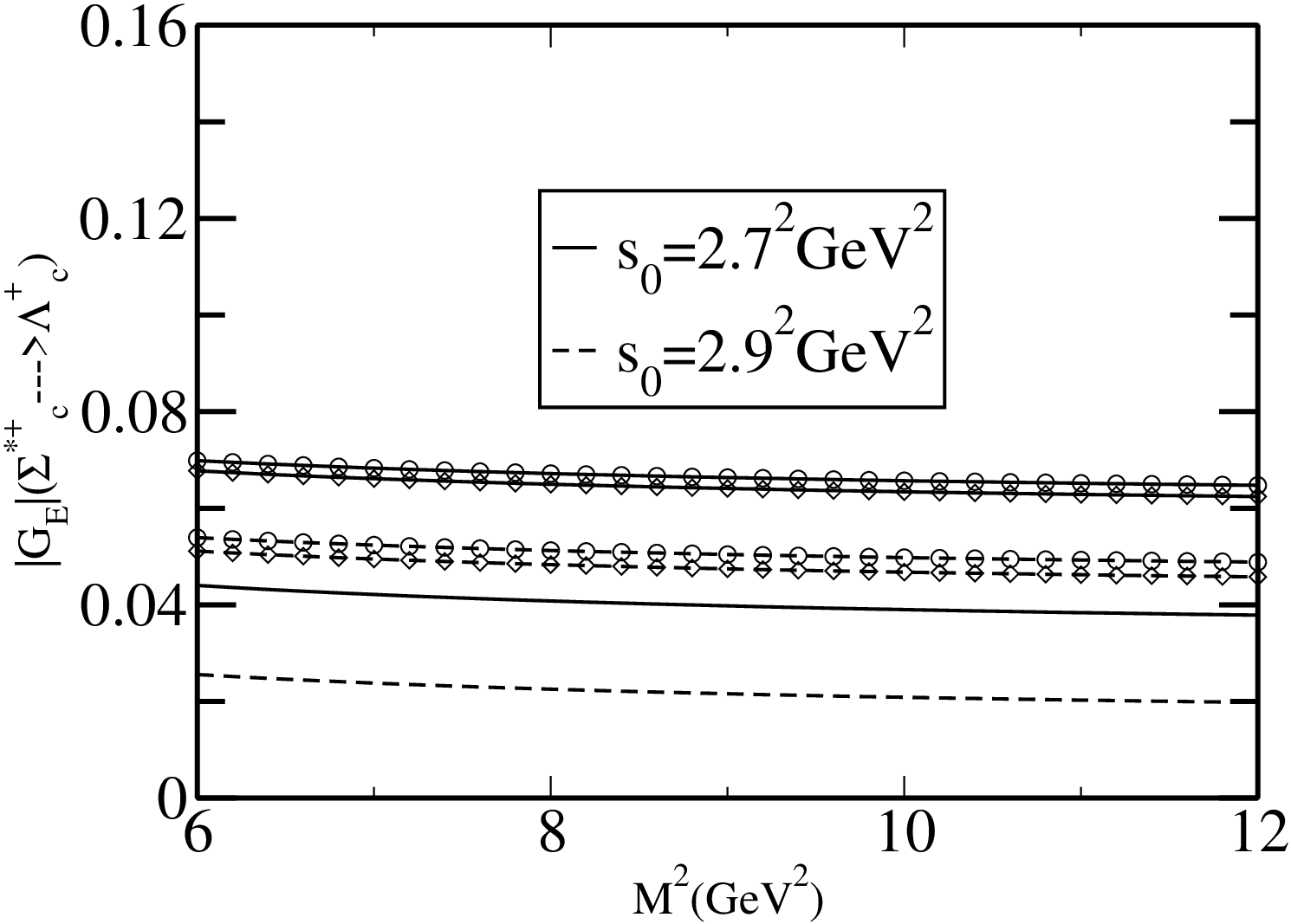}
\end{center}
\caption{The same as Fig. \ref{fig17}, but  for $\Sigma^{*+}_{c}\rightarrow \Lambda^{+}_{c}$.} \label{fig20}
\end{figure}

\begin{figure}[h!]
\begin{center}
\includegraphics[width=9cm]{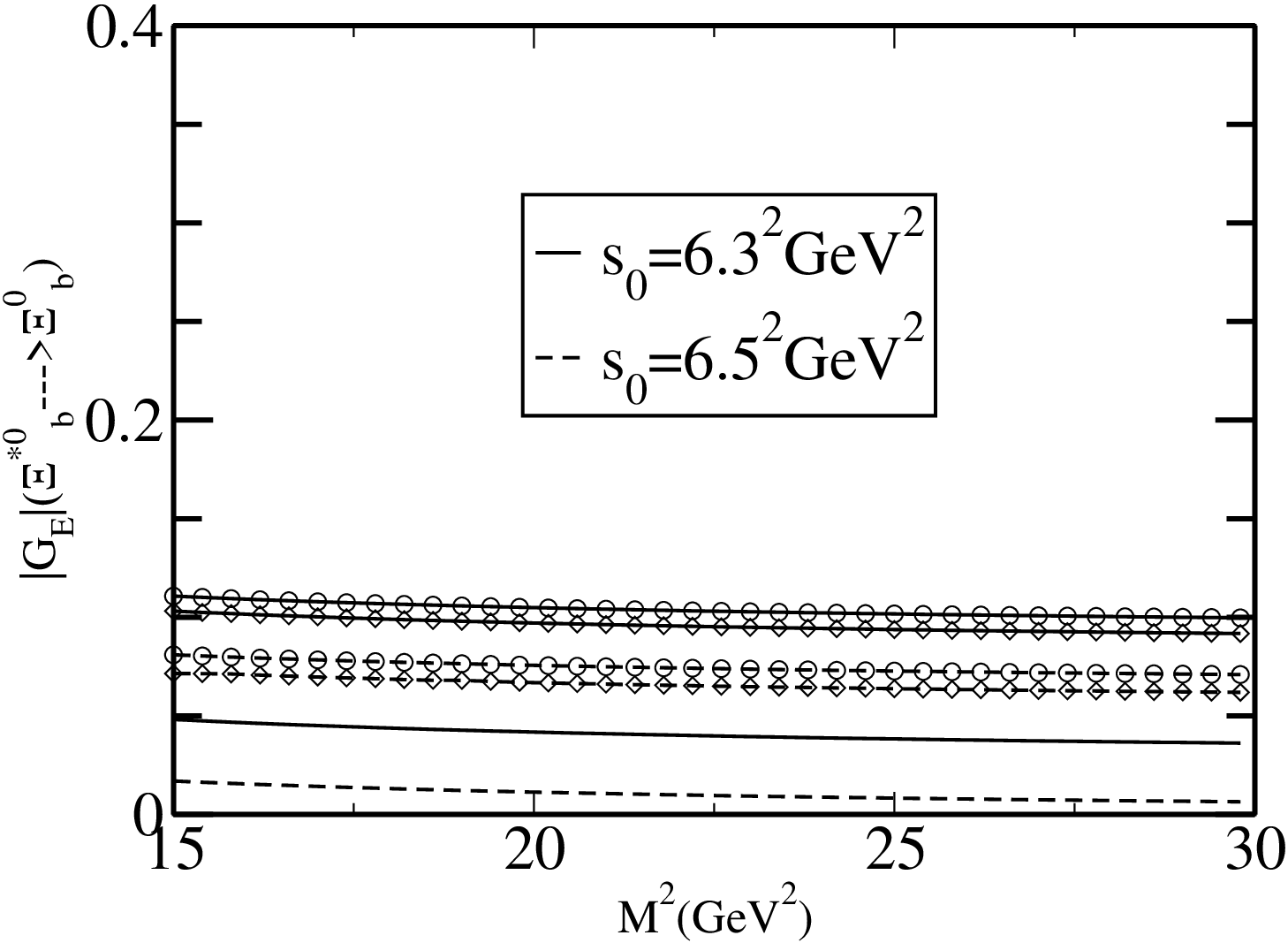}
\end{center}
\caption{The same as Fig. \ref{fig13}, but  for $\Xi^{*0}_{b}\rightarrow \Xi^{0}_{b}$.} \label{fig21}
\end{figure}

\begin{figure}[h!]
\begin{center}
\includegraphics[width=9cm]{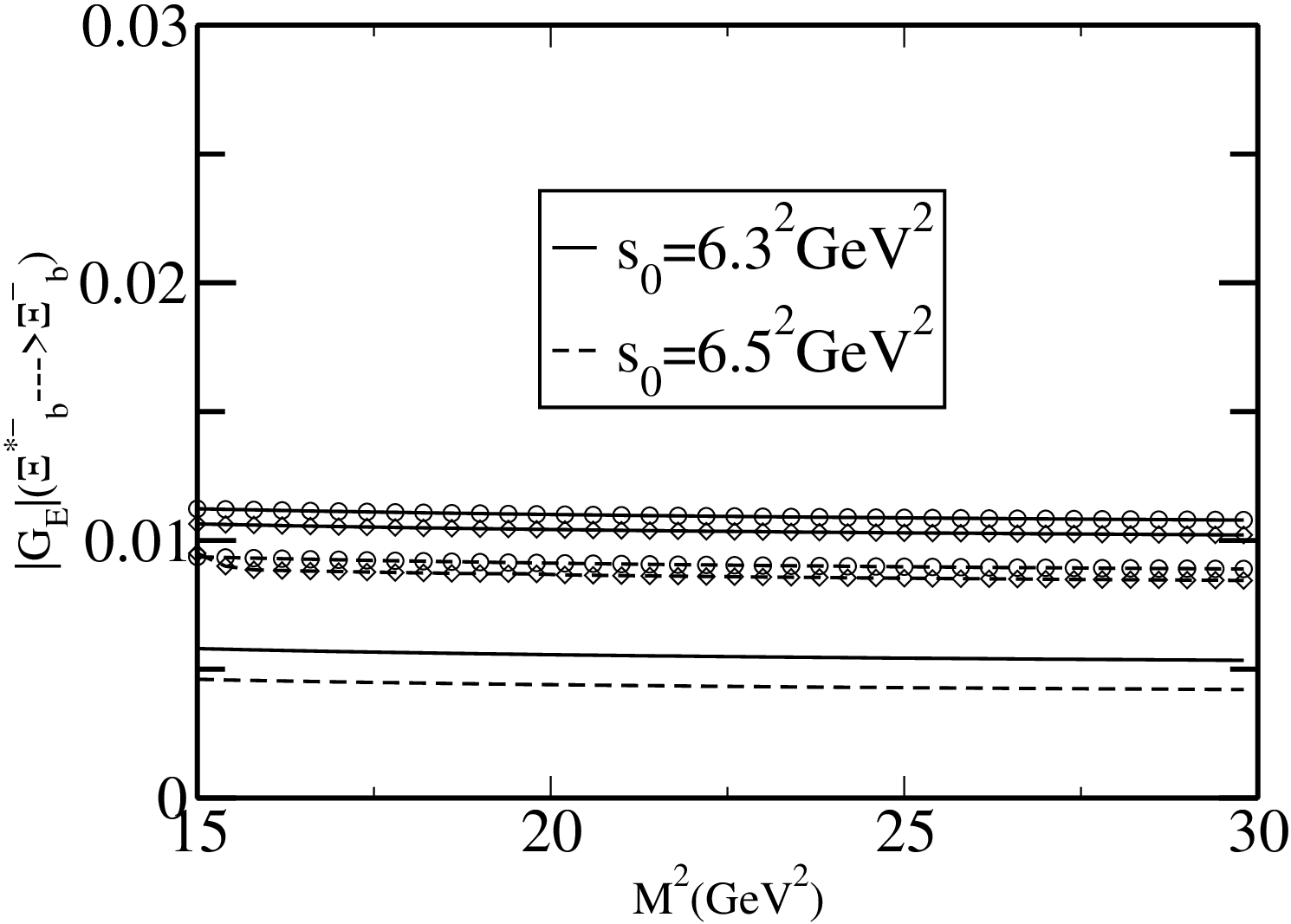}
\end{center}
\caption{The same as Fig. \ref{fig13}, but  for $\Xi^{*-}_{b}\rightarrow \Xi^{-}_{b}$.} \label{fig22}
\end{figure}

\begin{figure}[h!]
\begin{center}
\includegraphics[width=9cm]{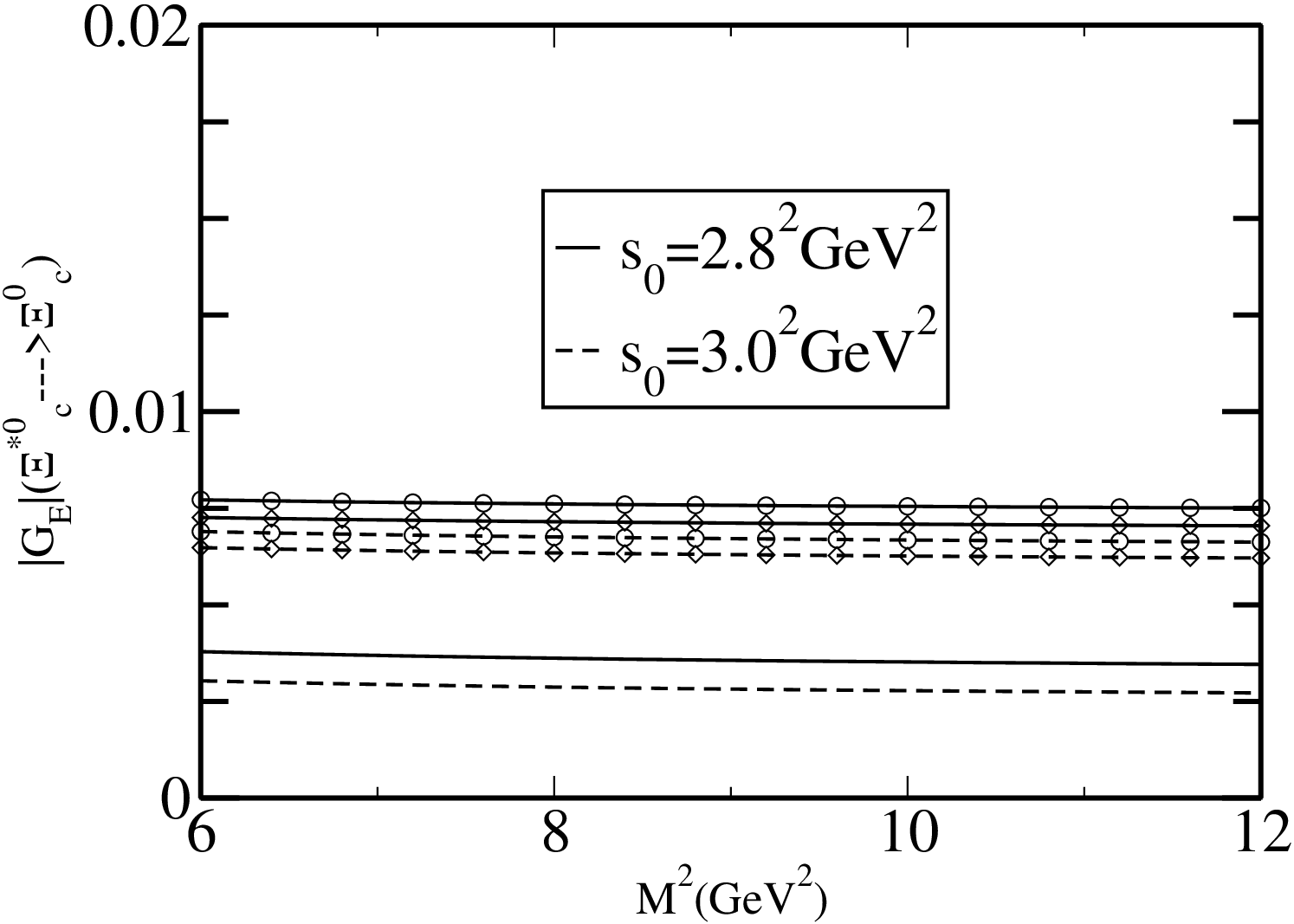}
\end{center}
\caption{The same as Fig. \ref{fig17}, but  for $\Xi^{*0}_{c}\rightarrow \Xi^{0}_{c}$.} \label{fig23}
\end{figure}

\begin{figure}[h!]
\begin{center}
\includegraphics[width=9cm]{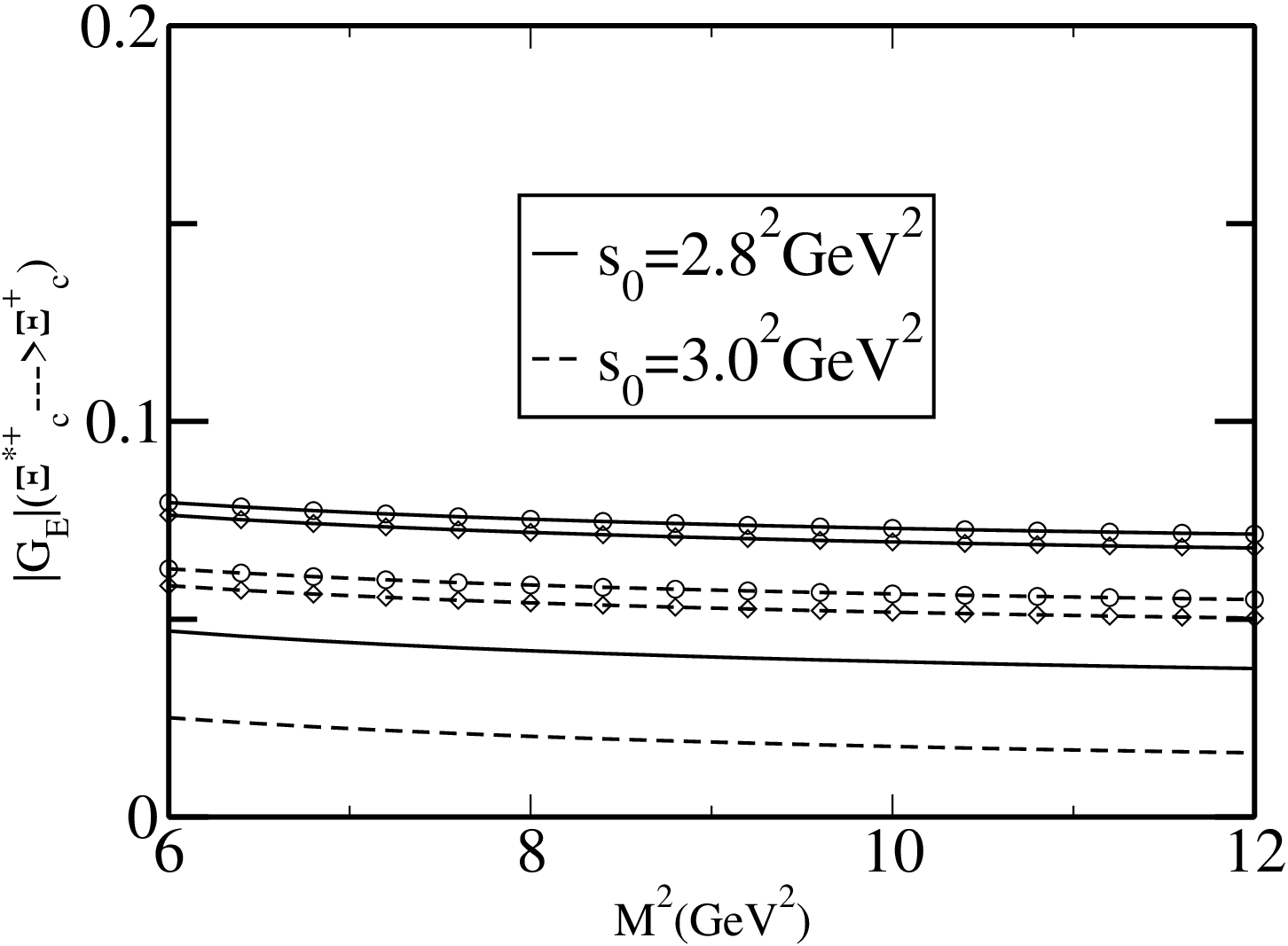}
\end{center}
\caption{The same as Fig. \ref{fig17}, but  for $\Xi^{*+}_{c}\rightarrow \Xi^{+}_{c}$.} \label{fig24}
\end{figure}

The sum rules   for the electromagnetic form factors also contain
three auxiliary parameters: the Borel mass parameter $M^2$, the
continuum threshold $s_{0}$ and the arbitrary  parameter $\beta$
entering the expression of the interpolating currents of the heavy
spin 1/2 baryons. The measurable physical quantities, i.e.  the
magnetic dipole and  electric quadrupole moments, should   be
independent of them. Therefore, we look for regions  for these
auxiliary parameters such that in these regions the $G_{M}$ and
$G_{E}$ are practically independent of them.  The working region for
$M^2$ are found requiring that not only  the contributions of the
higher states and continuum should be less than the ground state
contribution, but the highest power of $1/M^{2}$ be less than  say
$30^0/_{0}$ of the highest power of $M^{2}$. These two conditions
are both satisfied in the region $15~GeV^2\leq M^{2}\leq30~GeV^2 $
and $6~GeV^2\leq M^{2}\leq12~GeV^2 $ for baryons containing b and
c-quark, respectively. The working regions for the continuum
threshold $s_{0}$ and the general parameter $\beta$  are obtained
considering the fact that the results of the physical quantities are
approximately unchanged. We also would like to note that in Figs. \ref{fig25}-\ref{fig24}, the absolute values are plotted since it is not possible to calculate the sign of the residue from the mass sum-rules. Hence, it is not possible to predict the signs of the $G_{M}$ or $G_{E}$. The relative sign of the  $G_{M}$ and $G_{E}$ can only be predicted using the QCD sum rules.
The  dependence of the magnetic dipole moment $G_{M}$ and the
electric quadropole  moment $G_{E}$ on $cos\theta$,  where
$\beta=tan\theta$ at two fixed values of the continuum threshold
$s_{0}$ and  Borel mass square $M^2$  are depicted in Figs.
\ref{fig25}-\ref{fig48}. All presented figures have two following
common behavior: a) They becomes very large near to the end points,
i.e, $cos\theta=\pm1$ and they have zero points at some finite
values of $cos\theta$. Note that similar behavior was obtained in
analysis of the radiative decays of the light decuplet to octet
baryons  (see \cite{onbes}).  Explanation of these properties is as
follows. From the explicit forms of the interpolating currents, one
can see that the correlation, hence,
$\lambda_{1}\lambda_{2}(\beta)G_{M(E)}$  as well as
$\lambda_{2}(\beta)$ is a linear function of the $\beta$. In
general, zeros of these quantities do not coincide due to the fact
that the OPE is truncated, i.e., the calculations are not exact.
These points and any region between them   are not suitable regions
for $\beta$ and  hence the suitable regions for $\beta$ should be
far from these regions. It should be noted that in many cases, the
Ioffe current which corresponds to $cos\theta\simeq-0.71$, is out of
the working region of $\beta$.

 We also show the dependence of the magnetic dipole moment $G_{M}$ and the electric quadropole
  moment $G_{E}$ on $M^{2}$ at two fixed values of the continuum threshold $s_{0}$ and three values of the arbitrary parameter $\beta$ in
   Figs. \ref{fig1}-\ref{fig24}. From all these figures, it follows that the sum rules for $G_{M}$ and $G_{E}$  exhibit
   very good stability with respect to the $M^{2}$ in the working region.
   From these figures, we also see that the results   depend on $s_{0}$ weakly.   We should also
stress that our results practically do not change considering three
values of $\chi$ which we presented at the beginning of this
section.  Our final results on the absolute values of the  magnetic dipole and electric quadrupole moments of the considered
 radiative transitions are presented in Table 3.
The quoted errors in Table 3 are due to the uncertainties in
$m_{0}^2$, variation of $s_{0}$, $\beta$ and $M^2$ as well as errors
in the determination of the input parameters. Here, we would like to
make the following remark. From Eq. (\ref{acayip}), it follows that
the $G_{E}$ is proportional to the mass difference $m_{1}-m_{2}$ and
these masses are close to each other. Therefore, reliability of
predictions for $G_{E}$ are questionable and one can consider them
as order of magnitude. For this reason, we consider only the central
values for $G_{E}$ in Table 3.

\begin{table}[h]
\centering
\begin{tabular}{|c||c|c|}\hline
 & $|G_{M}|$ & $|G_{E}|$ \\\cline{1-3}
\hline\hline $\Sigma^{*0}_{b}\rightarrow \Sigma^{0}_{b}\gamma$
&$1.0\pm0.4$&$0.005$ \\\cline{1-3} $\Sigma^{*-}_{b}\rightarrow
\Sigma^{-}_{b}\gamma$&$2.1\pm0.7$ &$0.016$ \\\cline{1-3}
 $\Sigma^{*+}_{b}\rightarrow \Sigma^{+}_{b}\gamma$&$4.2\pm1.4$&$0.026$\\\cline{1-3}
 $\Sigma^{*+}_{c}\rightarrow \Sigma^{+}_{c}\gamma$&$1.2\pm0.2$&$0.014$\\\cline{1-3}
 $\Sigma^{*0}_{c}\rightarrow \Sigma^{0}_{c}\gamma$&$0.5\pm0.1$&$0.003$\\\cline{1-3}
$\Sigma^{*++}_{c}\rightarrow
\Sigma^{++}_{c}\gamma$&$2.8\pm0.8$&$0.030$\\\cline{1-3}
$\Xi^{*0}_{b}\rightarrow
\Xi^{0}_{b}\gamma$&$8.5\pm3.0$&$0.085$\\\cline{1-3}
$\Xi^{*-}_{b}\rightarrow
\Xi^{-}_{b}\gamma$&$0.9\pm0.3$&$0.011$\\\cline{1-3}
$\Xi^{*+}_{c}\rightarrow
\Xi^{+}_{c}\gamma$&$4.0\pm1.5$&$0.075$\\\cline{1-3}
 $\Xi^{*0}_{c}\rightarrow \Xi^{0}_{c}\gamma$&$0.45\pm0.15$&$0.007$\\\cline{1-3}
$\Sigma^{*0}_{b}\rightarrow
\Lambda^{0}_{b}\gamma$&$7.3\pm2.8$&$0.075$\\\cline{1-3}
$\Sigma^{*+}_{c}\rightarrow
\Lambda^{+}_{c}\gamma$&$3.8\pm1.0$&$0.060$\\\cline{1-3}
 \end{tabular}
 \vspace{0.8cm}
\caption{The   results for the  magnetic dipole moment $|G_{M}|$ and electric quadrupole  moment $|G_{E}|$ for the corresponding radiative decays  in units of their natural magneton.
}\label{tab:2}
\end{table}

At the end of this section, we would like to calculate the decay rate of the considered radiative transitions in terms of the multipole moments $G_E$ and $G_M$:
\begin{eqnarray}
\Gamma= 3 \frac{\alpha}{32} \frac{(m_{1}^2-m_{2}^2)^3}{m_{1}^3 m_{2}^2} \left( G_M^2 + 3 G_E^2\right)
\end{eqnarray}
The results for the decay rates are given in Table 4. In comparison,
we also present the  predictions of the constituent quark model in
$SU(2N _{f})\otimes O(3)$ symmetry \cite{oniki},  the relativistic
three-quark model \cite{onuc} and heavy quark effective theory
(HQET) \cite{ondort} in this Table. 
\begin{table}[h]
\centering
\begin{tabular}{|c||c|c|c|c|}\hline
 & $\Gamma$ (present work) & $\Gamma$ \cite{oniki}& $\Gamma$ \cite{onuc} & $\Gamma$ \cite{ondort}\\\cline{1-5}
\hline\hline $\Sigma^{*0}_{b}\rightarrow \Sigma^{0}_{b}\gamma$
&$0.028\pm0.016$&0.15&-&0.08 \\\cline{1-5}
$\Sigma^{*-}_{b}\rightarrow \Sigma^{-}_{b}\gamma$&$0.11\pm0.06$
&-&-& 0.32\\\cline{1-5}
 $\Sigma^{*+}_{b}\rightarrow \Sigma^{+}_{b}\gamma$&$0.46\pm0.22$&-&-&1.26\\\cline{1-5}
 $\Sigma^{*+}_{c}\rightarrow \Sigma^{+}_{c}\gamma$&$0.40\pm0.16$&0.22&$0.14\pm0.004$&-\\\cline{1-5}
 $\Sigma^{*0}_{c}\rightarrow \Sigma^{0}_{c}\gamma$&$0.08\pm0.03$&-&-&-\\\cline{1-5}
$\Sigma^{*++}_{c}\rightarrow
\Sigma^{++}_{c}\gamma$&$2.65\pm1.20$&-&-&-\\\cline{1-5}
$\Xi^{*0}_{b}\rightarrow
\Xi^{0}_{b}\gamma$&$135\pm65$&-&-&-\\\cline{1-5}
$\Xi^{*-}_{b}\rightarrow
\Xi^{-}_{b}\gamma$&$1.50\pm0.75$&-&-&-\\\cline{1-5}
$\Xi^{*+}_{c}\rightarrow
\Xi^{+}_{c}\gamma$&$52\pm25$&-&$54\pm3$&-\\\cline{1-5}
 $\Xi^{*0}_{c}\rightarrow \Xi^{0}_{c}\gamma$&$0.66\pm0.32$&-&$0.68\pm0.04$&-\\\cline{1-5}
$\Sigma^{*0}_{b}\rightarrow
\Lambda^{0}_{b}\gamma$&$114\pm45$&251&-&344\\\cline{1-5}
$\Sigma^{*+}_{c}\rightarrow
\Lambda^{+}_{c}\gamma$&$130\pm45$&233&$151\pm4$&-\\\cline{1-5}
 \end{tabular}
 \vspace{0.8cm}
\caption{The   results for the   decay rates of the corresponding radiative transitions  in KeV.
}\label{tab:3}
\end{table}

In summary, the transition magnetic dipole moment $G_{M}(q^2=0)$ and
electric quadrupole  moment $G_{E}(q^2=0)$ of  the  radiative decays
of the sextet heavy flavored spin $\frac{3}{2}$  to the heavy spin
$\frac{1}{2}$ baryons were calculated  within the light cone QCD
  sum rules approach. Using the obtained results for the electromagnetic  moments $G_{M}$ and $G_{E}$,
  the decay rate for these  transitions were also calculated. The
  comparison of our results on these decay rates
with the  predictions of the other approaches is also presented.
\section{Acknowledgment}
Two of the authors (K. A. and A. O.), would like to thank TUBITAK,
Turkish Scientific and Research Council, for their partial financial
support through the project number 106T333 and also through the PhD
scholarship program (K. A.).

\clearpage




\end{document}